\documentclass[aps,prd,10pt,showpacs,amsmath,amssymb,nofootinbib]{revtex4-1}
\usepackage[latin1]{inputenc}
\usepackage{axodraw4j}
\usepackage{color}
\usepackage{graphicx}
\usepackage{dcolumn}
\usepackage{bm}
\usepackage{slashed}

\begin{document}

\def\k{{\mathbf k}}
\def\q{{\mathbf q}}
\def\P{{\mathbf P}}
\def\K{{\mathbf K}}
\def\L{{\mathbf L}}
\def\x{{\mathbf x}}
\def\r{{\mathbf r}}

\title{Dipole factorization for DIS at NLO:\\
 Combining the $q\bar{q}$ and $q\bar{q}g$ contributions}
\author{Guillaume Beuf}
\email{beuf@ectstar.eu}
\affiliation{European Centre for Theoretical Studies in Nuclear Physics and Related Areas (ECT*)
and Fondazione Bruno Kessler, Strada delle Tabarelle 286, \\
 I-38123 Villazzano (TN), Italy\\}

\begin{abstract}
The NLO corrections to the DIS structure functions $F_2$ and $F_L$ (or equivalently the photon-target cross sections $\sigma^{\gamma^*}_{T}$ and $\sigma^{\gamma^*}_{L}$) at low $x_{Bj}$ are obtained, as a generalization of the dipole factorization formula. For the first time, the contributions of both the $q\bar{q}$ and the $q\bar{q}g$ Fock states in the photon are directly calculated, using earlier results \cite{Beuf:2016wdz} for the $q\bar{q}$ light-front wave-functions at one loop inside a dressed virtual photon.
Both the $q\bar{q}$ and the $q\bar{q}g$ contributions have UV divergences, which are shown to cancel each other, using conventional dimensional regularization as UV regulator.
Finally, the resummation of high-energy logarithms on top of the NLO results for $\sigma^{\gamma^*}_{T}$ and $\sigma^{\gamma^*}_{L}$ is discussed.
\end{abstract}


\maketitle


\section{Introduction}

Deep inelastic scattering (DIS) on a proton or nucleus at low Bjorken $x$ ($x_{Bj}$) provides the opportunity to study in a clean environment the transition from the dilute and purely perturbative regime of QCD into a fully nonlinear regime of QCD, while staying at weak coupling all the way if the photon virtuality $Q$ is large enough. Indeed, the evolution towards low $x_{Bj}$ is driven by the BFKL \cite{Lipatov:1976zz,Kuraev:1977fs,Balitsky:1978ic} resummation of high-energy  leading logarithms (LLs), provided that the parton density stays small. However, that evolution implies a rapid increase of the parton density, leading eventually to the nonlinear regime of gluon saturation \cite{Gribov:1984tu,Mueller:1985wy,McLerran:1993ni,McLerran:1993ka,McLerran:1994vd}, in which the BFKL evolution has to be replaced by the B-JIMWLK evolution \cite{Balitsky:1995ub,Jalilian-Marian:1997jx,Jalilian-Marian:1997gr,Jalilian-Marian:1997dw,Kovner:2000pt,Weigert:2000gi,Iancu:2000hn,Iancu:2001ad,Ferreiro:2001qy}
or, in a mean field approximation, by the Balitsky-Kovchegov (BK) equation \cite{Balitsky:1995ub,Kovchegov:1999yj,Kovchegov:1999ua}.
This transition can be most conveniently studied within the dipole factorization formalism \cite{Bjorken:1970ah,Nikolaev:1990ja,Nikolaev:1991et,Kopeliovich:1991pu,Mueller:1993rr,Mueller:1994jq,Mueller:1994gb}, valid both in the dilute and in the dense regimes, and
which provides a unified description of various observables related to DIS: not only inclusive DIS, but also semi-inclusive, diffractive or exclusive observables.

In particular, various fits \cite{Albacete:2009fh,Albacete:2010sy,Lappi:2013zma} have been performed on  the HERA data for DIS structure functions, using the leading order (LO) dipole factorization formula together with the LL resummation provided by the BK equation, including running coupling effects \cite{Balitsky:2006wa,Kovchegov:2006vj}. Most of the free parameters in these fits are associated with the shape of the dipole-target scattering amplitude at moderate energy, which is used as initial condition for the BK equation. These fits have then been used to make predictions (or postdictions) for many observables which can be expressed using the dipole-target scattering amplitude, not only in the case of DIS at HERA or at future colliders but also in the case of pp, pA or AA collisions at LHC or RHIC.
The obtained results are quite encouraging, but often lack precision, by comparison to the data.

In order to increase the precision of such phenomenological studies, the first step is to include next-to-leading order (NLO) corrections in $\alpha_s$ to the observables of interest, thus obtaining results at NLO+LL accuracy instead of LO+LL. The second step would be to push the resummation of high-energy logarithms to next-to-leading-logarithmic (NLL) accuracy,  thus obtaining results at NLO+NLL accuracy for the observables.

The priority is obviously to obtain the NLO corrections to inclusive DIS structure functions, due to the central role of these observables for the fits. These NLO corrections to DIS have been first calculated, including gluon saturation effects, in Ref.~\cite{Balitsky:2010ze}. However, the results were not provided in the standard dipole factorized form, which complicates their use in practical studies. They were also presented after linearization in the form of a NLO photon impact factor\footnote{Earlier results for the NLO photon impact factor are available in Refs.~\cite{Bartels:2000gt,Bartels:2001mv,Bartels:2002uz,Bartels:2004bi}.} \cite{Balitsky:2012bs}, suitable to use with the BFKL equation but not in the gluon saturation regime.
Then, another calculation was performed in Ref.~\cite{Beuf:2011xd}, providing directly the DIS structure functions at NLO in the dipole factorization form. However, in that paper, only the contribution from the $q\bar{q}g$ Fock state interacting with the target was explicitly calculated. By contrast, the NLO corrections to the $q\bar{q}$ Fock state contribution were guessed using a unitarity argument, which turned out to be wrong \cite{Beuf:2016wdz}. For that reason, the explicit calculation of the one-loop correction to the light-front wave-functions (LFWFs) for the $q\bar{q}$ Fock state inside a dressed virtual photon (transverse or longitudinal) has been performed in Ref.~\cite{Beuf:2016wdz}. The present article is a follow-up of that study, aiming at combining both the $q\bar{q}$ Fock state contribution at one-loop and the $q\bar{q}g$ Fock state contribution, in order to obtain the full result for the NLO corrections to the DIS structure functions in the dipole factorization formalism.

Other relevant observables for which the NLO corrections have been calculated include diffractive dijet production in DIS \cite{Boussarie:2014lxa,Boussarie:2016ogo}, exclusive light vector meson production in DIS \cite{Boussarie:2016bkq}, single inclusive forward hadron production in pp or pA collisions \cite{Chirilli:2011km,Chirilli:2012jd,Altinoluk:2015vax}, and inclusive photon production at central rapidities in pA collisions \cite{Benic:2016yqt,Benic:2016uku}.

In principle, the NLL resummation could be attempted in practice, since all the relevant evolution equations have been derived at that accuracy: not only the BFKL equation at NLL \cite{Fadin:1998py,Ciafaloni:1998gs}, but also the BK equation \cite{Balitsky:2008zz,Balitsky:2009xg} and even the B-JIMWLK evolution at NLL \cite{Balitsky:2013fea,Lublinsky:2016meo}.
However, it is well known that the naive perturbative expansion of the high-energy evolution equations is not reliable, and require collinear resummations \cite{Salam:1998tj}. The dominant part of the collinear resummation amounts to treat the kinematics in a more consistent way already at LL accuracy. This part of the collinear resummation has been performed for the BK equation both in Refs.~\cite{Beuf:2014uia} and \cite{Iancu:2015vea}, after a preliminary study in Ref.~\cite{Motyka:2009gi}.
As an application, fits to DIS have been redone at LO+LL accuracy with these kinematically consistent versions of the BK equation \cite{Iancu:2015joa,Albacete:2015xza}.
Moreover, a simple prescription providing a partial treatment of the subdominant part of the collinear resummation has been proposed in Ref.~\cite{Iancu:2015joa}. Including all of these modifications, the numerical simulations of the BK equation at NLL are now stable \cite{Lappi:2016fmu}, by contrast to the simulations at naive NLL accuracy without collinear resummations, which are giving absurd results \cite{Lappi:2015fma}. Hence, NLL BK resummations should now be feasible in practice, provided the required computing power is available.


The plan of this paper is as follows. In sec.~\ref{sec:DIS_NLO_general},
the DIS structure functions (or equivalently the photon-target cross sections) are shown to admit, within the eikonal approximation, a generalized dipole factorization formula beyond LO, Eq.~\eqref{sigma_gamma_master_eq}, involving the LFWFs for the Fock states inside the incoming dressed photon and multipole-target scattering amplitudes. The NLO $q\bar{q}$ contribution to the photon-target cross sections is then obtained from the result of Ref.~\cite{Beuf:2016wdz} for the one-loop $q\bar{q}$ LFWFs inside a dressed photon.
In sec.~\ref{sec:qqbarg_WF}, the $q\bar{q}g$ LFWFs at tree-level are rederived within light-front perturbation theory (LFPT), but this time in arbitrary dimension, in order to facilitate their combination with the one-loop calculations of Ref.~\cite{Beuf:2016wdz} which are done in conventional dimensional regularization (CDR).
Taking the square of the LFWFs and combining the $q\bar{q}$ and the $q\bar{q}g$ contributions together, the full NLO results for the photon-target cross sections are obtained in sec.~\ref{sec:sigma_L_NLO} in the longitudinal photon case (see Eqs.~\eqref{sigma_L_NLO_fixed}, \eqref{sigma_L_dipole_frac}, \eqref{sigma_L_q_to_g_frac} and \eqref{sigma_L_qbar_to_g_frac}), and in sec.~\ref{sec:sigma_T_NLO} in the transverse photon case (see Eqs.~\eqref{sigma_T_NLO_fixed}, \eqref{sigma_T_dipole_frac}, \eqref{sigma_T_q_to_g_frac} and \eqref{sigma_T_qbar_to_g_frac}). Special care is taken to demonstrate the cancellation of the UV divergence between the $q\bar{q}$ and $q\bar{q}g$ contributions to the cross sections.
The issue of the LL (and even NLL) resummation for the NLO photon-target cross sections is discussed in sec.~\ref{sec:LL_resum}. Various possible prescriptions to implement that high-energy resummation are proposed, see Eqs.~\eqref{sigma_TL_NLO_LL_unsub} and \eqref{sigma_TL_NLO_LL_sub}.
Conclusions are presented in sec.~\ref{sec:conclusion}. Finally, the integrals required to perform the Fourier transform of the $q\bar{q}g$ LFWFs from transverse momentum to transverse position space are studied in appendix~\ref{sec:FTs}.


\section{DIS cross-section at NLO: general structure and $q\bar{q}$ contributions\label{sec:DIS_NLO_general}}


\subsection{Photon-target total cross-sections and structure functions\label{sec:struct_funct}}

From the momenta $P^{\mu}$ of the target, $k^{\mu}$ of the incoming lepton, and the momentum $q^{\mu}$ lost by the lepton, one constructs the usual Lorentz-invariant variables to describe the DIS process: the virtuality $Q^2=-q^2$, the Mandelstam $s$ variable of the lepton-target collision defined as $s=(P+k)^2$, the Bjorken $x_{Bj}$ variable defined as $x_{Bj}=Q^2/(2 P\!\cdot\! q)$ and the inelasticity $y=(2 P\!\cdot\! q)/s=Q^2/(s\: x_{Bj})$.

In the one-photon exchange approximation and neglecting the lepton mass, the neutral current DIS cross section can then be written as
\begin{eqnarray}
\frac{d\sigma^{l+p\rightarrow l+X}}{dx_{Bj}\, d ^2Q}
&=& \frac{\alpha_{em}}{\pi x_{Bj} Q^2}\left[\left(1\!-\!y\!+\!\frac{y^2}{2}\right) \; \sigma^{\gamma^*}_{T}\!(x_{Bj},Q^2)
+ \left(1\!-\!y\right)\; \sigma^{\gamma^*}_{L}\!(x_{Bj},Q^2) \right]
\label{DIS_xsect_one_photon}
\, ,
\end{eqnarray}
where $\sigma^{\gamma^*}_{T,L}$ is the total cross section for the scattering of a transverse or longitudinal virtual photon on the target. The purpose of the present study is to derive the NLO QCD corrections to $\sigma^{\gamma^*}_{T,L}$, which correspond to the terms of order $\alpha_{em} \alpha_{s}$ in $\sigma^{\gamma^*}_{T,L}$.

The parametrization \eqref{DIS_xsect_one_photon} of the DIS cross-section is fully equivalent to the more familiar ones in terms of structure functions. Indeed, the total photon cross-sections are related to the transverse and longitudinal structure functions as
\begin{eqnarray}
\sigma^{\gamma^*}_{T,L}(x_{Bj},Q^2)&=& \frac{(2\pi)^2 \alpha_{em}}{Q^2}\; F_{T,L}(x_{Bj},Q^2)
\, ,
\label{rel_sigmaTL_FTL}
\end{eqnarray}
the latter being related to the $F_1$ and $F_2$ structure functions as
\begin{eqnarray}
F_{1} &=& \frac{1}{2\, x_{Bj}}\: F_{T} \\
F_{2}&=&F_{T}+F_{L}
\label{rel_FTL_F12}
\, .
\end{eqnarray}


\subsection{Dipole factorization for the photon total cross-sections\label{sec:dipole_fact}}

At low $x_{Bj}$, the DIS process is driven by the gluons with low momentum fraction in the target, which form a semi-classical gluons field shrinking to a shockwave due to Lorentz contraction. In that regime, the DIS cross-section on the target can be obtained by first calculating the scattering on a arbitrary classical gluon field shockwave, and then taking a particular statistical average over the random classical gluon field \cite{McLerran:1993ni,McLerran:1993ka,McLerran:1994vd}, in such a way that the scattering cross section on  the physical target is reproduced, up to power suppressed corrections at high-energy. The statistical averaging over the target field will be included in sec.~\ref{sec:LL_resum}, and until then, only the scattering on a given classical gluon shockwave field is considered.

According to the optical theorem, the total cross-section for a given photon polarization\footnote{$\lambda$ either corresponds to the longitudinal polarization, or to one of the transverse polarizations. The transverse photon cross section $\sigma^{\gamma^*}_{T}$ entering in Eq.~\eqref{DIS_xsect_one_photon} is the average of $\sigma^{\gamma^*}_{\lambda}$ over the transverse polarizations. Note that physically, there are two of them, but in the conventional dimensional regularization (CDR) used in the present calculation, there are $D\!-\!2$ of them. That number can be different in other variants of dimensional regularization.} $\lambda$ is related to the imaginary part of the forward elastic scattering amplitude as
\begin{eqnarray}
\sigma^{\gamma^*}_{\lambda}
= 2\, \textrm{Im}\: {\cal M}^{\textrm{fwd}}_{\gamma_{\lambda}^*\rightarrow \gamma_{\lambda}^*}
= 2\, \textrm{Re}\: (-i)\, {\cal M}^{\textrm{fwd}}_{\gamma_{\lambda}^*\rightarrow \gamma_{\lambda}^*}
\, ,
\end{eqnarray}
where the forward elastic amplitude for the photon scattering on the gluon shockwave is defined within in light-front quantization as
\begin{eqnarray}
\Big\langle\gamma_{\lambda}^*({q'}^+, \q'=0; Q^2)_{H}\Big|
\left(\hat{S}_E - {\mathbf{1}}\right)
\Big|\gamma_{\lambda}^*(q^+, \q=0; Q^2)_{H}\Big\rangle
&=& (2q^+)\, 2\pi \delta({q'}^+\!-\!q^+) \: i\,
{\cal M}^{\textrm{fwd}}_{\gamma_{\lambda}^*\rightarrow\gamma_{\lambda}^*}
\, ,
\label{def_forward_ampl}
\end{eqnarray}
up to power-suppressed corrections at high-energy \cite{Bjorken:1970ah}. In the definition \eqref{def_forward_ampl},
$\Big|\gamma_{\lambda}^*(q^+, \q; Q^2)_{H}\Big\rangle$ correspond to the one-virtual-photon dressed state in the Heisenberg picture for QCD+QED in the absence of external classical field \cite{Beuf:2011xd}.
The operator $\hat{S}_E$ describes the eikonal scattering on the classical gluon shockwave. Its action is best described on Fock-states in mixed-space, in which the kinematics of a particle is described by its light-cone momentum $k^+$ and transverse position $\x$. Indeed, upon eikonal scattering on the shockwave, each parton in mixed-space simply picks up a Wilson line, so that for gluons, quarks and antiquarks creations operators, one has
\begin{eqnarray}
\hat{S}_E\; \tilde{a}^{\dagger}(k^+,\x,\lambda,a)
&=& U_{A}(\x)_{b a}
\; \tilde{a}^{\dagger}(k^+,\x,\lambda,b)\; \hat{S}_E
\label{com_a_dag_SE}
\\
\hat{S}_E\; \tilde{b}^{\dagger}(k^+,\x,h,\alpha)
&=&  U_{F}(\x)_{\beta \alpha}\;
\tilde{b}^{\dagger}(k^+,\x,h,\beta)\; \hat{S}_E
\label{com_b_dag_SE}
\\
\hat{S}_E\; \tilde{d}^{\dagger}(k^+,\x,h,\alpha)
&=& \left[U_{F}^{\dag}(\x)\right]_{\alpha \beta}\;
\tilde{d}^{\dagger}(k^+,\x,h,\beta)\; \hat{S}_E
\label{com_d_dag_SE}
\end{eqnarray}
respectively. Moreover, $\hat{S}_E$ commutes with creation operators of colorless partons, such as bare photons or leptons, and leaves the Fock vacuum invariant: $\hat{S}_E \big|0\big\rangle
= \big|0\big\rangle$.
The Wilson lines appearing in the equations \eqref{com_a_dag_SE}, \eqref{com_b_dag_SE} and \eqref{com_d_dag_SE} are defined as\footnote{In this paper, QCD covariant derivatives are defined as $D_{\mu}\equiv \partial_{\mu} +i g\, T^a_R\, A_{\mu}^a(\x)$. Indeed, it is the convention used in Ref.~\cite{Brodsky:1997de}, from which the light-front perturbation theory rules presented in Ref.~\cite{Beuf:2016wdz} and used here have been obtained. By consistency, there is a minus in the exponent in the definition \eqref{def_Wilson_line} of the Wilson line. This convention is also used for example the books of Sterman \cite{Sterman:1994ce} and of Collins \cite{Collins_TMD_book}. By contrast, the opposite convention $D_{\mu}\equiv \partial_{\mu} -i g\, T^a_R\, A_{\mu}^a(\x)$, also very common in the literature, would require in particular a change of sign for the Feynman rules for all the vertices involving an odd number of gluons as well as a change of sign of the exponent in the Wilson lines \eqref{def_Wilson_line}. That other prescription is used for example in the book of Peskin and Schroeder \cite{Peskin:1995ev}.}
\begin{eqnarray}
U_{R}(\x) &=& {\cal P} \exp \bigg[-i g \int dx^+\, T^a_R\, {\cal A}^-_a(\underline{x}) \bigg]
\label{def_Wilson_line}
\,
\end{eqnarray}
with the notation $\underline{x}\equiv (x^+, \x)$ and where $R=A$ or $F$ is the appropriate representation of $SU(N_c)$, and ${\cal P}$ indicates the path ordering of the $T^a_R$ color generators. ${\cal A}^-_a(\underline{x})$ is the classical gluon field shockwave taken as target for the scattering.

In the appendix A of Ref. \cite{Beuf:2016wdz}, the definition in light-front quantization and the calculation in light-front perturbation theory of dressed states such as the one-photon dressed state present in Eq.~\eqref{def_forward_ampl} are explained in depth. In particular, one can expand such dressed state on a Fock state basis in mixed space as\footnote{The compact notation $\tilde{b}_{0}^{\dag}\equiv \tilde{b}^{\dagger}(k^+_0,\x_0,h_0,\alpha_0)$ and so on is introduced to avoid unnecessary cluttering.}
\begin{eqnarray}
\Big|\gamma_{\lambda}^*(q^+, \q; Q^2)_{H}\Big\rangle
&=& \sqrt{Z_{\gamma_{\lambda}^*}}
\Bigg\{
 \textrm{Non-QCD Fock states}
+\widetilde{\sum_{q_0 \bar{q}_1\textrm{ F. states}}}
\widetilde{\Psi}_{\gamma_{\lambda}^*\rightarrow q_0 \bar{q_1}}\;
 \tilde{b}_{0}^{\dag}\tilde{d}_{1}^{\dag} |0\rangle
\nonumber\\
&& \hspace{2cm}
+ \widetilde{\sum_{q_0 \bar{q}_1g_2\textrm{ F. states}}}
\widetilde{\Psi}_{\gamma_{\lambda}^*\rightarrow q_0 \bar{q_1}g_2}\;
 \tilde{b}_{0}^{\dag}\tilde{d}_{1}^{\dag}\tilde{a}_{2}^{\dag} |0\rangle
+ \cdots
\Bigg\}
\, .
\label{Fock_expand_dressed_photon}
\end{eqnarray}
The non-QCD Fock states are the ones containing only colorless partons, such as photon or leptons. They give a vanishing contribution to the forward scattering amplitude in Eq.~\eqref{def_forward_ampl} and thus to the DIS cross section, since $\hat{S}_E$ leave them invariant. The perturbative expansion of the quark-antiquark LFWF starts at order $e$, the one of the quark-antiquark-gluon LFWF  starts at order $e\, g$. By contrast, the perturbative expansion of LFWFs for the higher Fock states (represented by the dots in Eq.~\eqref{Fock_expand_dressed_photon}) start at order $e\, g^2$ or later, and thus will not contribute to the DIS cross section at NLO. Moreover, $Z_{\gamma_{\lambda}^*}=1+O(\alpha_{em})$, so that the photon wave-function renormalization constant can be dropped in the one-photon exchange approximation.

The two relevant LFWFs can be written (to all orders in perturbation theory, and beyond) as
\begin{eqnarray}
\widetilde{\Psi}_{\gamma_{\lambda}^*\rightarrow q_0 \bar{q_1}}
&=&
2\pi \delta(k_0^+\!+\!k_1^+\!-\!q^+)\,
e^{i\frac{\q}{q^+}\cdot(k^+_0 \x_0 + k^+_1 \x_1)}\;
 {\mathbf{1}}_{\alpha_0\alpha_1}\;
\widetilde{\psi}_{\gamma_{\lambda}^*\rightarrow q_0 \bar{q_1}}
\label{def_reduced_LFWF_mixed_qqbar}
\\
\widetilde{\Psi}_{\gamma_{\lambda}^*\rightarrow q_0 \bar{q_1}g_2}
&=&
2\pi \delta(k_0^+\!+\!k_1^+\!+\!k_2^+\!-\!q^+)\,
e^{i\frac{\q}{q^+}\cdot(k^+_0 \x_0 + k^+_1 \x_1 + k^+_2 \x_2)}\;
t^{a_2}_{\alpha_0\alpha_1}\;
\widetilde{\psi}_{\gamma_{\lambda}^*\rightarrow q_0 \bar{q_1}g_2}
\label{def_reduced_LFWF_mixed_qqbarg}
\end{eqnarray}
for the following reasons. First, the delta function is required by light-cone momentum $k^+$ conservation. Second, the phase factor is the dependence on $\q$ of the LFWF required when comparing the action of the transverse Galilean boosts\footnote{When using light-front quantization for a relativistic field theory in $D$ dimensions, the Poincar\'e algebra gets broken to a subalgebra which includes the $D\!-\!1$ Galilean algebra, see for example Ref.~\cite{Kogut:1969xa}.} on the momentum-space  dressed-state $|\gamma_{\lambda}^*(q^+, \q; Q^2)_{H}\rangle$ on the one hand, and on the mixed-space Fock states
$\tilde{b}_{0}^{\dag}\tilde{d}_{1}^{\dag}|0\rangle$ and $\tilde{b}_{0}^{\dag}\tilde{d}_{1}^{\dag}\tilde{a}_{2}^{\dag} |0\rangle$ on the other hand. The reduced LFWFs $\widetilde{\psi}_{\gamma_{\lambda}^*\rightarrow q_0 \bar{q_1}}$
and $\widetilde{\psi}_{\gamma_{\lambda}^*\rightarrow q_0 \bar{q_1}g_2}$ are then independent on the photon transverse momentum $\q$. Moreover, due to transverse momentum conservation, the reduced LFWFs cannot depend on the absolute transverse positions of the Fock state partons, just on their differences. Finally, ${\mathbf{1}}_{\alpha_0\alpha_1}$ and $t^{a_2}_{\alpha_0\alpha_1}$ are the only invariant color tensors available for the $q\bar{q}$ Fock state and for the $q\bar{q}g$ Fock state respectively, so that the corresponding LFWF have to be proportional to them.

Inserting the Fock state expansion \eqref{Fock_expand_dressed_photon} into Eq.~\eqref{def_forward_ampl}, and using the relations given in the present section or in the appendix A of Ref.~\cite{Beuf:2016wdz}, one arrives at the general expression\footnote{In formulas like Eq.~\eqref{sigma_gamma_master_eq}, the real part operator ${\textrm{Re}}$ is very often dropped. Indeed, when evaluated in the MV model \cite{McLerran:1993ni,McLerran:1993ka,McLerran:1994vd,Jeon:2004rk}, the dipole amplitude ${\cal S}_{01}$ averaged over the background field is real and symmetric by exchange of $\x_0$ and $\x_1$. Moreover, these properties are preserved by the B-JIMWLK and BK evolutions. But in general, in QCD, the averaged dipole amplitude can have an imaginary part antisymmetric by exchange of $\x_0$ and $\x_1$, which correspond to an Odderon exchange with the target \cite{Kovchegov:2003dm,Hatta:2005as,Jeon:2005cf,Lappi:2016gqe}.
Here, the real part operator ${\textrm{Re}}$ coming from the optical theorem is kept thoroughly, in order to insist on the fact the Odderon exchange does not contribute to the DIS cross section.}
\begin{eqnarray}
\sigma^{\gamma^*}_{\lambda}
&=& 2 N_c\widetilde{\sum_{q_0 \bar{q}_1\textrm{ F. states}}}
\frac{2\pi \delta(k_0^+\!+\!k_1^+\!-\!q^+)}{2q^+}\;
\left|\widetilde{\psi}_{\gamma_{\lambda}^*\rightarrow q_0 \bar{q_1}}\right|^2\;
{\textrm{Re}}\left[1-{\cal S}_{01}\right]
\nonumber\\
&& +2 N_c C_F
 \widetilde{\sum_{q_0 \bar{q}_1g_2\textrm{ F. states}}}
\frac{2\pi \delta(k_0^+\!+\!k_1^+\!+\!k_2^+\!-\!q^+)}{2q^+}\;
\left|\widetilde{\psi}_{\gamma_{\lambda}^*\rightarrow q_0 \bar{q_1}g_2}\right|^2\;
{\textrm{Re}}\left[1-{\cal S}^{(3)}_{012}\right]\;\;
+ O(\alpha_{em}\, \alpha_{s}^2)
\label{sigma_gamma_master_eq}
\end{eqnarray}
for the total cross-section of photon-shockwave scattering, introducing the notations
\begin{eqnarray}
{\cal S}_{01}
&\equiv&
\frac{1}{N_c} \textrm{Tr} \left(U_{F}(\mathbf{x}_{0})\, U_{F}^\dag(\mathbf{x}_{1}) \right)
\label{def_dipole}
\\
{\cal S}^{(3)}_{012}
&\equiv&
\frac{1}{N_c\, C_F} \textrm{Tr} \left(t^{b}U_{F}(\mathbf{x}_{0})\, t^{a} U_{F}^\dag(\mathbf{x}_{1}) \right)
U_{A}(\mathbf{x}_{2})_{b a}
\label{def_tripole}
\end{eqnarray}
for the dipole and tripole amplitudes\footnote{Strictly speaking, ${\cal S}_{01}$ and ${\cal S}^{(3)}_{012}$ correspond to S-matrices rather than scattering amplitudes, but will be called simply amplitudes, for simplicity.}.


The expression \eqref{sigma_gamma_master_eq} already displays a separation between the internal dynamics of the photon projectile, encoded in the LFWF, and the scattering on the shockwave target, encoded in the multipole amplitudes. The first term in Eq.~\eqref{sigma_gamma_master_eq} corresponds to the usual dipole factorization formula, valid at LO. Obviously, each order in perturbation theory adds new terms to the formula \eqref{sigma_gamma_master_eq}, by allowing the contribution of higher Fock states. By a slight abuse of language, Eq.~\eqref{sigma_gamma_master_eq} will still be called dipole factorization even when other Fock states than $q\bar{q}$ are included.


\subsection{Quark-antiquark contributions to the photon total cross-sections at NLO\label{sec:qqbar_NLO}}

Let us focus first on the $q\bar{q}$ contribution to $\sigma^{\gamma^*}_{\lambda}$, corresponding to the first term in Eq.~\eqref{sigma_gamma_master_eq}. Writing more explicitly the sum over $q\bar{q}$ Fock states, one has
\begin{eqnarray}
\sigma^{\gamma^*}_{\lambda} \Big|_{q\bar{q}}
&=& 2 N_c \sum_{f}
\int \frac{dk^+_0}{2\pi}\; \frac{\theta(k^+_0)}{2k^+_0}
\int \frac{dk^+_1}{2\pi}\; \frac{\theta(k^+_1)}{2k^+_1}\;
\frac{2\pi \delta(k_0^+\!+\!k_1^+\!-\!q^+)}{2q^+}\;
\nonumber\\
&& \times
\int d^{D-2} \x_0 \int d^{D-2} \x_1\;
{\textrm{Re}}\left[1-{\cal S}_{01}\right]\;
\sum_{h_0, h_1 = \pm 1/2}
\left|\widetilde{\psi}_{\gamma_{\lambda}^*\rightarrow q_0 \bar{q_1}}\right|^2\;
\, .
\end{eqnarray}
The $\gamma_{\lambda}^*\rightarrow q \bar{q}$ LFWFs have been calculated in Ref.~\cite{Beuf:2016wdz} including one QCD loop, regularizing the UV divergences with CDR, and the small $k^+$ divergences with a cut-off $k^+_{\min}$. It has been found that such NLO corrections factorize\footnote{These NLO corrections factorize at least within the regularization scheme used. But extra non-factorizing terms might appear in other regularization schemes \cite{Lappi:2016oup}.}
from the tree-level LFWFs, as
\begin{eqnarray}
\widetilde{\psi}_{\gamma_{\lambda}^{*}\rightarrow q_0 \bar{q}_1}
 &=&
\left[1+\left(\frac{\alpha_s\, C_F}{2\pi}\right)\;
\widetilde{{\cal V}}\;\right]\;
\widetilde{\psi}^{\textrm{tree}}_{\gamma_{\lambda}^{*}\rightarrow q_0 \bar{q}_1}
+{ O}(e\, \alpha_s^2)
 \, , \label{WF_qqbar_gen_NLO}
\end{eqnarray}
where the coefficient $\widetilde{{\cal V}}$ carrying the NLO corrections is given by
\begin{eqnarray}
\widetilde{{\cal V}}
&=&
2\left[\log\left(\frac{k^+_{\min}}{\sqrt{k_0^+k_1^+}}\right)
+\frac{3}{4} \right]
\left[\frac{1}{\left(2\!-\!\frac{D}{2}\right)}-\Psi(1)+\log\left(\pi\, {\x_{01}}^2\, \mu^2\right)
\right]
+\frac{1}{2} \left[\log\left(\frac{k_0^+}{k_1^+}\right)\right]^2
-\frac{\pi^2}{6}
 +3
 + O\left(D\!-\!4\right)
\label{mixed_V_result}
\end{eqnarray}
for $D\rightarrow 4$.
In $D$ dimensions, the tree-level LFWFs read \cite{Beuf:2016wdz}
\begin{eqnarray}
\widetilde{\psi}^{\textrm{tree}}_{\gamma_{L}^{*}\rightarrow q_0 \bar{q}_1}
 &=&(-1)\;
\overline{u_G}(0)\,\gamma^+ v_G(1)\;
  \frac{e\, e_f}{2\pi}\;
\frac{k_0^+ k_1^+}{(q^+)^2}\; 2 Q\;
\left(\frac{\overline{Q}}{2\pi\, |\x_{01}|\, \mu}\right)^{\frac{D}{2}-2}\;\;
 \textrm{K}_{\frac{D}{2}-2}\Big(|\x_{01}|\, \overline{Q}\Big)\;\;
  \label{WF_gamma_L_qqbar_tree}
\end{eqnarray}
for the longitudinal photon polarization, and
\begin{eqnarray}
\widetilde{\psi}^{\textrm{tree}}_{\gamma_{\lambda}^{*}\rightarrow q_0 \bar{q}_1}
 &=&
(-i)\,  e\, e_f\;
\varepsilon_{\lambda}^i\, \x_{01}^j\;
\overline{u_G}(0) \gamma^+ \Bigg\{
 \left(\frac{k^+_0\!-\!k^+_1}{q^+}\right)\; \delta^{ij}\;
-\frac{1}{2}\;
\left[\gamma^i, \gamma^j\right]
\Bigg\}\, v_G(1)
\nonumber\\
&& \hspace{1cm}\times\;
\mu^{2-\frac{D}{2}}\,
\left(\frac{\overline{Q}}{2\pi\, |\x_{01}|}\right)^{\frac{D}{2}-1}
 \textrm{K}_{\frac{D}{2}-1}\Big(|\x_{01}|\, \overline{Q}\Big)\;\;
 \label{WF_gamma_T_qqbar_tree}
\end{eqnarray}
and for transverse photon polarizations, introducing the notation
\begin{eqnarray}
\overline{Q}^2= \frac{k_0^+ (q^+\!-\!k_0^+)}{(q^+)^2}\; Q^2= \frac{k_1^+ (q^+\!-\!k_1^+)}{(q^+)^2}\; Q^2
\label{def_Qbar}
\, .
\end{eqnarray}

Using the definitions and properties of light-front spinors given in the appendix A of Ref.~\cite{Beuf:2016wdz}, it is straightforward to obtain
\begin{eqnarray}
\sum_{h_0, h_1 = \pm 1/2}
\left|\widetilde{\psi}_{\gamma_{L}^*\rightarrow q_0 \bar{q_1}}\right|^2
 &=& 2 (2k_0^+)\, (2k_1^+)\,
  \frac{e^2\, e_f^2}{(2\pi)^2}\; 4 Q^2
\frac{(k_0^+)^2 (k_1^+)^2}{(q^+)^4}\;
\left(\frac{\overline{Q}^2}{(2\pi)^2\, \x_{01}^2\, \mu^2}\right)^{\frac{D}{2}-2}\;\;
\left[ \textrm{K}_{\frac{D}{2}-2}\Big(|\x_{01}|\, \overline{Q}\Big)\right]^2\;\;
\nonumber\\
&& \hspace{2cm} \times
\left[1+\left(\frac{\alpha_s\, C_F}{\pi}\right)\;
\widetilde{{\cal V}}\;\right]\;\;
+{\cal O}(\alpha_{em}\, \alpha_s^2)
  \label{averaged WF_gamma_L_qqbar_squared}
\, ,
\end{eqnarray}
so that the $q\bar{q}$ contribution to longitudinal photon total cross section at NLO is
\begin{eqnarray}
\sigma^{\gamma^*}_{L} \Big|_{q\bar{q}}
&=& 4 N_c \alpha_{em} \sum_{f} e_f^2\;
\int_{0}^{+\infty}\!\!\!\!\!\! dk^+_0\; \int_{0}^{+\infty}\!\!\!\!\!\! dk^+_1\;
\frac{\delta(k_0^+\!+\!k_1^+\!-\!q^+)}{q^+}\;
\int \frac{d^{D-2} \x_0}{2\pi} \int \frac{d^{D-2} \x_1}{2\pi}\;
{\textrm{Re}}\left[1-{\cal S}_{01}\right]\;
\left[1+\left(\frac{\alpha_s\, C_F}{\pi}\right)
\widetilde{{\cal V}}\right]\;
\nonumber\\
& &\hspace{0.5cm} \times 4 Q^2
\frac{(k_0^+)^2 (k_1^+)^2}{(q^+)^4}\;
\left(\frac{\overline{Q}^2}{(2\pi)^2\, \x_{01}^2\, \mu^2}\right)^{\frac{D}{2}-2}
\left[ \textrm{K}_{\frac{D}{2}-2}\Big(|\x_{01}|\, \overline{Q}\Big)\right]^2
+{\cal O}(\alpha_{em}\, \alpha_s^2)\;
\, .
\label{sigma_L_qqbar_result}
\end{eqnarray}

In the transverse photon case, the numerator algebra can be performed for example thanks to the identities\footnote{${\cal P}_{G}\equiv \gamma^- \gamma^+/2$ is the projector onto the so-called good components of the spinors.}
\begin{eqnarray}
\textrm{Tr}\left\{{\cal P}_{G} \right\} &=& 2 \label{trace_gamma_1}\\
\textrm{Tr}\left\{{\cal P}_{G}\; [\gamma^i,\gamma^j]\right\} &=& 0\label{trace_gamma_2}\\
\delta^{i'i}\, \textrm{Tr}\left\{{\cal P}_{G}\; \frac{[\gamma^{i'},\gamma^{j'}]}{2}\; \frac{[\gamma^i,\gamma^j]}{2}\right\}
&=&-2 (D\!-\!3)\, \delta^{j'j}\label{trace_gamma_3}
\, ,
\end{eqnarray}
valid for arbitrary dimension. One finds
\begin{eqnarray}
\frac{1}{(D\!-\!2)}\sum_{T\textrm{ pol. }\lambda}\;\;  \sum_{h_0, h_1 = \pm 1/2}
\left|\widetilde{\psi}_{\gamma_{\lambda}^*\rightarrow q_0 \bar{q_1}}\right|^2
&=&\frac{(2k_0^+)(2k_1^+)}{(D\!-\!2)}\, \frac{4 \alpha_{em}}{2\pi}\, e_f^2\,
\bigg[(D\!-\!3)+\left(\frac{k_{0}^+\!-\!k_{1}^+}{q^+}\right)^2\bigg]\,
\left[1+\left(\frac{\alpha_s\, C_F}{\pi}\right) \widetilde{{\cal V}}\,\right]
\nonumber\\
&& \times
\Big[(2\pi)^2\, \mu^2\, \x_{01}^2\Big]^{2-\frac{D}{2}}\,
\left(\overline{Q}^2\right)^{\frac{D}{2}-1}
\Big[\textrm{K}_{\frac{D}{2}-1}\Big(|\x_{01}|\, \overline{Q}\Big)\Big]^2
\end{eqnarray}
and thus
\begin{eqnarray}
\sigma^{\gamma^*}_{T} \Big|_{q\bar{q}}
&=& 4 N_c \alpha_{em} \sum_{f} e_f^2\;
\int_{0}^{+\infty}\!\!\!\!\!\! dk^+_0\; \int_{0}^{+\infty}\!\!\!\!\!\! dk^+_1\;
\frac{\delta(k_0^+\!+\!k_1^+\!-\!q^+)}{q^+}\;
\int \frac{d^{D-2} \x_0}{2\pi} \int \frac{d^{D-2} \x_1}{2\pi}\;
{\textrm{Re}}\left[1-{\cal S}_{01}\right]\;
\left[1+\left(\frac{\alpha_s\, C_F}{\pi}\right)
\widetilde{{\cal V}}\right]\;
\nonumber\\
& &
\times
\frac{\big[(D\!-\!2)(q^+)^2\!-\!4k_{1}^+(q^+\!-\!k_{1}^+)\big]}{(D\!-\!2)\, (q^+)^2}\;
\Big[(2\pi)^2\, \mu^2\, \x_{01}^2\Big]^{\frac{D}{2}-2}
\Big(\overline{Q}^2\Big)^{\frac{D}{2}-1}
\left[\textrm{K}_{\frac{D}{2}-1}\Big(|\x_{01}|\, \overline{Q}\Big)\right]^2
+{\cal O}(\alpha_{em}\, \alpha_s^2)
\, .
\label{sigma_T_qqbar_result}
\end{eqnarray}

In summary, the $q\bar{q}$ contributions to $\sigma^{\gamma^*}_{L}$ and $\sigma^{\gamma^*}_{T}$ at NLO are given respectively by Eqs.~\eqref{sigma_L_qqbar_result} and \eqref{sigma_T_qqbar_result}, with the loop correction $\widetilde{{\cal V}}$ given by Eq.~\eqref{mixed_V_result}. It remains to calculate the $q\bar{q}g$ contributions, obtained from the $\gamma_{\lambda}^*\rightarrow q \bar{q} g$ LFWFs according to Eq.~\eqref{sigma_gamma_master_eq}, and combine both types of contributions.


\section{Quark-antiquark-gluon LFWFs\label{sec:qqbarg_WF}}

The LFWFs for the quark-antiquark-gluon Fock states inside a transverse or longitudinal virtual photon dressed state have already been calculated at tree level in Ref.~\cite{Beuf:2011xd}. However, that calculation was done in $D=4$ strictly, whereas the result in generic dimension is needed for a consistent combination with the $q\bar{q}$ contributions \eqref{sigma_L_qqbar_result} and \eqref{sigma_T_qqbar_result}. It is the purpose of the present section is to close this gap.

\subsection{Transverse photon case}

\subsubsection{Momentum space results}

\begin{figure}
\setbox1\hbox to 10cm{
\includegraphics{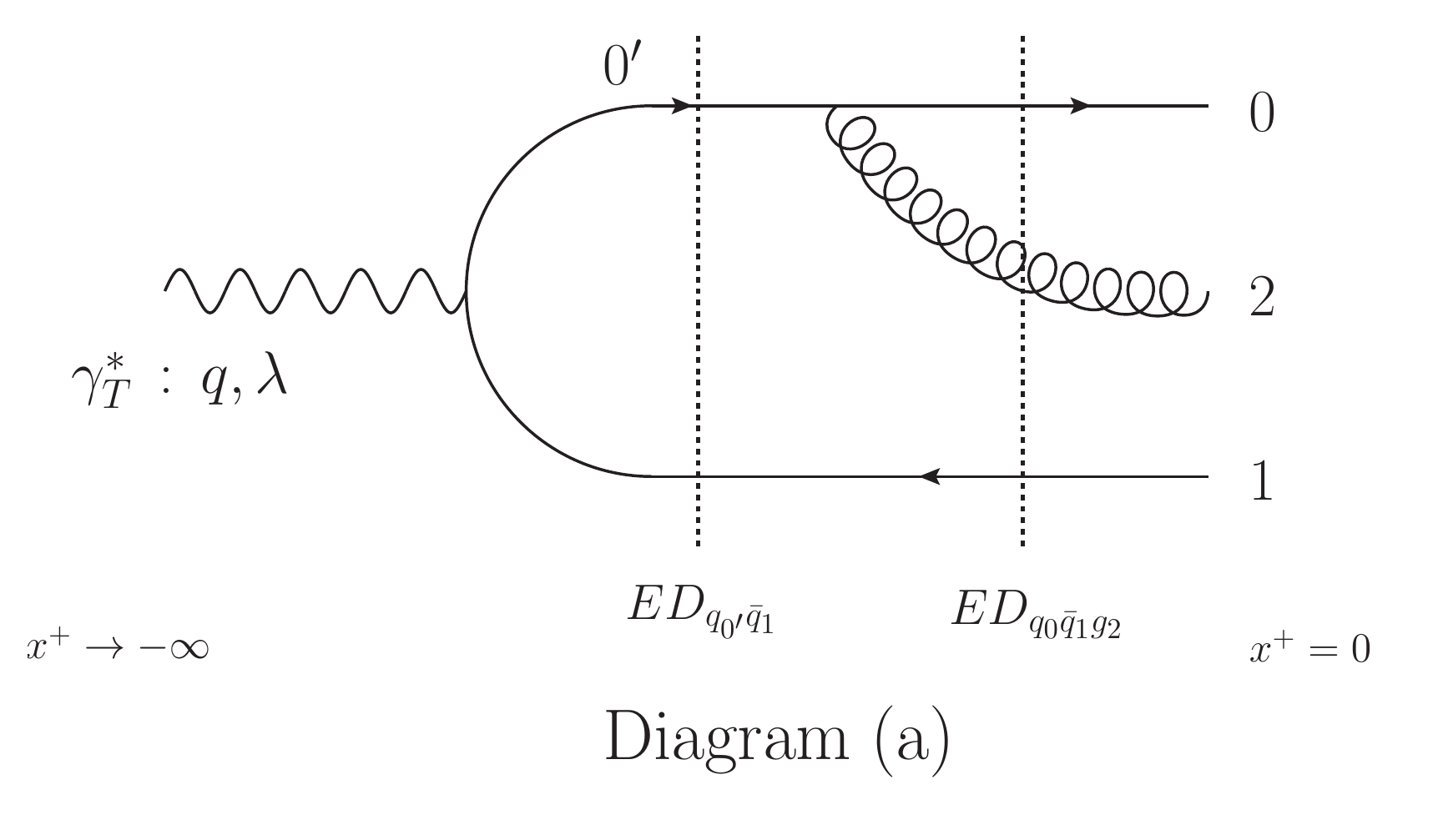}
}
\setbox2\hbox to 10cm{
\includegraphics{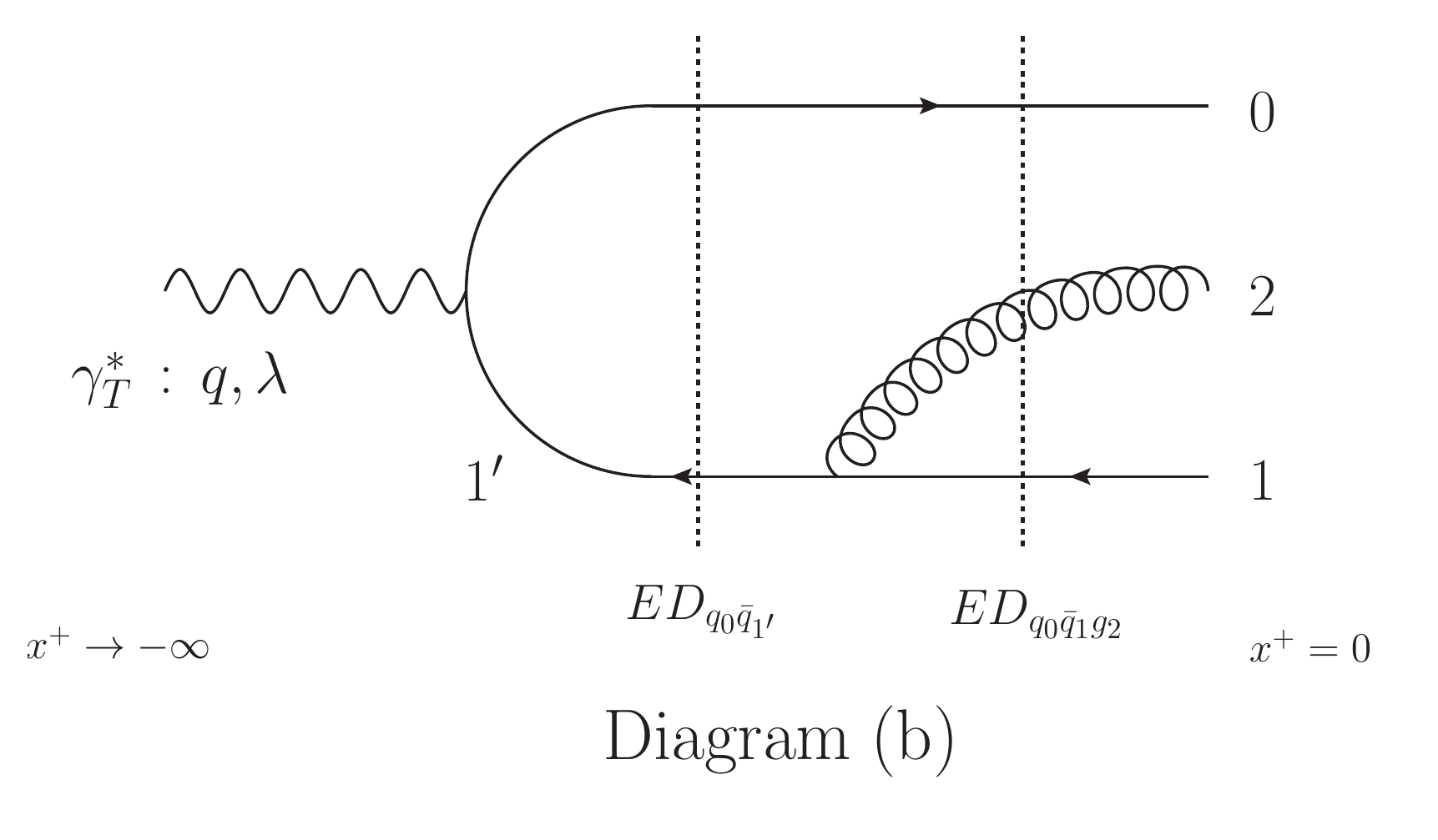}
}
\setbox3\hbox to 10cm{
\includegraphics{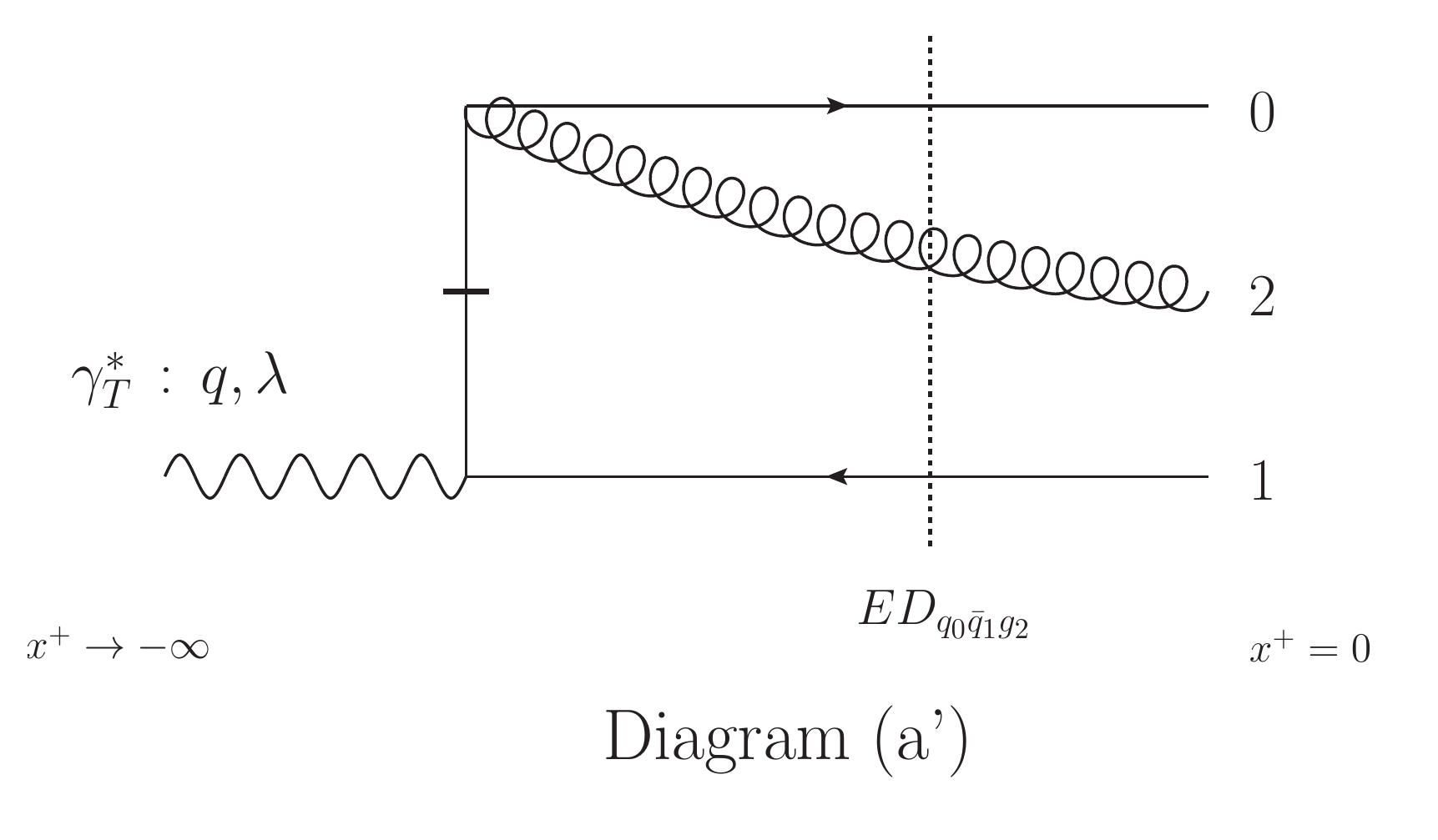}
}
\setbox4\hbox to 10cm{
\includegraphics{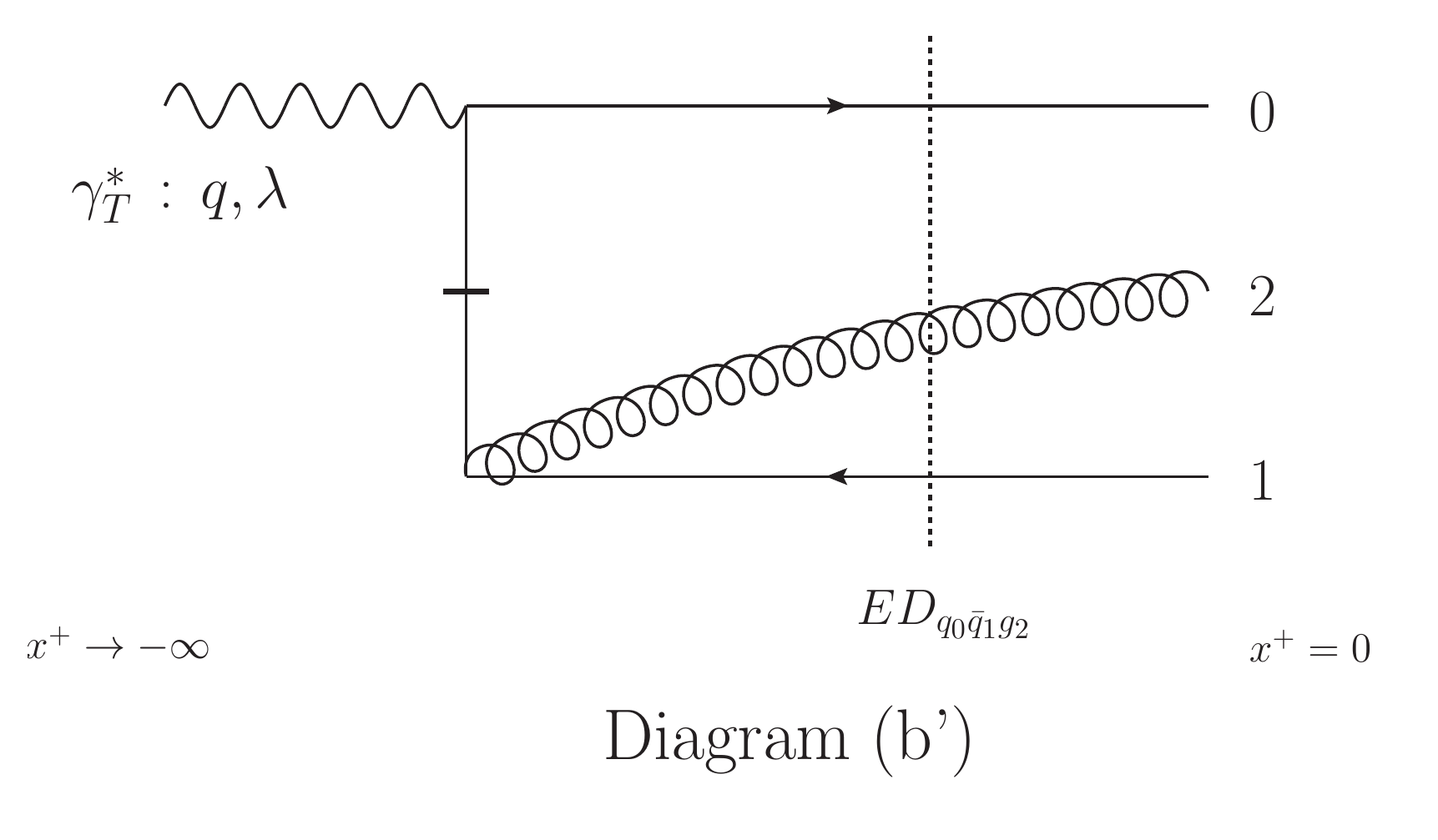}
}
\begin{center}
\resizebox*{10cm}{!}{\hspace{-7cm}\mbox{\box1 \hspace{9cm} \box2}}
\resizebox*{10cm}{!}{\hspace{-7cm}\mbox{\box3 \hspace{9cm} \box4}}
\caption{\label{Fig:gammaT_qqbarg}Tree-level diagrams contributing to the LFWF of the $q\bar{q}g$ Fock component inside a dressed transverse photon.}
\end{center}
\end{figure}

In the case of a transverse photon, the four light-front perturbation theory diagrams shown in Fig.\ref{Fig:gammaT_qqbarg} contribute at tree level to the $q\bar{q}g$ LFWF, so that in momentum space
\begin{eqnarray}
\Psi_{\gamma_{\lambda}^{*}\rightarrow q_0\bar{q}_1g_2}^{\textrm{Tree}}
&=& \sum_{q_{0'} \textrm{ states}}
\frac{
\langle 0| a_{2} b_0\, V_I(0)\, b_{0'}^{\dag} |0 \rangle\;
\langle 0| d_{1} b_{0'}\, V_I(0)\, a_{\gamma}^{\dag} |0 \rangle
}{(ED_{q_0\bar{q}_1g_2})\; (ED_{q_{0'}\bar{q}_1})}
\nonumber\\
& &
+ \sum_{\bar{q}_{1'} \textrm{ states}}
\frac{
\langle 0| a_{2} d_{1}\, V_I(0)\, d_{1'}^{\dag} |0 \rangle\;
\langle 0| d_{1'} b_{0}\, V_I(0)\, a_{\gamma}^{\dag} |0 \rangle
}{(ED_{q_0\bar{q}_1g_2})\; (ED_{q_{0}\bar{q}_{1'}})}
+ \frac{
\langle 0| a_{2} d_{1} b_{0}\, V_I(0)\, a_{\gamma}^{\dag} |0 \rangle
}{(ED_{q_{0}\bar{q}_{1'}})}
\label{qqbarg_WF_T_1}
\\
&=&\frac{(2\pi)^{D\!-\!1} \delta^{(D\!-\!1)}(\underline{k_0}\!+\!\underline{k_{1}}\!+\!\underline{k_{2}}\!-\!\underline{q})}{(ED_{q_0\bar{q}_1g_2})}
\; (\mu^2)^{2-\frac{D}{2}}\, e\, e_f\, g\; t^{a_2}_{\alpha_{0}\alpha_{1}}
\nonumber\\
& & \times
\Bigg\{
 \int \frac{d^{D-1} \underline{k_{0'}}}{(2\pi)^{D-1}}\; \frac{\theta(k_{0'}^+)}{2 k_{0'}^+}\;
 \frac{(2\pi)^{D\!-\!1} \delta^{(D\!-\!1)}(\underline{k_0}\!+\!\underline{k_{2}}\!-\!\underline{k_{0'}})}{(ED_{q_{0'}\bar{q}_1})}\;
 \sum_{h_{0'}=\pm 1/2} \overline{u}(0)\, \slashed{\epsilon}_{\lambda_2}^*\!(\underline{k_2})\, u(0')\;\;
 \overline{u}(0')\, \slashed{\epsilon}_{\lambda}\!(\underline{q})\, v(1)
\nonumber\\
& &
\hspace{0.3cm}
-  \int \frac{d^{D-1} \underline{k_{1'}}}{(2\pi)^{D-1}}\; \frac{\theta(k_{1'}^+)}{2 k_{1'}^+}\;
 \frac{(2\pi)^{D\!-\!1} \delta^{(D\!-\!1)}(\underline{k_1}\!+\!\underline{k_{2}}\!-\!\underline{k_{1'}})}{(ED_{q_{0}\bar{q}_{1'}})}\;
 \sum_{h_{1'}=\pm 1/2} \overline{u}(0)\, \slashed{\epsilon}_{\lambda}\!(\underline{q})\, v(1')\;\;
 \overline{v}(1')\, \slashed{\epsilon}_{\lambda_2}^*\!(\underline{k_2})\,  v(1)
\nonumber\\
& &
\hspace{0.3cm}
+ \frac{1}{2 (k_{0}^+\!+\!k_{2}^+)}\;
\overline{u}(0)\, \slashed{\epsilon}_{\lambda_2}^*\!(\underline{k_2})\, \gamma^+ \slashed{\epsilon}_{\lambda}\!(\underline{q})\, v(1)
-\frac{1}{2 (k_{1}^+\!+\!k_{2}^+)}\;
\overline{u}(0)\, \slashed{\epsilon}_{\lambda}\!(\underline{q})\, \gamma^+ \slashed{\epsilon}_{\lambda_2}^*\!(\underline{k_2})\, v(1)
\;\Bigg\}
\, .
\label{qqbarg_WF_T_2}
\end{eqnarray}
The three terms in the expression \eqref{qqbarg_WF_T_1} corresponds respectively to the diagram $(a)$, the diagram $(b)$, and the sum $(a')+(b')$ of the diagrams with an instantaneous fermion leg. The expression \eqref{qqbarg_WF_T_2} is then obtained using the rules presented in the appendix A of Ref.~\cite{Beuf:2016wdz}. The last two terms in that expression now correspond to the diagrams $(a')$ and $(b')$ separately.
The energy denominators appearing in the expressions \eqref{qqbarg_WF_T_1} and \eqref{qqbarg_WF_T_2} are defined as
\begin{eqnarray}
(ED_{q_0\bar{q}_1g_2})&\equiv & \frac{\q^2\!-\!Q^2}{2q^+} - \frac{\k_{0}^2}{2k^+_{0}}
- \frac{\k_{1}^2}{2k^+_{1}} - \frac{\k_{2}^2}{2k^+_{2}}
\label{ED_qqbarg}
\\
(ED_{q_{0'}\bar{q}_1})&\equiv & \frac{\q^2\!-\!Q^2}{2q^+} - \frac{\k_{0'}^2}{2k^+_{0'}}
- \frac{\k_{1}^2}{2k^+_{1}}
\label{ED_a}
\\
(ED_{q_{0}\bar{q}_{1'}})&\equiv & \frac{\q^2\!-\!Q^2}{2q^+} - \frac{\k_{0}^2}{2k^+_{0}}
- \frac{\k_{1'}^2}{2k^+_{1'}}
\label{ED_b}
\, ,
\end{eqnarray}
including the photon virtuality $Q^2$ as discussed in the appendix A.3 of Ref.~\cite{Beuf:2011xd}.

In order to perform the Fourier transform to mixed space at the LFWF level, one should extract the transverse momentum dependence out of the spinors $u$ and $v$ and out of the photon and gluon polarization vectors.
This can be done\footnote{Remember that the good components of the spinors $u$ and $v$ only depend on the light-front momentum $k^+$ and the light-front helicity $h$ of the considered parton, and that the dependence on its transverse momentum $\k$ arises only through the bad components.} thanks to the identity\footnote{See for example the derivation of Eq.~(B5) in Ref.~\cite{Beuf:2016wdz}.}
\begin{eqnarray}
\overline{\chi}(n)\: \slashed{\epsilon}_{\lambda}^{(*)}\!\!(\underline{q})\: \eta(m)
&=& \varepsilon_{\lambda}^{i}\, \left[\frac{\q^i}{q^+}\!-\!\frac{\k^i_n}{2k^+_n}\!-\!\frac{\k^i_m}{2k^+_m}\right]\,
       \overline{\chi_G}(n)\: \gamma^+\: \eta_G(m)
+ \varepsilon_{\lambda}^{i}\, \frac{1}{4}\,  \left[\frac{\k^j_m}{k^+_m}\!-\!\frac{\k^j_n}{k^+_n}\right]\,
   \overline{\chi_G}(n)\: \gamma^+\, [\gamma^i,\gamma^j]\: \eta_G(m)
\, ,
\label{spinor_bilin_mom_extract}
\end{eqnarray}
where each of $\chi$ and $\eta$ is either a $u$ or $v$ massless spinor. That identity is valid for arbitrary $D$, and does not require momentum conservation to be valid.

Inserting the relation \eqref{spinor_bilin_mom_extract} at each vertex, one can rewrite the contribution of the diagrams $(a)$ and $(a')$ to Eq.\eqref{qqbarg_WF_T_2} as
\begin{eqnarray}
\Psi_{\gamma_{\lambda}^{*}\rightarrow q_0\bar{q}_1g_2}^{(a)+(a')}
&=&\frac{(2\pi)^{D\!-\!1} \delta^{(D\!-\!1)}(\underline{k_0}\!+\!\underline{k_{1}}\!+\!\underline{k_{2}}\!-\!\underline{q})\; (\mu^2)^{2-\frac{D}{2}}\, e\, e_f\, g\; t^{a_2}_{\alpha_{0}\alpha_{1}}
\varepsilon_{\lambda}^{i}\, \varepsilon_{\lambda_2}^{j *}}
{\left\{\left(\k_{2}\!-\!\frac{k_{2}^+}{k_{0}^+}\, \k_{0}\right)^2
+ \frac{q^+ k_{2}^+}{k_{0}^+ k_{1}^+} \left[ \left(\k_{1}\!-\!\frac{k_{1}^+}{q^+}\, \q\right)^2
 +\overline{Q}_{(a)}^2\right] \right\}}\;
\Bigg\{ \frac{\left[\k_{2}^m\!-\!\frac{k_{2}^+}{k_{0}^+}\, \k_{0}^m\right]\, \left[\k_{1}^l\!-\!\frac{k_{1}^+}{q^+}\, \q^l\right]}{\left[\left(\k_{1}\!-\!\frac{k_{1}^+}{q^+}\, \q\right)^2
+\overline{Q}_{(a)}^2\right]}
 \;
\nonumber\\
& & \hspace{0.3cm} \times\,
  \overline{u_G}(0)\, \gamma^+ \left[\frac{(2k_{0}^+\!+\!k_{2}^+)}{k_{0}^+} \delta^{jm}
   + \frac{k_{2}^+}{2k_{0}^+}\, [\gamma^j,\gamma^m]\right]
   \left[\frac{(2k_{1}^+\!-\!q^+)}{q^+} \delta^{il}
   + \frac{1}{2}\, [\gamma^i,\gamma^l]\right]  v_G(1)
\nonumber\\
& &
\hspace{0.3cm}
+ \frac{k_{2}^+}{k_{0}^+}\;  \overline{u_G}(0)\, \gamma^+ \gamma^j \gamma^i  v_G(1)
\;\Bigg\}
\label{qqbarg_WF_T_a_plus_a_prime}
\end{eqnarray}
and the one of the diagrams $(b)$ and $(b')$ as
\begin{eqnarray}
\Psi_{\gamma_{\lambda}^{*}\rightarrow q_0\bar{q}_1g_2}^{(b)+(b')}
&=&-\, \frac{(2\pi)^{D\!-\!1} \delta^{(D\!-\!1)}(\underline{k_0}\!+\!\underline{k_{1}}\!+\!\underline{k_{2}}\!-\!\underline{q})\; (\mu^2)^{2-\frac{D}{2}}\, e\, e_f\, g\; t^{a_2}_{\alpha_{0}\alpha_{1}}
\varepsilon_{\lambda}^{i}\, \varepsilon_{\lambda_2}^{j *}}
{\left\{\left(\k_{2}\!-\!\frac{k_{2}^+}{k_{1}^+}\, \k_{1}\right)^2
+ \frac{q^+ k_{2}^+}{k_{0}^+ k_{1}^+} \left[ \left(\k_{0}\!-\!\frac{k_{0}^+}{q^+}\, \q\right)^2
 +\overline{Q}_{(b)}^2\right] \right\}}\;
\Bigg\{ \frac{\left[\k_{2}^m\!-\!\frac{k_{2}^+}{k_{1}^+}\, \k_{1}^m\right]\, \left[\k_{0}^l\!-\!\frac{k_{0}^+}{q^+}\, \q^l\right]}{\left[\left(\k_{0}\!-\!\frac{k_{0}^+}{q^+}\, \q\right)^2
+\overline{Q}_{(b)}^2\right]}
 \;
\nonumber\\
& & \hspace{0.3cm} \times\,
  \overline{u_G}(0)\, \gamma^+ \left[\frac{(2k_{0}^+\!-\!q^+)}{q^+} \delta^{il}
   - \frac{1}{2}\, [\gamma^i,\gamma^l]\right]
  \left[\frac{(2k_{1}^+\!+\!k_{2}^+)}{k_{1}^+} \delta^{jm}
   - \frac{k_{2}^+}{2k_{1}^+}\, [\gamma^j,\gamma^m]\right]
     v_G(1)
\nonumber\\
& &
\hspace{0.3cm}
+ \frac{k_{2}^+}{k_{1}^+}\;  \overline{u_G}(0)\, \gamma^+ \gamma^i \gamma^j  v_G(1)
\;\Bigg\}
\, ,
\label{qqbarg_WF_T_b_plus_b_prime}
\end{eqnarray}
using the notations
\begin{eqnarray}
\overline{Q}_{(a)}^2&=& \frac{k_1^+ (q^+\!-\!k_1^+)}{(q^+)^2}\; Q^2
\label{def_Qa}
\\
\overline{Q}_{(b)}^2&=& \frac{k_0^+ (q^+\!-\!k_0^+)}{(q^+)^2}\; Q^2
\label{def_Qb}
\, .
\end{eqnarray}


\subsubsection{Mixed space results}

The Fourier transform to mixed space of the $q\bar{q}g$ LFWF is defined by
\begin{equation}
\widetilde{\Psi}_{\gamma_{\lambda}^{*}\rightarrow q_0\bar{q}_1g_2}
= \int \frac{d^{D-2} \k_0}{(2\pi)^{D-2}}\;
\int \frac{d^{D-2} \k_1}{(2\pi)^{D-2}}\;
\int \frac{d^{D-2} \k_2}{(2\pi)^{D-2}}\;
e^{i\k_0 \cdot \x_0 + i\k_1 \cdot \x_1 + i\k_2 \cdot \x_2}
\; \Psi_{\gamma_{\lambda}^{*}\rightarrow q_0\bar{q}_1g_2}
 \, .\label{def_LFWF_mix}
\end{equation}

For the contribution of the diagrams $(a)$ and $(a')$ given in Eq.~\eqref{qqbarg_WF_T_a_plus_a_prime}, it is convenient to make the change of variables $(\k_1,\k_2)\mapsto (\P,\K)$ defined as
\begin{eqnarray}
\P &=& -\k_1 + \frac{k_1^+}{q^+}\, \q \nonumber\\
\K &=& \k_2 -\frac{k_2^+}{k_0^+\!+\!k_2^+}\, (\k_0\!+\!\k_2)
= \left(\frac{k_0^+}{k_0^+\!+\!k_2^+}\right) \left[\k_2 -\frac{k_2^+}{k_0^+}\, \k_0\right]
\label{cv_Fourier_a}
\end{eqnarray}
and use the transverse delta function to eliminate the integration over $\k_0$. With this change of variables, one has
\begin{eqnarray}
d^{D-2} \k_2 = \left(\frac{k_0^+\!+\!k_2^+}{k_0^+}\right)^{D-2}\, d^{D-2} \K
\end{eqnarray}
and
\begin{eqnarray}
\delta^{(D\!-\!1)}(\underline{k_0}\!+\!\underline{k_{1}}\!+\!\underline{k_{2}}\!-\!\underline{q})
&=& \delta^{(D\!-\!1)}\left(\left(\frac{k_0^+\!+\!k_2^+}{k_0^+}\right) \k_0
 - \left(\frac{k_0^+\!+\!k_2^+}{q^+}\right) \q + \frac{k_0^+\!+\!k_2^+}{k_0^+}\K -\P\right) \nonumber\\
&=&  \left(\frac{k_0^+}{k_0^+\!+\!k_2^+}\right)^{D-2}\,
 \delta^{(D\!-\!1)}\left(\k_0
 - \frac{k_0+}{q^+} \q + \K -\left(\frac{k_0^+}{k_0^+\!+\!k_2^+}\right)\P\right)
 \, ,
\end{eqnarray}
so that the Jacobian cancels. Moreover, the phase factor becomes
\begin{eqnarray}
e^{i\k_0 \cdot \x_0 + i\k_1 \cdot \x_1 + i\k_2 \cdot \x_2}
&=&  e^{i\frac{\q}{q^+}\cdot (k_0^+\x_0+k_1^+\x_1+k_2^+\x_2)}\,
 e^{i\K\cdot \x_{20}}\,
e^{i\P\cdot \x_{0+2;1}}\,
\, ,
\end{eqnarray}
introducing the notation
\begin{eqnarray}
\x_{n+m;p}&= & -\x_{p;n+m}
\equiv \left(\frac{k_n^+\x_n + k_m^+\x_m }{k_n^+\!+\!k_m^+}\right) -\x_p
\label{def_pre-split_parent_dipole}
\end{eqnarray}
in addition to the more standard one $\x_{nm}\equiv \x_n\!-\!\x_m$.

For the contribution of the diagrams $(b)$ and $(b')$, the calculation is analog, except that
 the relevant change of variables is
$(\k_0,\k_2)\mapsto (\P,\K)$, defined as
\begin{eqnarray}
\P &=& \k_0 - \frac{k_0^+}{q^+}\, \q \nonumber\\
\K &=& \k_2 -\frac{k_2^+}{k_1^+\!+\!k_2^+}\, (\k_1\!+\!\k_2)
= \left(\frac{k_1^+}{k_1^+\!+\!k_2^+}\right) \left[\k_2 -\frac{k_2^+}{k_1^+}\, \k_1\right]
\label{cv_Fourier_b}
\,
\end{eqnarray}
and that the transverse delta function is used to eliminate $\k_1$.

After straightforward calculations, one can then write the $q\bar{q}g$ LFWF in mixed space as a linear combination of Fourier integrals of the type ${\cal I}$ and ${\cal I}^{lm}$, defined in Eqs.~\eqref{def_int_I} and \eqref{def_int_Ilm} respectively, and extract universal factors like in Eq.~\eqref{def_reduced_LFWF_mixed_qqbarg}, to get the $q\bar{q}g$ reduced LFWF in mixed space

\begin{eqnarray}
\widetilde{\psi}_{\gamma_{\lambda}^{*}\rightarrow q_0\bar{q}_1g_2}^{\textrm{Tree}}
&=&  e\, e_f\, g\;
\varepsilon_{\lambda}^{i}\, \varepsilon_{\lambda_2}^{j *}\;
\Bigg\{-\frac{1}{(k_{0}^+\!+\!k_{2}^+)q^+}\;
   {\cal I}^{lm}\!\left(\x_{0+2;1},\x_{20};\overline{Q}_{(a)}^2,{\cal C}_{(a)}\right)\;
 \nonumber\\
& &
\hspace{2.7cm}
\times
  \overline{u_G}(0) \gamma^+\!\! \left[(2k_{0}^+\!+\!k_{2}^+) \delta^{jm}
   + \frac{k_{2}^+}{2}\, [\gamma^j,\gamma^m]\right]\!
   \left[(2k_{1}^+\!-\!q^+) \delta^{il}
   + \frac{q^+}{2}\, [\gamma^i,\gamma^l]\right]\!  v_G(1)
\nonumber\\
& &
\hspace{2.25cm}
-\frac{1}{(k_{1}^+\!+\!k_{2}^+)q^+}\;
   {\cal I}^{lm}\!\left(\x_{0;1+2},\x_{21};\overline{Q}_{(b)}^2,{\cal C}_{(b)}\right)\;
 \nonumber\\
& &
\hspace{2.7cm}
 \times
  \overline{u_G}(0) \gamma^+\!\! \left[(2k_{0}^+\!-\!q^+) \delta^{il}
   - \frac{q^+}{2}\, [\gamma^i,\gamma^l]\right]\! \left[(2k_{1}^+\!+\!k_{2}^+) \delta^{jm}
   - \frac{k_{2}^+}{2}\, [\gamma^j,\gamma^m]\right]
   \!  v_G(1)
\nonumber\\
& &
\hspace{2.25cm}
+ \frac{k_{2}^+ k_{0}^+}{(k_{0}^+\!+\!k_{2}^+)^2}\;
{\cal I}\!\left(\x_{0+2;1},\x_{20};\overline{Q}_{(a)}^2,{\cal C}_{(a)}\right)\;
 \overline{u_G}(0)\, \gamma^+ \gamma^j \gamma^i  v_G(1)
 \nonumber\\
& &
\hspace{2.25cm}
- \frac{k_{2}^+ k_{1}^+}{(k_{1}^+\!+\!k_{2}^+)^2}\;   {\cal I}\!\left(\x_{0;1+2},\x_{21};\overline{Q}_{(b)}^2,{\cal C}_{(b)}\right)\;  \overline{u_G}(0)\, \gamma^+ \gamma^i \gamma^j  v_G(1)\;
\Bigg\}
\, ,
\label{qqbarg_WF_T_mixed}
\end{eqnarray}
introducing the notations
\begin{eqnarray}
{\cal C}_{(a)}&\equiv & \frac{q^+\, k_0^+\,k_2^+}{k_1^+ (k_0^+\!+\!k_2^+)^2}
\\
{\cal C}_{(b)}&\equiv & \frac{q^+\, k_1^+\,k_2^+}{k_0^+ (k_1^+\!+\!k_2^+)^2}
\, .
\end{eqnarray}

In the following, the compact notations
\begin{eqnarray}
{\cal I}^{lm}\!\left(a\right) &\equiv & {\cal I}^{lm}\!\left(\x_{0+2;1},\x_{20};\overline{Q}_{(a)}^2,{\cal C}_{(a)}\right)
\\
{\cal I}^{lm}\!\left(b\right) &\equiv & {\cal I}^{lm}\!\left(\x_{0;1+2},\x_{21};\overline{Q}_{(b)}^2,{\cal C}_{(b)}\right)
\\
{\cal I}\left(a\right) &\equiv & {\cal I}\left(\x_{0+2;1},\x_{20};\overline{Q}_{(a)}^2,{\cal C}_{(a)}\right)
\\
{\cal I}\left(b\right) &\equiv & {\cal I}\left(\x_{0;1+2},\x_{21};\overline{Q}_{(b)}^2,{\cal C}_{(b)}\right)
\end{eqnarray}
will be used.


\subsubsection{D=4 result}

The expression \eqref{qqbarg_WF_T_mixed} for the $q\bar{q}g$ LFWF is valid for arbitrary dimension $D$. However, it remains to evaluate the Fourier integrals and the spinor bilinears structures.
Each of these two tasks is more complicated for generic $D$ than for $D=4$. Let us comment now on the spinor part.
In arbitrary $D$, one has the identity
\begin{eqnarray}
\overline{u_G}(0) \gamma^+ v_G(1)= \sqrt{2 k_0^+}\, \sqrt{2 k_1^+}\; \delta_{h_1,-h_0}
\, .
\label{u_gammaplus_v}
\end{eqnarray}
However, for generic $D$, there is in general no such simple expression for spinor bilinears containing at least a commutator like $[\gamma^i,\gamma^j]$. Instead, one should continue the Dirac algebra calculations at the cross-section level.

For the particular case $D=4$, due to the structure of the Lorentz group in $D=4$, one has the identity
\begin{eqnarray}
\frac{i}{4}\, \left[\gamma^i,\gamma^j\right]=  \epsilon^{ij}\, S^3
\label{rel_gamma_i_com_and_helicity}
\end{eqnarray}
where $\epsilon^{ij}$ is antisymmetric with $\epsilon^{12}=+1$, and $S^3$
is the light-front helicity operator, acting on the good components of spinors as
\begin{eqnarray}
S^3\, u_G(p^+,h)&=& h\: u_G(p^+,h)
\\
S^3\, v_G(p^+,h)&=& -h\: v_G(p^+,h)
\end{eqnarray}
Hence, for $D=4$, there is a drastic simplification of the Dirac algebra: one can get rid of the commutators $[\gamma^i,\gamma^j]$ at the LFWF level, and finally use the identity \eqref{u_gammaplus_v}. In this way, one obtains for the $q\bar{q}g$ LFWF the expression
\begin{eqnarray}
&&\widetilde{\psi}_{\gamma_{\lambda}^{*}\rightarrow q_0\bar{q}_1g_2}^{\textrm{Tree}}
=  e\, e_f\, g\;
\varepsilon_{\lambda}^{i}\, \varepsilon_{\lambda_2}^{j *}\;
 \sqrt{2 k_0^+}\, \sqrt{2 k_1^+}\; \delta_{h_1,-h_0}
\nonumber\\
 & &\hspace{1.2cm} \times
\Bigg\{-\frac{1}{(k_{0}^+\!+\!k_{2}^+)q^+}\;
 \Big[(2k_{0}^+\!+\!k_{2}^+) \delta^{jm}
   -i\, k_{2}^+\, (2h_0)\,  \epsilon^{jm} \Big]\!
   \Big[(2k_{1}^+\!-\!q^+) \delta^{il}
   -i\, q^+\, (2h_0)\, \epsilon^{il} \Big]\;
   {\cal I}^{lm}\!\left(a\right)
\nonumber\\
& &
\hspace{1.9cm}
-\frac{1}{(k_{1}^+\!+\!k_{2}^+)q^+}\;
  \Big[(2k_{0}^+\!-\!q^+) \delta^{il}
  + i\, q^+\, (2h_0)\, \epsilon^{il}\Big]\! \Big[(2k_{1}^+\!+\!k_{2}^+) \delta^{jm}
   +i\, k_{2}^+(2h_0)\,  \epsilon^{jm}\Big]\;
   {\cal I}^{lm}\!\left(b\right)
\nonumber\\
& &
\hspace{1.9cm}
- \frac{k_{2}^+ k_{0}^+}{(k_{0}^+\!+\!k_{2}^+)^2}\;
\Big[\delta^{ij}  - i\, (2h_0)\, \epsilon^{ij}\Big]\;
{\cal I}\!\left(a\right)
+ \frac{k_{2}^+ k_{1}^+}{(k_{1}^+\!+\!k_{2}^+)^2}\;
 \Big[\delta^{ij} + i\, (2h_0)\, \epsilon^{ij}\Big]\;
{\cal I}\!\left(b\right)
\Bigg\}
\, .
\label{qqbarg_WF_T_mixed_4D_1}
\end{eqnarray}
Moreover, in $D=4$, both of the Fourier integrals ${\cal I}$ and ${\cal I}^{lm}$ can be calculated analytically, see appendix \ref{sec:FTs}. Inserting the results \eqref{int_I_result} and \eqref{int_Ilm_result_4D}, one gets
\begin{eqnarray}
&&\widetilde{\psi}_{\gamma_{\lambda}^{*}\rightarrow q_0\bar{q}_1g_2}^{\textrm{Tree}}
=    \frac{ e\, e_f\, g}{(2\pi)^2}\;
\varepsilon_{\lambda}^{i}\, \varepsilon_{\lambda_2}^{j *}\;
 \sqrt{2 k_0^+}\, \sqrt{2 k_1^+}\; \delta_{h_1,-h_0}\;
 \frac{Q}{X_{012}}\; \textrm{K}_1\!\left(Q X_{012}\right)
\nonumber\\
& &\hspace{1.2cm} \times
\Bigg\{\frac{k_{1}^+}{(q^+)^3}\;
 \Big[(2k_{0}^+\!+\!k_{2}^+) \delta^{jm}
   -i\, (2h_0)\, k_{2}^+\,  \epsilon^{jm} \Big]\!
   \Big[(2k_{1}^+\!-\!q^+) \delta^{il}
   -i\, (2h_0)\, q^+\, \epsilon^{il} \Big]\;
   \x_{0+2;1}^l\, \left(\frac{\x_{20}^m}{\x_{20}^2}\right)
\nonumber\\
& &
\hspace{1.9cm}
+\frac{k_{0}^+}{(q^+)^3}\;
  \Big[(2k_{1}^+\!+\!k_{2}^+) \delta^{jm}
   +i(2h_0)\, k_{2}^+\,  \epsilon^{jm}\Big]\! \Big[(2k_{0}^+\!-\!q^+) \delta^{il}
  + i\, (2h_0)\, q^+\, \epsilon^{il}\Big] \;
  \x_{0;1+2}^l\, \left(\frac{\x_{21}^m}{\x_{21}^2}\right)
\nonumber\\
& &
\hspace{1.9cm}
- \frac{k_{0}^+ k_{1}^+ k_{2}^+}{(k_{0}^+\!+\!k_{2}^+)(q^+)^2}\;
\Big[\delta^{ij}  - i\, (2h_0)\, \epsilon^{ij}\Big]
+ \frac{k_{0}^+ k_{1}^+ k_{2}^+}{(k_{1}^+\!+\!k_{2}^+)(q^+)^2}\;
 \Big[\delta^{ij} + i\, (2h_0)\, \epsilon^{ij}\Big]
\Bigg\}
\, ,
\label{qqbarg_WF_T_mixed_4D_fin}
\end{eqnarray}
where the variable $X_{012}$ is defined by
\begin{eqnarray}
X_{012}^2 &=& \frac{1}{(q^+)^2}\, \Big[k_{0}^+ k_{1}^+ \x_{01}^2 + k_{0}^+ k_{2}^+ \x_{02}^2+ k_{1}^+ k_{2}^+ \x_{12}^2 \Big]
\, .
\end{eqnarray}
As discussed in the section II.C of Ref.~\cite{Beuf:2011xd}, the quantity $Q^2 X_{012}^2$ corresponds to the ratio of the formation time of the $q\bar{q}g$ Fock state over the lifetime of the virtual photon out of which it is formed.
The dependence of the LFWF \eqref{qqbarg_WF_T_mixed_4D_fin} on the virtuality $Q$ is limited to the factor $Q\; \textrm{K}_1\!\left(Q X_{012}\right)$, which cuts off exponentially the contribution from Fock states with formation time larger than the photon lifetime, and which has a finite limit in the real photon case:
\begin{eqnarray}
Q\; \textrm{K}_1\!\left(Q X_{012}\right) &\rightarrow & \frac{1}{X_{012}}
\end{eqnarray}
for $Q\rightarrow 0$.


\subsection{Longitudinal photon case}

\subsubsection{Momentum space results}

In the case of a longitudinal photon, there are only two light-front perturbation theory diagrams contributing to the $q\bar{q}g$ LFWF at tree level, which are the analog of the diagrams $(a)$ and $(b)$ from Fig.~\ref{Fig:gammaT_qqbarg}, and there is no diagram with an instantaneous fermion line. Indeed, in light-front perturbation theory, the longitudinal photons (or gluons) correspond to the instantaneous Coulomb exchange contained inside non-local four points vertices. As explained in the appendix A.3 of Ref.~\cite{Beuf:2011xd}, one can effectively define LFWFs inside a longitudinal photon by slicing the $l\rightarrow lq\bar{q}$ vertex initiating each diagram contributing to the LFWF for lepton$+$colored partons Fock states inside a dressed lepton $l$.
Then, one can calculate the LFWF for QCD Fock states inside an longitudinal dressed photon by using the effective $\gamma^*_{L}\rightarrow q\bar{q}$ vertex
\begin{eqnarray}
V_{\gamma_L \rightarrow q_{0} \bar{q}_{1}}
&=& (2\pi)^{D-1}\delta^{(D-1)}(\underline{k_{0}}+\underline{k_{1}}\!-\!
\underline{q})\; \delta_{\alpha_{0},\, \alpha_{1}}\;
\mu^{2-\frac{D}{2}}\, e\, e_f\,  \frac{Q}{q^+}\;
\overline{u}(0)\,  \gamma^+\, v(1)
\, .
\label{gammaL_to_qqbar_vertex}
\end{eqnarray}
Hence, the tree-level $q\bar{q}g$ LFWF inside a longitudinal photon can be written in momentum space as
\begin{eqnarray}
\Psi_{\gamma_L^{*}\rightarrow q_0\bar{q}_1g_2}^{\textrm{Tree}}
&=& \sum_{q_{0'} \textrm{ states}}
\frac{
\langle 0| a_{2} b_0\, V_I(0)\, b_{0'}^{\dag} |0 \rangle\;
V_{\gamma_L \rightarrow q_{0'} \bar{q}_{1}}
}{(ED_{q_0\bar{q}_1g_2})\; (ED_{q_{0'}\bar{q}_1})}
+ \sum_{\bar{q}_{1'} \textrm{ states}}
\frac{
\langle 0| a_{2} d_{1}\, V_I(0)\, d_{1'}^{\dag} |0 \rangle\;
V_{\gamma_L \rightarrow q_{0} \bar{q}_{1'}}
}{(ED_{q_0\bar{q}_1g_2})\; (ED_{q_{0}\bar{q}_{1'}})}
\nonumber\\
&=&\frac{(2\pi)^{D\!-\!1} \delta^{(D\!-\!1)}(\underline{k_0}\!+\!\underline{k_{1}}\!+\!\underline{k_{2}}\!-\!\underline{q})}{(ED_{q_0\bar{q}_1g_2})}
\; (\mu^2)^{2-\frac{D}{2}}\, e\, e_f\, g\; t^{a_2}_{\alpha_{0}\alpha_{1}}\, \frac{Q}{q^+}
\nonumber\\
& & \times
\Bigg\{
 \int \frac{d^{D-1} \underline{k_{0'}}}{(2\pi)^{D-1}}\; \frac{\theta(k_{0'}^+)}{2 k_{0'}^+}\;
 \frac{(2\pi)^{D\!-\!1} \delta^{(D\!-\!1)}(\underline{k_0}\!+\!\underline{k_{2}}\!-\!\underline{k_{0'}})}{(ED_{q_{0'}\bar{q}_1})}\;
 \sum_{h_{0'}=\pm 1/2} \overline{u}(0)\, \slashed{\epsilon}_{\lambda_2}^*\!(\underline{k_2})\, u(0')\;\;
 \overline{u}(0') \gamma^+ v(1)
\nonumber\\
& &
\hspace{0.3cm}
-  \int \frac{d^{D-1} \underline{k_{1'}}}{(2\pi)^{D-1}}\; \frac{\theta(k_{1'}^+)}{2 k_{1'}^+}\;
 \frac{(2\pi)^{D\!-\!1} \delta^{(D\!-\!1)}(\underline{k_1}\!+\!\underline{k_{2}}\!-\!\underline{k_{1'}})}{(ED_{q_{0}\bar{q}_{1'}})}\;
 \sum_{h_{1'}=\pm 1/2} \overline{u}(0) \gamma^+ v(1')\;\;
 \overline{v}(1')\, \slashed{\epsilon}_{\lambda_2}^*\!(\underline{k_2})\,  v(1)
\Bigg\}\, ,
\label{qqbarg_WF_L_1}
\end{eqnarray}
using the same energy denominators \eqref{ED_qqbarg}, \eqref{ED_a} and \eqref{ED_b} as in the transverse photon case.
Using the relation \eqref{spinor_bilin_mom_extract} to make the full transverse momentum dependence explicit, one finds
\begin{eqnarray}
\Psi_{\gamma_L^{*}\rightarrow q_0\bar{q}_1g_2}^{\textrm{Tree}}
&=& (2\pi)^{D\!-\!1} \delta^{(D\!-\!1)}(\underline{k_0}\!+\!\underline{k_{1}}\!+\!\underline{k_{2}}\!-\!\underline{q})
\;  e\, e_f\, g\; t^{a_2}_{\alpha_{0}\alpha_{1}}\, \frac{2Q}{(q^+)^2}\, \varepsilon_{\lambda_2}^{j *}\: (\mu^2)^{2-\frac{D}{2}}
\nonumber\\
& & \times
\Bigg\{\frac{k_1^+(k_{0}^+\!+\!k_{2}^+)}{k_{0}^+}\; \frac{\left[\k_{2}^m\!-\!\frac{k_{2}^+}{k_{0}^+}\, \k_{0}^m\right]\:
 \overline{u_G}(0)\, \gamma^+ \left[(2k_{0}^+\!+\!k_{2}^+) \delta^{jm}
   + \frac{k_{2}^+}{2}\, [\gamma^j,\gamma^m]\right]  v_G(1)
}{
\left[\left(\k_{1}\!-\!\frac{k_{1}^+}{q^+}\, \q\right)^2
+\overline{Q}_{(a)}^2\right]
\left\{\left(\k_{2}\!-\!\frac{k_{2}^+}{k_{0}^+}\, \k_{0}\right)^2
+ \frac{q^+ k_{2}^+}{k_{0}^+ k_{1}^+} \left[ \left(\k_{1}\!-\!\frac{k_{1}^+}{q^+}\, \q\right)^2
 +\overline{Q}_{(a)}^2\right] \right\}}
\nonumber\\
& &
\hspace{0.3cm}
-  \frac{k_0^+(k_{1}^+\!+\!k_{2}^+)}{k_{1}^+}\; \frac{\left[\k_{2}^m\!-\!\frac{k_{2}^+}{k_{1}^+}\, \k_{1}^m\right]\:
 \overline{u_G}(0)\, \gamma^+ \left[(2k_{1}^+\!+\!k_{2}^+) \delta^{jm}
   - \frac{k_{2}^+}{2}\, [\gamma^j,\gamma^m]\right]  v_G(1)
}{
\left[\left(\k_{0}\!-\!\frac{k_{0}^+}{q^+}\, \q\right)^2
+\overline{Q}_{(b)}^2\right]
\left\{\left(\k_{2}\!-\!\frac{k_{2}^+}{k_{1}^+}\, \k_{1}\right)^2
+ \frac{q^+ k_{2}^+}{k_{0}^+ k_{1}^+} \left[ \left(\k_{0}\!-\!\frac{k_{0}^+}{q^+}\, \q\right)^2
 +\overline{Q}_{(b)}^2\right] \right\}}
\Bigg\}\, .
\label{qqbarg_WF_L_2}
\end{eqnarray}


\subsubsection{Mixed space results}

In the longitudinal photon case, the Fourier transform to mixed space of the $q\bar{q}g$  LFWF is still defined by Eq.~\eqref{def_LFWF_mix}, and can be performed in a similar way as in the transverse photon case, using the change of variables defined by Eqs.~\eqref{cv_Fourier_a} for the diagram $(a)$, and by Eqs.~\eqref{cv_Fourier_b} for the diagram $(b)$. After removing the universal factors like in Eq.~\eqref{def_reduced_LFWF_mixed_qqbarg}, one obtains the reduced LFWF in mixed space
\begin{eqnarray}
\widetilde{\psi}_{\gamma_L^{*}\rightarrow q_0\bar{q}_1g_2}^{\textrm{Tree}}
&=&  e\, e_f\, g\;  \varepsilon_{\lambda_2}^{j *}\, 2Q\;
\Bigg\{\frac{k_{1}^+}{(q^+)^2}\; \;
\overline{u_G}(0) \gamma^+\! \left[(2k_{0}^+\!+\!k_{2}^+) \delta^{jm}
   + \frac{k_{2}^+}{2}\, [\gamma^j,\gamma^m]\right]\!  v_G(1)\;\;
   {\cal I}^{m}\!\left(\x_{0+2;1},\x_{20};\overline{Q}_{(a)}^2,{\cal C}_{(a)}\right)\;
\nonumber\\
& &
\hspace{1.5cm}
-\frac{k_{0}^+}{(q^+)^2}\; \;
\overline{u_G}(0) \gamma^+\! \left[(2k_{1}^+\!+\!k_{2}^+) \delta^{jm}
   - \frac{k_{2}^+}{2}\, [\gamma^j,\gamma^m]\right]\!  v_G(1)\;\;
   {\cal I}^{m}\!\left(\x_{0;1+2},\x_{21};\overline{Q}_{(b)}^2,{\cal C}_{(b)}\right)\;
\Bigg\}
\, ,
\label{qqbarg_WF_L_mixed}
\end{eqnarray}
as a linear combination of Fourier integrals of the type ${\cal I}^{m}$, defined in Eq.~\eqref{def_int_Im}.

In the following, the compact notations
\begin{eqnarray}
{\cal I}^{m}\!\left(a\right) &\equiv & {\cal I}^{m}\!\left(\x_{0+2;1},\x_{20};\overline{Q}_{(a)}^2,{\cal C}_{(a)}\right)
\\
{\cal I}^{m}\!\left(b\right) &\equiv & {\cal I}^{m}\!\left(\x_{0;1+2},\x_{21};\overline{Q}_{(b)}^2,{\cal C}_{(b)}\right)
\end{eqnarray}
will be used.


\subsubsection{D=4 result}

In the particular case of $D=4$, the spinor bilinears can be calculated using the relation \eqref{rel_gamma_i_com_and_helicity}. One obtains
\begin{eqnarray}
&&\widetilde{\psi}_{\gamma_L^{*}\rightarrow q_0\bar{q}_1g_2}^{\textrm{Tree}}
=   e\, e_f\, g\;  \varepsilon_{\lambda_2}^{j *}\, 2Q\;
 \sqrt{2 k_0^+}\, \sqrt{2 k_1^+}\; \delta_{h_1,-h_0}
\nonumber\\
& &
\times
\Bigg\{\frac{k_{1}^+}{(q^+)^2}\; \;
\Big[(2k_{0}^+\!+\!k_{2}^+) \delta^{jm}
   -i\, k_{2}^+(2h_0)\,  \epsilon^{jm}\Big]\;\;
   {\cal I}^{m}\!\left(a\right)\;
-\frac{k_{0}^+}{(q^+)^2}\; \;
\Big[(2k_{1}^+\!+\!k_{2}^+) \delta^{jm}
   +i\, k_{2}^+(2h_0)\,  \epsilon^{jm}\Big]\;\;
   {\cal I}^{m}\!\left(b\right)\;
\Bigg\}
\, .
\label{qqbarg_WF_L_mixed_4D_1}
\end{eqnarray}
Moreover, in $D=4$, the Fourier integrals of the type ${\cal I}^{m}$ can be calculated analytically, see Eq.~\eqref{int_Im_result_4D}. Hence, one gets
\begin{eqnarray}
&& \widetilde{\psi}_{\gamma_L^{*}\rightarrow q_0\bar{q}_1g_2}^{\textrm{Tree}}
=      e\, e_f\, g\; \frac{i}{(2\pi)^2} \varepsilon_{\lambda_2}^{j *}\,
2Q\; \textrm{K}_0\!\left(Q X_{012}\right)
 \sqrt{2 k_0^+}\, \sqrt{2 k_1^+}\; \delta_{h_1,-h_0}
\nonumber\\
& & 
\times
\Bigg\{\frac{k_{1}^+}{(q^+)^2}\:
\Big[(2k_{0}^+\!+\!k_{2}^+) \delta^{jm}
   -i(2h_0)\, k_{2}^+\,  \epsilon^{jm}\Big]\:
   \left(\frac{\x_{20}^m}{\x_{20}^2}\right)
-\frac{k_{0}^+}{(q^+)^2}\:
\Big[(2k_{1}^+\!+\!k_{2}^+) \delta^{jm}
   +i(2h_0)\, k_{2}^+\,  \epsilon^{jm}\Big]\:
   \left(\frac{\x_{21}^m}{\x_{21}^2}\right)
\Bigg\}
\, .
\label{qqbarg_WF_L_mixed_4D_fin}
\end{eqnarray}
Note that the $Q$ dependence occurs via the factor $Q\; \textrm{K}_0\!\left(Q X_{012}\right)$, which suppresses
exponentially the contribution from Fock states with formation time larger than the photon lifetime, and also vanishes in the opposite limit of real photon. This is to be expected, since a real photon has no longitudinal polarization.


\section{Longitudinal photon cross-section at NLO\label{sec:sigma_L_NLO}}

\subsection{Quark-antiquark-gluon contribution in D dimensions}
The $q\bar{q}g$ contribution to the longitudinal photon cross-section is provided by the second term in Eq.~\eqref{sigma_gamma_master_eq},
\begin{eqnarray}
\sigma^{\gamma^*}_{L} \Big|_{q\bar{q}g}
&=& 2 N_c C_F
\sum_{f}
\int \frac{[dk^+_0]}{2\pi}\; \frac{\theta(k^+_0)}{2k^+_0}
\int \frac{[dk^+_1]}{2\pi}\; \frac{\theta(k^+_1)}{2k^+_1}\;
\int \frac{[dk^+_2]}{2\pi}\; \frac{\theta(k^+_2)}{2k^+_2}\;
\frac{2\pi \delta(k_0^+\!+\!k_1^+\!+\!k_2^+\!-\!q^+)}{2q^+}\;
\nonumber\\
&& \times
\int d^{D-2} \x_0 \int d^{D-2} \x_1 \int d^{D-2} \x_2\;
{\textrm{Re}}\left[1-{\cal S}^{(3)}_{012}\right]\;
\sum_{T\textrm{ pol. }\lambda_2}\;\;  \sum_{h_0, h_1 = \pm 1/2}
\left|\widetilde{\psi}_{\gamma_{L}^*\rightarrow q_0 \bar{q_1} g_2}\right|^2\;
\nonumber\\
&=&  N_c C_F
\sum_{f}\;
\widetilde{\sum_{{\cal P.S.}\, q_0 \bar{q}_1g_2; D}}\;
{\textrm{Re}}\left[1-{\cal S}^{(3)}_{012}\right]\;
\frac{(2\pi)}{2(2k_0^+)(2k_1^+)}\:
\sum_{T\textrm{ pol. }\lambda_2}\;\;  \sum_{h_0, h_1 = \pm 1/2}
\left|\widetilde{\psi}_{\gamma_{L}^*\rightarrow q_0 \bar{q_1} g_2}\right|^2\;
\, ,\label{sigma_L_qqbarg_master}
\end{eqnarray}
introducing the notation
\begin{eqnarray}
&& \hspace{-0.5cm}
\widetilde{\sum_{{\cal P.S.}\, q_0 \bar{q}_1g_2; D}}\; \Big[\cdots\Big]
= \int_{0}^{+\infty} \hspace{-0.5cm}[dk^+_0] \int_{0}^{+\infty}  \hspace{-0.5cm}[dk^+_1]
\int_{0}^{+\infty}  
\frac{[dk^+_2]}{k^+_2}\; \frac{\delta(k_0^+\!+\!k_1^+\!+\!k_2^+\!-\!q^+)}{q^+}\;
\int \frac{d^{D-2} \x_0}{2\pi} \int \frac{d^{D-2} \x_1}{2\pi} \int \frac{d^{D-2} \x_2}{2\pi}\: \Big[\cdots\Big]
\, ,\label{phase_space_qqbarg}
\end{eqnarray}
for the phase-space integration for the $q\bar{q}g$ Fock state in $D$ dimensional mixed space\footnote{A regulator is needed for at least the $k_2^+$ integration, for example a cut-off at $k_2^+=k^+_{\min}$, as used in sec.~\ref{sec:qqbar_NLO} and in Ref.~\cite{Beuf:2016wdz}. A large part of the calculations presented in this paper is independent on the precise regularization procedure. Then, the square brackets appearing in the integration measures $[dk_n^+]$ in Eqs.~\eqref{sigma_L_qqbarg_master} and \eqref{phase_space_qqbarg} are there to indicate the use of an arbitrary regularization for these integrals.}.

In arbitrary dimension $D$, using the expression \eqref{qqbarg_WF_L_mixed} for the LFWF, one finds
\begin{eqnarray}
&&\sum_{T\textrm{ pol. }\lambda_2}\;\;  \sum_{h_0, h_1 = \pm 1/2}
\left|\widetilde{\psi}_{\gamma_{L}^*\rightarrow q_0 \bar{q_1} g_2}\right|^2
=
e^2\, e_f^2\, g^2\, 4Q^2\; \sum_{T\textrm{ pol. }\lambda_2}\varepsilon_{\lambda_2}^{j'} \varepsilon_{\lambda_2}^{j *}\, \!\!\!\sum_{h_0, h_1 = \pm 1/2} \!\!\!
\overline{v_G}(1) \gamma^+
\nonumber\\
& & 
\times
\bigg\{\frac{k_{1}^+}{(q^+)^2}
 \left[(2k_{0}^+\!+\!k_{2}^+) \delta^{j'm'}
   \!-\! \frac{k_{2}^+}{2}\, [\gamma^{j'},\gamma^{m'}]\right]
   {\cal I}^{m'}\!\left(a\right)^*
\!-\!\frac{k_{0}^+}{(q^+)^2}
 \left[(2k_{1}^+\!+\!k_{2}^+) \delta^{j'm'}
   \!+\! \frac{k_{2}^+}{2}\, [\gamma^{j'},\gamma^{m'}]\right]
   {\cal I}^{m'}\!\left(b\right)^*
\bigg\} u_G(0)
\nonumber\\
& & 
\times \hspace{0.1cm}
\overline{u_G}(0) \gamma^+
\bigg\{\frac{k_{1}^+}{(q^+)^2}
 \left[(2k_{0}^+\!+\!k_{2}^+) \delta^{jm}
   \!+\! \frac{k_{2}^+}{2}\, [\gamma^j,\gamma^m]\right]
   {\cal I}^{m}\!\left(a\right)
\!-\!\frac{k_{0}^+}{(q^+)^2}
 \left[(2k_{1}^+\!+\!k_{2}^+) \delta^{jm}
   \!-\! \frac{k_{2}^+}{2}\, [\gamma^j,\gamma^m]\right]
   {\cal I}^{m}\!\left(b\right)
\bigg\} v_G(1)
\nonumber\\
&=&
 2 (2k_0^+)(2k_1^+) (4\pi)^2 \alpha_{em} e_f^2 \alpha_s \frac{4Q^2}{(q^+)^4}
\Bigg\{
\bigg| k_{1}^+(2k_{0}^+\!+\!k_{2}^+) {\cal I}^{m}\!\left(a\right)
\!-\!k_{0}^+(2k_{1}^+\!+\!k_{2}^+) {\cal I}^{m}\!\left(b\right) \bigg|^2
\nonumber\\
& & \hspace{5.5cm}
\!+\! (D\!-\!3) (k_{2}^+)^2
\bigg| k_{1}^+ {\cal I}^{m}\!\left(a\right)
\!+\!k_{0}^+ {\cal I}^{m}\!\left(b\right) \bigg|^2
\Bigg\}
\end{eqnarray}
thanks to the identities \eqref{trace_gamma_1}, \eqref{trace_gamma_2} and \eqref{trace_gamma_3}, as well as others given in the appendix A of Ref.~\cite{Beuf:2016wdz}. Inserting this result into Eq.~\eqref{sigma_L_qqbarg_master}, one gets
\begin{eqnarray}
\sigma^{\gamma^*}_{L} \Big|_{q\bar{q}g}
&=& 4 N_c \alpha_{em}  \left(\frac{\alpha_s C_F}{\pi}\right)
\sum_{f}  e_f^2
\widetilde{\sum_{{\cal P.S.}\, q_0 \bar{q}_1g_2; D}}\;
{\textrm{Re}}\left[1-{\cal S}^{(3)}_{012}\right]\;
\frac{4 Q^2}{(q^+)^4}\; \frac{(2\pi)^4}{2}\;
\nonumber\\
&& \times
\Bigg\{
(k_{1}^+)^2\, \Big[4k_{0}^+(k_{0}^+\!+\!k_{2}^+) + (D\!-\!2)(k_{2}^+)^2\Big]\,
\Big|{\cal I}^{m}\!\left(a\right)\Big|^2
+
(k_{0}^+)^2\, \Big[4k_{1}^+(k_{1}^+\!+\!k_{2}^+) + (D\!-\!2)(k_{2}^+)^2\Big]\,
\Big|{\cal I}^{m}\!\left(b\right)\Big|^2
\nonumber\\
&& \hspace{2.5cm}
-2k_{0}^+ k_{1}^+ \Big[2k_{1}^+(k_{0}^+\!+\!k_{2}^+)
+2k_{0}^+(k_{1}^+\!+\!k_{2}^+)-(D\!-\!4)(k_{2}^+)^2\Big]\,
{\textrm{Re}}\Big({\cal I}^{m}\!\left(a\right)^*\, {\cal I}^{m}\!\left(b\right) \Big)
\Bigg\}
\, ,
\label{sigma_L_qqbarg_1}
\end{eqnarray}
where the three terms correspond respectively to the contributions from the square of the diagram $(a)$, the square of the diagram $(b)$, and the interferences between the $(a)$ and $(b)$ diagrams.


\subsection{Subtracting the UV divergences}

The expression \eqref{sigma_L_qqbarg_1} has UV divergences for $D\!=\!4$ coming from the integration over $\x_2$, which need to be taken care of. More precisely, the $|a|^2$ contribution has a UV divergence in the $\x_2\rightarrow\x_0$ regime and the $|b|^2$ contribution has one in the $\x_2\rightarrow\x_1$ regime, whereas the interference contribution is UV finite.

At the accuracy of the present calculation, UV renormalization is not relevant yet, so that the UV divergences have to cancel between the $q\bar{q}$ and the $q\bar{q}g$ contributions to the cross section. Due to the presence of the dipole or tripole amplitudes in the integrand, the phase-space integrals cannot be performed analytically. Hence, it is desirable to understand the cancellation of UV divergences at the integrand level.
Since the $q\bar{q}$ and $q\bar{q}g$ contributions have, by definition, phase-spaces of different dimensions, this require some care. This situation is very reminiscent to the cancellation of soft and of collinear divergences between real and virtual higher order corrections to hard scattering processes.

In high-energy QCD, this issue has already been encountered in particular in the calculation of NLO corrections to high-energy evolution equations, for example in the quark loop corrections to the BK equation \cite{Balitsky:2006wa,Kovchegov:2006vj}. The general idea used in these papers to deal with the UV divergence is to rely on the group theory properties of Wilson lines at coincident points, such as
\begin{eqnarray}
\lim_{\x_2\rightarrow \x_0} \Big[t^{b}U_{F}(\mathbf{x}_{0})\Big]\; U_{A}(\mathbf{x}_{2})_{b a}
= \Big[U_{F}(\mathbf{x}_{0})t^{a}\Big]
\end{eqnarray}
which are the mathematical origin of the color coherence of QCD branching processes. Such identities imply in particular
\begin{eqnarray}
\lim_{\x_2\rightarrow \x_0} {\cal S}^{(3)}_{012} = \lim_{\x_2\rightarrow \x_1} {\cal S}^{(3)}_{012}
= {\cal S}_{01}
\, .
\label{tripole_coincident_points}
\end{eqnarray}
Hence, one can regulate a $q\bar{q}g$ contribution which is logarithmically UV divergent for $\x_2\rightarrow\x_0$ or for $\x_2\rightarrow\x_1$ by making the replacement
\begin{eqnarray}
 {\cal S}^{(3)}_{012} \mapsto \Big[ {\cal S}^{(3)}_{012} - {\cal S}_{01} \Big]
 \label{UV_subtract_procedure_naive}
\end{eqnarray}
in its integrand. Then, the term subtracted from the $q\bar{q}g$ contribution has to be added to the $q\bar{q}$ contribution, in order to keep the total cross section unchanged. The ${\cal S}_{01}$ factor contained in the UV subtraction term can be pulled out of the $\x_2$ integral, which might then be evaluated explicitly in generic dimension $D$, before combination with the $q\bar{q}$ contribution. This is the general idea of the method used for example in Ref.~\cite{Balitsky:2006wa} to deal with the UV divergences.

In the present study, it is convenient to slightly modify this procedure. Indeed, the integral ${\cal I}^{m}\!\left(a\right)$ (as well as $ {\cal I}^{m}\!\left(b\right)$) is difficult to calculate fully analytically in generic dimension $D$, whereas this method would require the knowledge of the integral over $\x_2$ of its square, in dimension $D$. However, in the limit $\x_2\rightarrow \x_0$, the integral ${\cal I}^{m}\!\left(a\right)$ is equivalent to the integral ${\cal I}^{m}_{\textrm{UV}}(\x_{01},\x_{20};\overline{Q}_{(a)}^2)$, whose result is given in Eq.~\eqref{Im_UV_approx} for generic $D$. Hence, it is tempting to replace the UV subtraction procedure \eqref{UV_subtract_procedure_naive} by
\begin{eqnarray}
{\textrm{Re}}\left[1-{\cal S}^{(3)}_{012}\right]\; \Big|{\cal I}^{m}\!\left(a\right)\Big|^2
&\mapsto&
\Bigg\{{\textrm{Re}}\left[1-{\cal S}^{(3)}_{012}\right]\; \Big|{\cal I}^{m}\!\left(a\right)\Big|^2
- {\textrm{Re}}\left[1-{\cal S}_{01}\right]\;
\Big|{\cal I}^{m}_{\textrm{UV}}\!\left(\x_{01},\x_{20};\overline{Q}_{(a)}^2\right)\Big|^2
\Bigg\}
 \label{UV_subtract_procedure_1}
\end{eqnarray}
for the $|a|^2$ contribution, and a symmetric one for the $|b|^2$ contribution, in order to facilitate the analytical evaluation of the subtraction terms in dimension $D$ before their combination with the $q\bar{q}$ contribution.

However, in the prescription \eqref{UV_subtract_procedure_1}, the UV subtraction term has an IR divergence from the regime $|\x_{20}|\sim |\x_{21}|\gg|\x_{01}|$, which is absent from the original $|a|^2$ contribution. In that sense, the prescription \eqref{UV_subtract_procedure_1} provides a correct UV subtraction, but is overall a clear oversubtraction which would provide a pathological result. This issue can be corrected easily by including in the subtraction term another piece which cancels this IR divergence while being UV finite, changing the Coulomb behavior of the subtraction term in the IR into a dipolar behavior. Hence, one is lead to the UV subtraction prescription
\begin{eqnarray}
{\textrm{Re}}\left[1-{\cal S}^{(3)}_{012}\right]\; \Big|{\cal I}^{m}\!\left(a\right)\Big|^2
&\mapsto&
\Bigg\{{\textrm{Re}}\left[1-{\cal S}^{(3)}_{012}\right]\; \Big|{\cal I}^{m}\!\left(a\right)\Big|^2
- {\textrm{Re}}\left[1-{\cal S}_{01}\right]\; \bigg[
\Big|{\cal I}^{m}_{\textrm{UV}}\!\left(\x_{01},\x_{20};\overline{Q}_{(a)}^2\right)\Big|^2
\nonumber\\
&& \hspace{3cm}
-{\textrm{Re}}\bigg({\cal I}^{m}_{\textrm{UV}}\!\left(\x_{01},\x_{20};\overline{Q}_{(a)}^2\right)^*\;
{\cal I}^{m}_{\textrm{UV}}\!\left(\x_{01},\x_{21};\overline{Q}_{(a)}^2\right)
\bigg)
\bigg]
\Bigg\}
 \label{UV_subtract_procedure_final}
\end{eqnarray}
for the $|a|^2$ contribution, and a symmetric one for the $|b|^2$ contribution.
Thus, the full expression of the term used to subtract the UV divergence from the $|a|^2$ contribution is
\begin{eqnarray}
\sigma^{\gamma^*}_{L} \Big|_{UV, |a|^2}
&=& 4 N_c \alpha_{em}  \left(\frac{\alpha_s C_F}{\pi}\right)
\sum_{f}  e_f^2
\widetilde{\sum_{{\cal P.S.}\, q_0 \bar{q}_1g_2; D}}\,
{\textrm{Re}}\left[1-{\cal S}_{01}\right]\,
\frac{4 Q^2}{(q^+)^4}\, \frac{(2\pi)^4(k_{1}^+)^2}{2}
\, \Big[4k_{0}^+(k_{0}^+\!+\!k_{2}^+) + (D\!-\!2)(k_{2}^+)^2\Big]
\nonumber\\
&&  \hspace{1cm}
\times \hspace{0.3cm}
\bigg[
\Big|{\cal I}^{m}_{\textrm{UV}}\!\left(\x_{01},\x_{20};\overline{Q}_{(a)}^2\right)\Big|^2
-{\textrm{Re}}\bigg({\cal I}^{m}_{\textrm{UV}}\!\left(\x_{01},\x_{20};\overline{Q}_{(a)}^2\right)^*\;
{\cal I}^{m}_{\textrm{UV}}\!\left(\x_{01},\x_{21};\overline{Q}_{(a)}^2\right)
\bigg)
\bigg]\, .
\label{UV_subtr_term_a2_L}
\end{eqnarray}
Similarly, the UV subtraction term for $|b|^2$ is
\begin{eqnarray}
\sigma^{\gamma^*}_{L} \Big|_{UV, |b|^2}
&=& 4 N_c \alpha_{em}  \left(\frac{\alpha_s C_F}{\pi}\right)
\sum_{f}  e_f^2
\widetilde{\sum_{{\cal P.S.}\, q_0 \bar{q}_1g_2; D}}\,
{\textrm{Re}}\left[1-{\cal S}_{01}\right]\,
\frac{4 Q^2}{(q^+)^4}\, \frac{(2\pi)^4(k_{0}^+)^2}{2}
\, \Big[4k_{1}^+(k_{1}^+\!+\!k_{2}^+) + (D\!-\!2)(k_{2}^+)^2\Big]
\nonumber\\
&& \hspace{1cm}
\times \hspace{0.3cm}
\bigg[
\Big|{\cal I}^{m}_{\textrm{UV}}\!\left(\x_{01},\x_{21};\overline{Q}_{(b)}^2\right)\Big|^2
-{\textrm{Re}}\bigg({\cal I}^{m}_{\textrm{UV}}\!\left(\x_{01},\x_{21};\overline{Q}_{(b)}^2\right)^*\;
{\cal I}^{m}_{\textrm{UV}}\!\left(\x_{01},\x_{20};\overline{Q}_{(b)}^2\right)
\bigg)
\bigg]
\, .
\label{UV_subtr_term_b2_L}
\end{eqnarray}

Now, combining the expression for the contribution $|a|^2$ that can be read from Eq.~\eqref{sigma_L_qqbarg_1} with its UV subtraction term \eqref{UV_subtr_term_a2_L}, one can take the $D\rightarrow 4$ limit safely under the $\x_2$ integral, which allows to use the $D=4$ expression \eqref{int_Im_result_4D} for ${\cal I}^{m}\!\left(a\right)$. This way, one obtains
\begin{eqnarray}
&&\sigma^{\gamma^*}_{L} \Big|_{q\bar{q}g, |a|^2}-\sigma^{\gamma^*}_{L} \Big|_{UV; |a|^2}
= 4 N_c \alpha_{em}  \left(\frac{\alpha_s C_F}{\pi}\right)
\sum_{f}  e_f^2
\widetilde{\sum_{{\cal P.S.}\, q_0 \bar{q}_1g_2; 4}}\,
4 Q^2\, \frac{(k_{1}^+)^2}{(q^+)^4}\,
\Big[2k_{0}^+(k_{0}^+\!+\!k_{2}^+) + (k_{2}^+)^2\Big]
\nonumber\\
&&  \hspace{1cm}
\times
\Bigg\{\frac{1}{\x_{20}^2}\;
 \Big[\textrm{K}_{0}\left(Q X_{012} \right)\Big]^2\;
{\textrm{Re}}\left[1-{\cal S}^{(3)}_{012}\right]\;
-\, \frac{\x_{20}}{\x_{20}^2} \!\cdot\! \left(\frac{\x_{20}}{\x_{20}^2}\!-\!\frac{\x_{21}}{\x_{21}^2}\right)
\Big[\textrm{K}_{0}\left(\overline{Q}_{(a)}\, |\x_{01}|\right)\Big]^2\;
{\textrm{Re}}\left[1-{\cal S}_{01}\right]
\Bigg\}
\label{sigma_L_qqbarg_a2_minus_UV}
\, ,
\end{eqnarray}
and analogously
\begin{eqnarray}
&&\sigma^{\gamma^*}_{L} \Big|_{q\bar{q}g, |b|^2}-\sigma^{\gamma^*}_{L} \Big|_{UV; |b|^2}
= 4 N_c \alpha_{em}  \left(\frac{\alpha_s C_F}{\pi}\right)
\sum_{f}  e_f^2
\widetilde{\sum_{{\cal P.S.}\, q_0 \bar{q}_1g_2; 4}}\,
4 Q^2\, \frac{(k_{0}^+)^2}{(q^+)^4}\,
\Big[2k_{1}^+(k_{1}^+\!+\!k_{2}^+) + (k_{2}^+)^2\Big]
\nonumber\\
&&  \hspace{1cm}
\times
\Bigg\{\frac{1}{\x_{21}^2}\;
 \Big[\textrm{K}_{0}\left(Q X_{012} \right)\Big]^2\;
{\textrm{Re}}\left[1-{\cal S}^{(3)}_{012}\right]\;
-\, \frac{\x_{21}}{\x_{21}^2} \!\cdot\! \left(\frac{\x_{21}}{\x_{21}^2}\!-\!\frac{\x_{20}}{\x_{20}^2}\right)
\Big[\textrm{K}_{0}\left(\overline{Q}_{(b)}\, |\x_{01}|\right)\Big]^2\;
{\textrm{Re}}\left[1-{\cal S}_{01}\right]
\Bigg\}
\label{sigma_L_qqbarg_b2_minus_UV}
\, .
\end{eqnarray}
Finally, the contribution of the interference between the $(a)$ and $(b)$ graphs is both UV and IR finite, so that the $D\rightarrow 4$ limit can be taken without any problem. One finds
\begin{eqnarray}
\sigma^{\gamma^*}_{L} \Big|_{q\bar{q}g;\, (a),(b) \textrm{interf.}}
&=& 4 N_c \alpha_{em}  \left(\frac{\alpha_s C_F}{\pi}\right)
\sum_{f}  e_f^2
\widetilde{\sum_{{\cal P.S.}\, q_0 \bar{q}_1g_2; 4}}\,
4 Q^2\, \frac{(-2)k_{0}^+k_{1}^+}{(q^+)^4}\,
\Big[k_{1}^+(k_{0}^+\!+\!k_{2}^+) + k_{0}^+(k_{1}^+\!+\!k_{2}^+)\Big]
\nonumber\\
&&  \hspace{1cm}
\times
\,
 \left(\frac{\x_{20}\!\cdot\! \x_{21}}{\x_{20}^2 \x_{21}^2}\right)\;
 \Big[\textrm{K}_{0}\left(Q X_{012} \right)\Big]^2\;
{\textrm{Re}}\left[1-{\cal S}^{(3)}_{012}\right]
\label{sigma_L_qqbarg_ab_interf}
\, .
\end{eqnarray}
The expressions \eqref{sigma_L_qqbarg_a2_minus_UV}, \eqref{sigma_L_qqbarg_b2_minus_UV} and \eqref{sigma_L_qqbarg_ab_interf} are consistent with the $D=4$ result from Ref.~\cite{Beuf:2011xd} for the $q\bar{q}g$ contribution to the longitudinal photon cross section (see Eqs.~(43) and (50) there), but obviously with different UV subtraction terms.

In each of the three pieces \eqref{sigma_L_qqbarg_a2_minus_UV}, \eqref{sigma_L_qqbarg_b2_minus_UV} and \eqref{sigma_L_qqbarg_ab_interf} the transverse integrations are now convergent, and the only potential divergence is a logarithmic divergence at small $k^+_2$. In view of the resummation of the corresponding high-energy leading logs, performed in section~\ref{sec:LL_resum}, it is convenient to split the interference contribution \eqref{sigma_L_qqbarg_ab_interf} in two pieces, and group one of them with the UV-subtracted $|a|^2$ contribution \eqref{sigma_L_qqbarg_a2_minus_UV} and the other with the the UV-subtracted $|b|^2$ contribution \eqref{sigma_L_qqbarg_b2_minus_UV}. One then obtains
\begin{eqnarray}
\sigma^{\gamma^*}_{L} \Big|_{q\rightarrow g}
&=& 4 N_c \alpha_{em}  \left(\frac{\alpha_s C_F}{\pi}\right)
\sum_{f}  e_f^2
\widetilde{\sum_{{\cal P.S.}\, q_0 \bar{q}_1g_2; 4}}\,
4 Q^2\, \frac{(k_{1}^+)^2}{(q^+)^4}\,
\Bigg\{
(k_2^+)^2\,  \left(\frac{\x_{20}\!\cdot\! \x_{21}}{\x_{20}^2 \x_{21}^2}\right)\;
 \Big(\textrm{K}_{0}\left(Q X_{012} \right)\Big)^2\;
{\textrm{Re}}\left(1-{\cal S}^{(3)}_{012}\right)
\nonumber\\
&&  \hspace{0.5cm}
+
\Big[2k_{0}^+(k_{0}^+\!+\!k_{2}^+) + (k_{2}^+)^2\Big]\,
\frac{\x_{20}}{\x_{20}^2} \!\cdot\! \left(\frac{\x_{20}}{\x_{20}^2}\!-\!\frac{\x_{21}}{\x_{21}^2}\right)
\bigg[
 \Big(\textrm{K}_{0}\left(Q X_{012} \right)\Big)^2\;
{\textrm{Re}}\left(1-{\cal S}^{(3)}_{012}\right)\;
-\Big(\x_{2}\rightarrow \x_{0}\Big)\;
\bigg]
\Bigg\}
\label{sigma_L_q_to_g_plus}
\end{eqnarray}
and
\begin{eqnarray}
\sigma^{\gamma^*}_{L} \Big|_{\bar{q}\rightarrow g}
&=& 4 N_c \alpha_{em}  \left(\frac{\alpha_s C_F}{\pi}\right)
\sum_{f}  e_f^2
\widetilde{\sum_{{\cal P.S.}\, q_0 \bar{q}_1g_2; 4}}\,
4 Q^2\, \frac{(k_{0}^+)^2}{(q^+)^4}\,
\Bigg\{
(k_2^+)^2\,  \left(\frac{\x_{20}\!\cdot\! \x_{21}}{\x_{20}^2 \x_{21}^2}\right)\;
 \Big(\textrm{K}_{0}\left(Q X_{012} \right)\Big)^2\;
{\textrm{Re}}\left(1-{\cal S}^{(3)}_{012}\right)
\nonumber\\
&&  \hspace{0.5cm}
+
\Big[2k_{1}^+(k_{1}^+\!+\!k_{2}^+) + (k_{2}^+)^2\Big]\,
\frac{\x_{21}}{\x_{21}^2} \!\cdot\! \left(\frac{\x_{21}}{\x_{21}^2}\!-\!\frac{\x_{20}}{\x_{20}^2}\right)
\bigg[
\Big(\textrm{K}_{0}\left(Q X_{012} \right)\Big)^2\;
{\textrm{Re}}\left(1-{\cal S}^{(3)}_{012}\right)\;
-\Big(\x_{2}\rightarrow \x_{1}\Big)\;
\bigg]
\Bigg\}
\, ,
\label{sigma_L_qbar_to_g_plus}
\end{eqnarray}
where
\begin{eqnarray}
\sigma^{\gamma^*}_{L} \Big|_{q\rightarrow g}
+\sigma^{\gamma^*}_{L} \Big|_{\bar{q}\rightarrow g}
&=&\sigma^{\gamma^*}_{L} \Big|_{q\bar{q}g}-\sigma^{\gamma^*}_{L} \Big|_{UV; |a|^2}
-\sigma^{\gamma^*}_{L} \Big|_{UV; |b|^2}
\, .
\end{eqnarray}
Note that the potential divergence at small $k_2^+$ comes now entirely from the terms in the second line of Eqs.~\eqref{sigma_L_q_to_g_plus} and \eqref{sigma_L_qbar_to_g_plus}. Note also that the integrands in Eqs.~\eqref{sigma_L_q_to_g_plus} and \eqref{sigma_L_qbar_to_g_plus} are images of each other under the exchange of $(k_{0}^+,\x_0)$ and $(k_{1}^+,\x_1)$. Hence, one has the property
\begin{eqnarray}
\sigma^{\gamma^*}_{L} \Big|_{\bar{q}\rightarrow g} = \sigma^{\gamma^*}_{L} \Big|_{q\rightarrow g}
\, ,
\end{eqnarray}
thanks to the symmetry of the phase-space integration.


\subsection{Combining the UV subtraction terms and the quark-antiquark contribution}

It remains now to calculate the UV subtraction terms in $D$ dimensions, and combine them with the $q\bar{q}$ contribution to the cross section.
Inserting expression \eqref{Im_UV_approx} into \eqref{UV_subtr_term_a2_L} and making the change of variable $k_0^+\mapsto p_0^+= k_0^++k_2^+$ from the daughter to the parent quark LC momentum\footnote{Here, the regularization procedure for the $k_n^+$ integrals is taken to be the same as in Ref.~\cite{Beuf:2016wdz}: a cut-off $k^+_{\min}$ for the divergent $k_2^+$ integral only.}:
\begin{eqnarray}
\sigma^{\gamma^*}_{L} \Big|_{UV, |a|^2}
&=& 4 N_c \alpha_{em}  \left(\frac{\alpha_s C_F}{\pi}\right)
\sum_{f}  e_f^2
\int_{0}^{+\infty}\!\!\!\!\!\! dk^+_1\;
\int_{k^+_{\min}}^{+\infty}
\frac{dk^+_2}{k^+_2}\;\int_{k^+_2}^{+\infty}\!\!\!\!\!\! dp^+_0\;
\frac{\delta(p_0^+\!+\!k_1^+\!-\!q^+)}{q^+}\;
\int \frac{d^{D-2} \x_0}{2\pi} \int \frac{d^{D-2} \x_1}{2\pi}
\nonumber\\
&&  \hspace{1cm}
\times
\int \frac{d^{D-2} \x_2}{2\pi}\;
{\textrm{Re}}\left[1-{\cal S}_{01}\right]\;
4 Q^2\; \frac{(2\pi)^4}{2}\;
\frac{(k_{1}^+)^2}{(q^+)^4}\, \Big[4(p_{0}^+\!-\!k_{2}^+)p_{0}^+ + (D\!-\!2)(k_{2}^+)^2\Big]\,
\nonumber\\
&& \hspace{2cm}
\times \hspace{0.3cm}
\frac{1}{(2\pi)^4}\; \left[\Gamma\!\left(\frac{D}{2}\!-\!1\right)\right]^2\,
\left[\frac{4Q^2 k_{1}^+ (q^+\!-\!k_{1}^+)}{(2\pi)^4\mu^4 \x_{01}^2 (q^+)^2}\right]^{\frac{D}{2}-2}\,
\left[\textrm{K}_{\frac{D}{2}-2}\left(Q \frac{\sqrt{k_{1}^+ (q^+\!-\!k_{1}^+)}}{q^+}\, |\x_{01}|\right)\right]^2
\nonumber\\
&& \hspace{2cm}
\times \hspace{0.3cm}
\bigg[
{\x_{20}}^m\,  \left(\x_{20}^2\right)^{1-\frac{D}{2}}\Big(
{\x_{20}}^m\,  \left(\x_{20}^2\right)^{1-\frac{D}{2}}
\!-\!{\x_{21}}^m\,  \left(\x_{21}^2\right)^{1-\frac{D}{2}}
\Big)
\bigg]
\, .
\label{UV_subtr_term_a2_calc_1}
\end{eqnarray}

Relabelling the variable $p_0^+$ into $k_0^+$, and using the notation $\overline{Q}$ defined in Eq.~\eqref{def_Qbar}, one can rewrite this UV subtraction term as
\begin{eqnarray}
\sigma^{\gamma^*}_{L} \Big|_{UV, |a|^2}
&=& 4 N_c \alpha_{em} \sum_{f} e_f^2\;
\int_{k^+_{\min}}^{+\infty}\!\!\!\!\!\! dk^+_0\; \int_{0}^{+\infty}\!\!\!\!\!\! dk^+_1\;
\frac{\delta(k_0^+\!+\!k_1^+\!-\!q^+)}{q^+}\;
4 Q^2
\frac{(k_0^+)^2 (k_1^+)^2}{(q^+)^4}\;
\int \frac{d^{D-2} \x_0}{2\pi} \int \frac{d^{D-2} \x_1}{2\pi}\;
{\textrm{Re}}\left[1-{\cal S}_{01}\right]\;
\nonumber\\
& &\hspace{2cm} \times
\left(\frac{\overline{Q}^2}{(2\pi)^2\, \x_{01}^2\, \mu^2}\right)^{\frac{D}{2}-2}\;\;
\left[ \textrm{K}_{\frac{D}{2}-2}\Big(|\x_{01}|\, \overline{Q}\Big)\right]^2\;\;
\left(\frac{\alpha_s\, C_F}{\pi}\right)\;
\widetilde{{\cal V}}_{UV, |a|^2}
\label{UV_subtr_term_a2_calc_2}
\end{eqnarray}
with
\begin{eqnarray}
\widetilde{{\cal V}}_{UV, |a|^2}
&=& \Big[\pi^2 \mu^2\Big]^{2-\frac{D}{2}}\, \left[\Gamma\!\left(\frac{D}{2}\!-\!1\right)\right]^2\,
    \int_{k^+_{\min}}^{k^+_0}  \frac{dk^+_2}{k^+_2}\;
    \left[\frac{2(k^+_0\!-\!k^+_2)}{k^+_0}  +\frac{(D\!-\!2)}{2}\, \frac{(k^+_2)^2}{(k^+_0)^2} \right]
\nonumber\\
& &\hspace{0.5cm} \times
    \int \frac{d^{D-2} \x_2}{2\pi}\;
    \bigg[
{\x_{20}}^m\,  \left(\x_{20}^2\right)^{1-\frac{D}{2}}\Big(
{\x_{20}}^m\,  \left(\x_{20}^2\right)^{1-\frac{D}{2}}
\!-\!{\x_{21}}^m\,  \left(\x_{21}^2\right)^{1-\frac{D}{2}}
\Big)
\bigg]
\nonumber\\
&=& - \left[\log\left(\frac{k^+_{\min}}{k^+_0}\right)+ \frac{3}{4} +\frac{(2\!-\!\frac{D}{2})}{4}\right]\;
\frac{1}{(2\!-\!\frac{D}{2})}\;  \Gamma\!\left(\frac{D}{2}\!-\!1\right)\;
\Big[\pi \mu^2 \x_{01}^2\Big]^{2-\frac{D}{2}}
\nonumber\\
&=& -\left[\log\left(\frac{k^+_{\min}}{k^+_0}\right)+\frac{3}{4}\right]\;
\left[\frac{1}{(2\!-\!\frac{D}{2})} - \Psi(1) +\log\left(\pi \mu^2 \x_{01}^2\right)
\right]
-\frac{1}{4}
+ O(D\!-\!4)
\, .
\label{V_UV_subtr_term_a2}
\end{eqnarray}
The UV subtraction term for the contribution $|b|^2$ can be calculated in the same way, and one gets
\begin{eqnarray}
\sigma^{\gamma^*}_{L} \Big|_{UV, |b|^2}
&=& 4 N_c \alpha_{em} \sum_{f} e_f^2\;
\int_{0}^{+\infty}\!\!\!\!\!\! dk^+_0\; \int_{k^+_{\min}}^{+\infty}\!\!\!\!\!\! dk^+_1\;
\frac{\delta(k_0^+\!+\!k_1^+\!-\!q^+)}{q^+}\;
4 Q^2
\frac{(k_0^+)^2 (k_1^+)^2}{(q^+)^4}\;
\int \frac{d^{D-2} \x_0}{2\pi} \int \frac{d^{D-2} \x_1}{2\pi}\;
{\textrm{Re}}\left[1-{\cal S}_{01}\right]\;
\nonumber\\
& &\hspace{2cm} \times
\left(\frac{\overline{Q}^2}{(2\pi)^2\, \x_{01}^2\, \mu^2}\right)^{\frac{D}{2}-2}\;\;
\left[ \textrm{K}_{\frac{D}{2}-2}\Big(|\x_{01}|\, \overline{Q}\Big)\right]^2\;\;
\left(\frac{\alpha_s\, C_F}{\pi}\right)\;
\widetilde{{\cal V}}_{UV, |b|^2}
\label{UV_subtr_term_b2_calc_2}
\end{eqnarray}
with the coefficient
\begin{eqnarray}
\widetilde{{\cal V}}_{UV, |b|^2}
&=& - \left[\log\left(\frac{k^+_{\min}}{k^+_1}\right)+ \frac{3}{4} +\frac{(2\!-\!\frac{D}{2})}{4}\right]\;
\frac{1}{(2\!-\!\frac{D}{2})}\;  \Gamma\!\left(\frac{D}{2}\!-\!1\right)\;
\Big[\pi \mu^2 \x_{01}^2\Big]^{2-\frac{D}{2}}
\bigg]
\nonumber\\
&=& -\left[\log\left(\frac{k^+_{\min}}{k^+_1}\right)+\frac{3}{4}\right]\;
\left[\frac{1}{(2\!-\!\frac{D}{2})} - \Psi(1) +\log\left(\pi \mu^2 \x_{01}^2\right)
\right]
-\frac{1}{4}
+ O(D\!-\!4)
\, .
\label{V_UV_subtr_term_b2}
\end{eqnarray}
In Eq.~\eqref{UV_subtr_term_a2_calc_2} (resp. \eqref{UV_subtr_term_b2_calc_2}), the cut-off $k^+_{\min}$ on $k_0^+$ (resp. $k_1^+$) comes as a consequence of the cut-off on $k_2^+$, and sets the sign of the log present in Eq.~\eqref{V_UV_subtr_term_a2} (resp. \eqref{V_UV_subtr_term_b2}). Apart from that cut-off, Eqs.~\eqref{UV_subtr_term_a2_calc_2} and \eqref{UV_subtr_term_b2_calc_2} have the same structure as the NLO piece of Eq.~\eqref{sigma_L_qqbar_result}. Actually, in the calculation of the loop corrections to the LFWF for the $q\bar{q}$ Fock component in Ref.~\cite{Beuf:2016wdz}, each contribution should have come with a cut-off $k^+_{\min}$ either on $k_0^+$ or on $k_1^+$, as a consequence of the cut-off on $k_2^+$, but this has been implicitly neglected in Ref.~\cite{Beuf:2016wdz}, and accordingly in Eq.~\eqref{sigma_L_qqbar_result}.

Adding together the coefficients $\widetilde{{\cal V}}$, $\widetilde{{\cal V}}_{UV, |a|^2}$ and $\widetilde{{\cal V}}_{UV, |b|^2}$ (provided in Eqs.~\eqref{mixed_V_result}, \eqref{V_UV_subtr_term_a2} and \eqref{V_UV_subtr_term_b2}) lead to a cancellation of the logs of $k^+_{\min}$ as well as of the UV divergences, and of the UV regularization scheme dependent terms, such as the rational terms induced by the $D$ dependence of the Dirac traces in CDR, which correspond to a contribution $1/2$ in $\widetilde{{\cal V}}$, as discussed in Ref.~\cite{Beuf:2016wdz},
and to the term $-1/4$ in Eqs.~\eqref{V_UV_subtr_term_a2} and \eqref{V_UV_subtr_term_b2}. Importantly, all these cancellations stay valid when the different phase spaces are taken into account, with the cut-off $k^+_{\min}$ either on $k_0^+$ or on $k_1^+$. Once the logs of $k^+_{\min}$ have canceled, it is perfectly legitimate to drop the leftover cut-off on $k_0^+$ or $k_1^+$, since it corresponds to a power suppressed correction.
It is then convenient to define
\begin{eqnarray}
\widetilde{{\cal V}}_{\textrm{reg.}}
&\equiv& \lim_{D\rightarrow 4}\left[ \widetilde{{\cal V}}+\widetilde{{\cal V}}_{UV, |a|^2}+\widetilde{{\cal V}}_{UV, |b|^2}\right]
= \frac{1}{2} \left[\log\left(\frac{k_0^+}{k_1^+}\right)\right]^2
-\frac{\pi^2}{6}
 +\frac{5}{2}
 \, ,
\label{V_reg}
\end{eqnarray}
so that sum of the dipole-like contributions
\begin{eqnarray}
\sigma^{\gamma^*}_{L} \Big|_{\textrm{dipole}}
&\equiv &
\sigma^{\gamma^*}_{L} \Big|_{q\bar{q}} + \sigma^{\gamma^*}_{L} \Big|_{UV; |a|^2}
    + \sigma^{\gamma^*}_{L} \Big|_{UV; |b|^2}
\, ,
\end{eqnarray}
can be written as
\begin{eqnarray}
\sigma^{\gamma^*}_{L} \Big|_{\textrm{dipole}}
&=& 4 N_c \alpha_{em} \sum_{f} e_f^2\;
\int_{0}^{+\infty}\!\!\!\!\!\! dk^+_0\; \int_{0}^{+\infty}\!\!\!\!\!\! dk^+_1\;
\frac{\delta(k_0^+\!+\!k_1^+\!-\!q^+)}{q^+}\;
\int \frac{d^2 \x_0}{2\pi} \int \frac{d^2 \x_1}{2\pi}\;
{\textrm{Re}}\left[1-{\cal S}_{01}\right]\;
\nonumber\\
& &
\times
4 Q^2\frac{(k_0^+)^2 (k_1^+)^2}{(q^+)^4}\; \left[ \textrm{K}_{0}\Big(|\x_{01}|\, \overline{Q}\Big)\right]^2\;\;
\left[1+\left(\frac{\alpha_s\, C_F}{\pi}\right)\;
\widetilde{{\cal V}}_{\textrm{reg.}}\;\right]\;\;
\, .
\label{sigma_L_dip_plus}
\end{eqnarray}

Hence , the full longitudinal photon cross section at NLO is
\begin{eqnarray}
\sigma^{\gamma^*}_{L}&=&\sigma^{\gamma^*}_{L} \Big|_{q\bar{q}} +\sigma^{\gamma^*}_{L} \Big|_{q\bar{q}g}
=\sigma^{\gamma^*}_{L} \Big|_{\textrm{dipole}} + \sigma^{\gamma^*}_{L} \Big|_{q\rightarrow g}
+\sigma^{\gamma^*}_{L} \Big|_{\bar{q}\rightarrow g}
=\sigma^{\gamma^*}_{L} \Big|_{\textrm{dipole}} +2\, \sigma^{\gamma^*}_{L} \Big|_{q\rightarrow g}
\, ,
\label{sigma_L_NLO_fixed}
\end{eqnarray}
with the individual contributions given in Eqs.~\eqref{sigma_L_q_to_g_plus}, \eqref{sigma_L_qbar_to_g_plus} and
\eqref{sigma_L_dip_plus}.


\subsection{Non-minimal cut-off prescription \label{sec:non_mini_cut}}

As a remark, note that the cut-off procedure used in the previous section for the $k^+$ integrals is not unique. For example, it might make sense to impose a cut-off on the $k^+$ of all partons crossing the shockwave, in order to avoid the region where the eikonal approximation for the scattering of that parton off the shock breaks down. Such additional restrictions on regular integrals correspond to effects which are suppressed by powers of the cut-off, to be related later to inverse powers of the collision energy. Hence, these effects are in principle beyond the accuracy of our calculation. In practice, it might be useful to have the results in two different prescriptions for the regularization in $k^+$, in order to check in what range of the parameters the difference between the two prescriptions is indeed negligible.

One possibility would be, as suggested before, to impose a cut-off on the $k^+$ of each of the partons crossing the shockwave (and of the gluon running in the loop, for the NLO correction to the $q\bar{q}$ Fock state). Another possibility would be to impose a cut-off on the $k^+$ of each of the internal lines, in all of the diagrams contributing to the photon total cross sections. However, both of these possibility lead to very cumbersome results.

In each diagram for the one-loop correction to the $q\bar{q}$ Fock state LFWF, imposing the cut-off on $k_2^+$ for the gluon results indirectly in a cut-off in either $k_0^+$ or $k_1^+$, depending on the diagram.
And for each contribution to the $q\bar{q}g$ Fock state LFWF, a cut-off for the gluon imposes a restriction on its parent $k^+$. Hence, one can build a non-minimal regularization prescription as follows:
\begin{itemize}
  \item For the LO contribution, impose the cut-off at $k^+_{\min}$ for both $k_0^+$ and $k_1^+$.
  \item For each one-loop correction to the $q\bar{q}$ LFWF, impose the cut-off at $k^+_{\min}$ only for $k_2^+$ and $k_1^+$ (resp. $k_0^+$) a priori, if a restriction on $k_0^+$  (resp. $k_1^+$) can be obtained as a consequence of the cut-off in $k_2^+$\, .
  \item For the $q\rightarrow g$ contribution \eqref{sigma_L_q_to_g_plus} and the corresponding UV subtraction term \eqref{UV_subtr_term_a2_L}, impose the impose the cut-off only for $k_2^+$ and $k_1^+$ a priori.
  \item For the $\bar{q}\rightarrow g$ contribution \eqref{sigma_L_qbar_to_g_plus} and the corresponding UV subtraction term \eqref{UV_subtr_term_b2_L}, impose the impose the cut-off only for $k_2^+$ and $k_0^+$ a priori.
\end{itemize}
Hence the phase space integral for the $q\bar{q}g$ Fock state defined in Eq.~\eqref{phase_space_qqbarg} should be taken as
\begin{eqnarray}
\widetilde{\sum_{{\cal P.S.}\, q_0 \bar{q}_1g_2; D}}\; \Big[\cdots\Big]
&\mapsto &
\int_{0}^{+\infty} \hspace{-0.5cm}dk^+_0 \int_{k^+_{\min}}^{+\infty}  \hspace{-0.5cm}dk^+_1
\int_{k^+_{\min}}^{+\infty}  
\frac{dk^+_2}{k^+_2}\; \frac{\delta(k_0^+\!+\!k_1^+\!+\!k_2^+\!-\!q^+)}{q^+}\;
\int \frac{d^{D-2} \x_0}{2\pi} \int \frac{d^{D-2} \x_1}{2\pi} \int \frac{d^{D-2} \x_2}{2\pi}\: \Big[\cdots\Big]
\nonumber\\
&=& \frac{\theta(q^+\!-\!2 k^+_{\min})}{q^+}
 \int_{k^+_{\min}}^{q^+\!-k^+_{\min}}  \hspace{-0.2cm}
 dk^+_1
\int_{k^+_{\min}}^{q^+\!-k^+_1}  
\frac{dk^+_2}{k^+_2}\;
\int \frac{d^{D-2} \x_0}{2\pi} \int \frac{d^{D-2} \x_1}{2\pi} \int \frac{d^{D-2} \x_2}{2\pi}\: \Big[\cdots\Big]
\label{phase_space_qqbarg_q2g}
\end{eqnarray}
for the $q\rightarrow g$ contribution \eqref{sigma_L_q_to_g_plus}, and as
\begin{eqnarray}
\widetilde{\sum_{{\cal P.S.}\, q_0 \bar{q}_1g_2; D}}\; \Big[\cdots\Big]
&\mapsto &
\int_{k^+_{\min}}^{+\infty} \hspace{-0.5cm}dk^+_0 \int_{0}^{+\infty}  \hspace{-0.5cm}dk^+_1
\int_{k^+_{\min}}^{+\infty}  
\frac{dk^+_2}{k^+_2}\; \frac{\delta(k_0^+\!+\!k_1^+\!+\!k_2^+\!-\!q^+)}{q^+}\;
\int \frac{d^{D-2} \x_0}{2\pi} \int \frac{d^{D-2} \x_1}{2\pi} \int \frac{d^{D-2} \x_2}{2\pi}\: \Big[\cdots\Big]
\nonumber\\
&=& \frac{\theta(q^+\!-\!2 k^+_{\min})}{q^+}
 \int_{k^+_{\min}}^{q^+\!-k^+_{\min}}  \hspace{-0.2cm}
 dk^+_0
\int_{k^+_{\min}}^{q^+\!-k^+_0}  
\frac{dk^+_2}{k^+_2}\;
\int \frac{d^{D-2} \x_0}{2\pi} \int \frac{d^{D-2} \x_1}{2\pi} \int \frac{d^{D-2} \x_2}{2\pi}\: \Big[\cdots\Big]
\label{phase_space_qqbarg_qbar2g}
\end{eqnarray}
for the $\bar{q}\rightarrow g$ contribution \eqref{sigma_L_qbar_to_g_plus}.
Moreover, with that non-minimal cut-off prescription, Eq.~\eqref{sigma_L_dip_plus} is replaced by
\begin{eqnarray}
\sigma^{\gamma^*}_{L} \Big|_{\textrm{dipole}}
&=& 4 N_c \alpha_{em} \sum_{f} e_f^2\;
\int_{k^+_{\min}}^{+\infty}\!\!\!\!\!\! dk^+_0\; \int_{k^+_{\min}}^{+\infty}\!\!\!\!\!\! dk^+_1\;
\frac{\delta(k_0^+\!+\!k_1^+\!-\!q^+)}{q^+}\;
\int \frac{d^2 \x_0}{2\pi} \int \frac{d^2 \x_1}{2\pi}\;
{\textrm{Re}}\left[1-{\cal S}_{01}\right]\;
\nonumber\\
& &
\times
4 Q^2\frac{(k_0^+)^2 (k_1^+)^2}{(q^+)^4}\; \left[ \textrm{K}_{0}\Big(|\x_{01}|\, \overline{Q}\Big)\right]^2\;\;
\left[1+\left(\frac{\alpha_s\, C_F}{\pi}\right)\;
\widetilde{{\cal V}}_{\textrm{reg.}}\;\right]\;\;
\, .
\label{sigma_L_dip_plus_non_min}
\end{eqnarray}


\subsection{Final result for the longitudinal photon cross section at NLO before high-energy resummation\label{sec:final_res_fixed_NLO_L}}

A simpler expression for the longitudinal photon cross section at NLO can be obtained by switching from $k^+$ variables to light-cone momentum fractions. For the dipole-like contribution, taking $z=k_0^+/q^+$ and eliminating $k_1^+$ thanks to the delta function, one finds
\begin{eqnarray}
\sigma^{\gamma^*}_{L} \Big|_{\textrm{dipole}}
&=& 4 N_c \alpha_{em} \sum_{f} e_f^2\; \theta(1\!-\!2 z_c)
\int_{z_c}^{1-z_c}\!\! dz\; 4 z^2(1\!-\!z)^2
\int \frac{d^2 \x_0}{2\pi} \int \frac{d^2 \x_1}{2\pi}\;
Q^2 \bigg( \textrm{K}_{0}\Big(Q \sqrt{z(1\!-\!z)}|\x_{01}|\Big)\bigg)^2\;
\nonumber\\
& &\hspace{2.5cm}
\times\;
{\textrm{Re}}\big(1-{\cal S}_{01}\big)\;
\Bigg\{1+\left(\frac{\alpha_s\, C_F}{\pi}\right)\;
\bigg[\frac{1}{2} \left(\log\left(\frac{z}{1\!-\!z}\right)\right)^2
-\frac{\pi^2}{6}
 +\frac{5}{2}
\bigg]\Bigg\}\;\;
\label{sigma_L_dipole_frac}
\, ,
\end{eqnarray}
with $z_c=0$ if the minimal cut-off prescription (for $k_2^+$ only) is chosen, or $z_c=k^+_{\min}/q^+$ if the non-minimal cut-off prescription of sec.~\ref{sec:non_mini_cut} is used.

For the $q\rightarrow g$ contribution, the appropriate change of variable is $(k_{1}^+,k_{2}^+)\mapsto (z, \zeta)$, with $k_{1}^+=(1\!-\!z)q^+$ and $k_{2}^+=\zeta\, z\, q^+$, after eliminating $k_{0}^+$ with the delta function. Then, one gets
\begin{eqnarray}
\sigma^{\gamma^*}_{L} \Big|_{q\rightarrow g}
&=& 4 N_c \alpha_{em} \left(\frac{\alpha_s\, C_F}{\pi}\right)\sum_{f} e_f^2\;
\theta\!\left(1\!-\! \frac{k^+_{\min}}{q^+}\!-\! z_c\right)
\int_{\frac{k^+_{\min}}{q^+}}^{1- z_c}\! dz\; 4 z^2(1\!-\!z)^2  \int_{\frac{k^+_{\min}}{z q^+}}^{1}
d \zeta\;
\int \frac{d^2 \x_0}{2\pi} \int \frac{d^2 \x_1}{2\pi}\int \frac{d^2 \x_2}{2\pi}\;
\nonumber\\
&&\hspace{0.5cm}
\times \Bigg\{
\left[\frac{1+(1\!-\!\zeta)^2}{\zeta}\right]\;
\frac{\x_{20}}{\x_{20}^2} \!\cdot\! \left(\frac{\x_{20}}{\x_{20}^2}\!-\!\frac{\x_{21}}{\x_{21}^2}\right)\,
\bigg[Q^2 \Big(\textrm{K}_0\!\left(Q X_{012}\right)\Big)^2\;
{\textrm{Re}}\left(1-{\cal S}^{(3)}_{012}\right)-\Big(\x_{2}\rightarrow \x_{0}\Big)\bigg]
\nonumber\\
&&  \hspace{3cm}
+\zeta\, \frac{(\x_{20}\!\cdot\!\x_{21})}{\x_{20}^2 \x_{21}^2}
\,
Q^2 \Big(\textrm{K}_0\!\left(Q X_{012}\right)\Big)^2\;
{\textrm{Re}}\left(1-{\cal S}^{(3)}_{012}\right)
\Bigg\}
\label{sigma_L_q_to_g_frac}
\, ,
\end{eqnarray}
with now
\begin{eqnarray}
X_{012}^2= z(1\!-\!z)(1\!-\!\zeta)\, \x_{01}^2 + z^2 \zeta (1\!-\!\zeta)\, \x_{20}^2
+ z(1\!-\!z)\zeta\, \x_{21}^2
\label{def_X012_q2g}
\, .
\end{eqnarray}

For the $\bar{q}\rightarrow g$ contribution, taking the change of variable $(k_{0}^+,k_{2}^+)\mapsto (z, \zeta)$, with $k_{0}^+=z q^+$ and $k_{2}^+=\zeta\, (1\!-\!z)\, q^+$ after eliminating $k_{1}^+$, one finds
\begin{eqnarray}
\sigma^{\gamma^*}_{L} \Big|_{\bar{q}\rightarrow g}
&=& 4 N_c \alpha_{em} \left(\frac{\alpha_s\, C_F}{\pi}\right)\sum_{f} e_f^2\;
\theta\!\left(1\!-\! \frac{k^+_{\min}}{q^+}\!-\! z_c\right)
\int_{z_c}^{1- \frac{k^+_{\min}}{q^+}}\! dz\; 4 z^2(1\!-\!z)^2  \int_{\frac{k^+_{\min}}{(1\!-\!z) q^+}}^{1}
d \zeta\;
\int \frac{d^2 \x_0}{2\pi} \int \frac{d^2 \x_1}{2\pi}\int \frac{d^2 \x_2}{2\pi}\;
\nonumber\\
&&\hspace{0.5cm}
\times \Bigg\{
\left[\frac{1+(1\!-\!\zeta)^2}{\zeta}\right]\;
\frac{\x_{21}}{\x_{21}^2} \!\cdot\! \left(\frac{\x_{21}}{\x_{21}^2}\!-\!\frac{\x_{20}}{\x_{20}^2}\right)\,
\bigg[Q^2 \Big(\textrm{K}_0\!\left(Q X_{012}\right)\Big)^2\;
{\textrm{Re}}\left(1-{\cal S}^{(3)}_{012}\right)-\Big(\x_{2}\rightarrow \x_{0}\Big)\bigg]
\nonumber\\
&&  \hspace{3cm}
+\zeta\, \frac{(\x_{20}\!\cdot\!\x_{21})}{\x_{20}^2 \x_{21}^2}
\,
Q^2 \Big(\textrm{K}_0\!\left(Q X_{012}\right)\Big)^2\;
{\textrm{Re}}\left(1-{\cal S}^{(3)}_{012}\right)
\Bigg\}
\label{sigma_L_qbar_to_g_frac}
\, ,
\end{eqnarray}
with, in this case,
\begin{eqnarray}
X_{012}^2= z(1\!-\!z)(1\!-\!\zeta)\, \x_{01}^2 + z(1\!-\!z) \zeta\, \x_{20}^2
+ (1\!-\!z)^2 \zeta(1\!-\!\zeta)\, \x_{21}^2
\label{def_X012_qbar2g}
\, .
\end{eqnarray}


\section{Transverse photon cross-section at NLO\label{sec:sigma_T_NLO}}

In the transverse photon case, among the $q\bar{q}g$ contributions, only the ones corresponding to the squares of the diagrams $(a)$ and $(b)$ from Fig.~\ref{Fig:gammaT_qqbarg}, noted $|a|^2$ and $|b|^2$, have UV divergences. The other terms can be calculated directly in $D=4$. This is an important simplification in particular for the interference between the diagrams $(a)$ and $(b)$. It is then convenient to decompose the transverse photon cross section at NLO as
\begin{eqnarray}
\sigma^{\gamma^*}_{T}
=\sigma^{\gamma^*}_{T} \Big|_{q\bar{q}} +\sigma^{\gamma^*}_{T} \Big|_{q\bar{q}g}
=\sigma^{\gamma^*}_{T} \Big|_{q\bar{q}}
+\sigma^{\gamma^*}_{T} \Big|_{q\bar{q}g, |a|^2}+\sigma^{\gamma^*}_{T} \Big|_{q\bar{q}g, |b|^2}
+\sigma^{\gamma^*}_{T} \Big|_{q\bar{q}g;\, \textrm{UV finite terms}}
\, .
\end{eqnarray}

\subsection{Quark-antiquark-gluon contribution: UV finite terms}

The contribution from the UV finite terms is calculated in $D=4$ as
\begin{eqnarray}
\sigma^{\gamma^*}_{T} \Big|_{q\bar{q}g;\, \textrm{UV finite terms}}
&=&  N_c C_F
\sum_{f}
\widetilde{\sum_{{\cal P.S.}\, q_0 \bar{q}_1g_2; 4}}\;
\frac{(2\pi) }{2(2k^+_0)(2k^+_1)}\;
{\textrm{Re}}\left[1-{\cal S}^{(3)}_{012}\right]\;
\nonumber\\
&& \hspace{1cm} \times
\;
 \frac{1}{2}\,
\bigg\{\sum_{T\textrm{ pol. }\lambda,\, \lambda_2}\;\;  \sum_{h_0, h_1 = \pm 1/2}
\left|\widetilde{\psi}_{\gamma_{\lambda}^*\rightarrow q_0 \bar{q_1} g_2}\right|^2 \bigg\}_{\textrm{UV finite terms}}
\, ,
\end{eqnarray}
using the expression \eqref{qqbarg_WF_T_mixed_4D_fin} for the LFWF, and discarding the $|a|^2$ and $|b|^2$ terms from the squared LFWF.

The contribution from the interference between the instantaneous diagrams $(a')$ and $(b')$ is shown to vanish as
\begin{eqnarray}
&& \sum_{T\textrm{ pol. }\lambda,\, \lambda_2}\;\;  \sum_{h_0, h_1 = \pm 1/2}
 2 {\textrm{Re}} \left(\widetilde{\psi}^{(a')}_{\gamma_{\lambda}^*\rightarrow q_0 \bar{q_1} g_2}\right)^*
 \widetilde{\psi}^{(b')}_{\gamma_{\lambda}^*\rightarrow q_0 \bar{q_1} g_2}
=  2 {\textrm{Re}}\, \frac{ e^2\, e_f^2\, g^2}{(2\pi)^4}\, \delta^{i'i}\, \delta^{j'j}\,
  (2 k_0^+)\, (2 k_1^+)\; \frac{Q^2}{X_{012}^2}\; \Big[\textrm{K}_1\!\left(Q X_{012}\right)\Big]^2\,
\nonumber\\
&& \hspace{3cm} \times
  \frac{(-1)(k_{0}^+)^2 (k_{1}^+)^2 (k_{2}^+)^2}{(k_{0}^+\!+\!k_{2}^+)(k_{1}^+\!+\!k_{2}^+)(q^+)^4}\:
   \sum_{h_0 = \pm 1/2} \Big[\delta^{i'j'}  + i\, (2h_0)\, \epsilon^{i'j'}\Big]
   \Big[\delta^{ij}  + i\, (2h_0)\, \epsilon^{ij}\Big]
\nonumber\\
&& =  2\, \frac{ e^2\, e_f^2\, g^2}{(2\pi)^4}\,
  (2 k_0^+)\, (2 k_1^+)\; \frac{Q^2}{X_{012}^2}\; \Big[\textrm{K}_1\!\left(Q X_{012}\right)\Big]^2\,
   \frac{(-1)(k_{0}^+)^2 (k_{1}^+)^2 (k_{2}^+)^2}{(k_{0}^+\!+\!k_{2}^+)(k_{1}^+\!+\!k_{2}^+)(q^+)^4}\:
   2\Big[\delta^{ij} \delta^{ij} - \epsilon^{ij}\epsilon^{ij}\Big] =0
\, .
\end{eqnarray}
Note that this is true only in $D=4$. For general $D$, that contribution would be linear in $(D\!-\!4)$ instead.

Then, the contributions involving the diagram $(a')$ are obtained as
\begin{eqnarray}
&&  \sum_{T\textrm{ pol. }\lambda,\, \lambda_2}\;\;  \sum_{h_0, h_1 = \pm 1/2} {\textrm{Re}}\:
\bigg\{\left(\widetilde{\psi}^{(a')}_{\gamma_{\lambda}^*\rightarrow q_0 \bar{q_1} g_2}\right)^* \bigg[\widetilde{\psi}^{(a')}_{\gamma_{\lambda}^*\rightarrow q_0 \bar{q_1} g_2}
 +  2  \widetilde{\psi}^{(a)}_{\gamma_{\lambda}^*\rightarrow q_0 \bar{q_1} g_2}
 +  2   \widetilde{\psi}^{(b)}_{\gamma_{\lambda}^*\rightarrow q_0 \bar{q_1} g_2}
 \bigg]\bigg\}
\nonumber\\
&& = (2 k_0^+)\, (2 k_1^+)\; \frac{4 \alpha_{em}}{2\pi}\, e_f^2\, \left(\frac{\alpha_s}{\pi}\right)\,
\frac{Q^2}{X_{012}^2}\; \Big[\textrm{K}_1\!\left(Q X_{012}\right)\Big]^2\,
\bigg\{
 \frac{4(k_{0}^+)^2 (k_{1}^+)^2 (k_{2}^+)^2}{(k_{0}^+\!+\!k_{2}^+)^2(q^+)^4}
 \nonumber\\
&&
\hspace{2cm} -\frac{8(k_{0}^+)^2 (k_{1}^+)^3 k_{2}^+}{(k_{0}^+\!+\!k_{2}^+)(q^+)^5}\;
 \frac{\left(\x_{0+2;1}\!\cdot\! \x_{20}\right)}{\x_{20}^2}
+\frac{8(k_{0}^+)^2 k_{1}^+(k_{1}^+\!+\!k_{2}^+)^2 k_{2}^+}{(k_{0}^+\!+\!k_{2}^+)(q^+)^5}\;
 \frac{\left(\x_{0;1+2}\!\cdot\! \x_{21}\right)}{\x_{21}^2}
\bigg\}
\, ,
\end{eqnarray}
the ones involving the diagram $(b')$ as
\begin{eqnarray}
&& \sum_{T\textrm{ pol. }\lambda,\, \lambda_2}\;\;  \sum_{h_0, h_1 = \pm 1/2} {\textrm{Re}}\:
\bigg\{\left(\widetilde{\psi}^{(b')}_{\gamma_{\lambda}^*\rightarrow q_0 \bar{q_1} g_2}\right)^* \bigg[\widetilde{\psi}^{(b')}_{\gamma_{\lambda}^*\rightarrow q_0 \bar{q_1} g_2}
 +  2  \widetilde{\psi}^{(a)}_{\gamma_{\lambda}^*\rightarrow q_0 \bar{q_1} g_2}
 +  2  \widetilde{\psi}^{(b)}_{\gamma_{\lambda}^*\rightarrow q_0 \bar{q_1} g_2}
 \bigg]\bigg\}
\nonumber\\
&& = (2 k_0^+)\, (2 k_1^+)\; \frac{4 \alpha_{em}}{2\pi}\, e_f^2\, \left(\frac{\alpha_s}{\pi}\right)\,
\frac{Q^2}{X_{012}^2}\; \Big[\textrm{K}_1\!\left(Q X_{012}\right)\Big]^2\,
\bigg\{
 \frac{4(k_{0}^+)^2 (k_{1}^+)^2 (k_{2}^+)^2}{(k_{1}^+\!+\!k_{2}^+)^2(q^+)^4}
 \nonumber\\
&&
\hspace{2cm} -\frac{8 k_{0}^+ (k_{1}^+)^2 (k_{0}^+\!+\!k_{2}^+)^2 k_{2}^+}{(k_{1}^+\!+\!k_{2}^+)(q^+)^5}\;
 \frac{\left(\x_{0+2;1}\!\cdot\! \x_{20}\right)}{\x_{20}^2}
+\frac{8(k_{0}^+)^3 (k_{1}^+)^2 k_{2}^+}{(k_{1}^+\!+\!k_{2}^+)(q^+)^5}\;
 \frac{\left(\x_{0;1+2}\!\cdot\! \x_{21}\right)}{\x_{21}^2}
\bigg\}
\end{eqnarray}
and the interference between the diagrams $(a)$ and $(b)$ as
\begin{eqnarray}
&&  \sum_{T\textrm{ pol. }\lambda,\, \lambda_2}\;\;  \sum_{h_0, h_1 = \pm 1/2}
 2 {\textrm{Re}} \left(\widetilde{\psi}^{(a)}_{\gamma_{\lambda}^*\rightarrow q_0 \bar{q_1} g_2}\right)^*
 \widetilde{\psi}^{(b)}_{\gamma_{\lambda}^*\rightarrow q_0 \bar{q_1} g_2}
 = (2 k_0^+)\, (2 k_1^+)\; \frac{4 \alpha_{em}}{2\pi}\, e_f^2\, \left(\frac{\alpha_s}{\pi}\right)\,
\frac{Q^2}{X_{012}^2}\; \Big[\textrm{K}_1\!\left(Q X_{012}\right)\Big]^2
\nonumber\\
&& \hspace{1cm} \times  \frac{8 k_{0}^+ k_{1}^+}{(q^+)^6}\,
\bigg\{ q^+ k_{2}^+ (k_{0}^+\!-\!k_{1}^+)^2\:
 \frac{\left(\x_{20}\wedge \x_{21}\right)}{\x_{20}^2\, \x_{21}^2}\,
 \left(\x_{0+2;1}\wedge \x_{0;1+2}\right)
 \nonumber\\
&&
\hspace{2cm}
- \Big[k_{1}^+(k_{0}^+\!+\!k_{2}^+)+k_{0}^+(k_{1}^+\!+\!k_{2}^+)\Big]
\Big[k_{0}^+(k_{0}^+\!+\!k_{2}^+)+k_{1}^+(k_{1}^+\!+\!k_{2}^+)\Big]
 \frac{\left(\x_{20}\!\cdot\! \x_{21}\right)}{\x_{20}^2\, \x_{21}^2}\,
 \left(\x_{0+2;1}\!\cdot\! \x_{0;1+2}\right)
\bigg\}
 \nonumber\\
&=& (2 k_0^+)\, (2 k_1^+)\; \frac{4 \alpha_{em}}{2\pi}\, e_f^2\, \left(\frac{\alpha_s}{\pi}\right)\,
\frac{Q^2}{X_{012}^2}\; \Big[\textrm{K}_1\!\left(Q X_{012}\right)\Big]^2\;
\frac{8 k_{0}^+ k_{1}^+}{(q^+)^4}\,
\bigg\{  \frac{(k_{2}^+)^2 (k_{0}^+\!-\!k_{1}^+)^2}{(k_{0}^+\!+\!k_{2}^+)(k_{1}^+\!+\!k_{2}^+)}\:
 \frac{\left(\x_{20}\wedge \x_{21}\right)^2}{\x_{20}^2\, \x_{21}^2}
 \nonumber\\
&& \hspace{1cm}
-\frac{1}{(q^+)^2}\, \Big[k_{1}^+(k_{0}^+\!+\!k_{2}^+)+k_{0}^+(k_{1}^+\!+\!k_{2}^+)\Big]
\Big[k_{0}^+(k_{0}^+\!+\!k_{2}^+)+k_{1}^+(k_{1}^+\!+\!k_{2}^+)\Big]
 \frac{\left(\x_{20}\!\cdot\! \x_{21}\right)}{\x_{20}^2\, \x_{21}^2}\,
 \left(\x_{0+2;1}\!\cdot\! \x_{0;1+2}\right)
 \bigg\}
 \, .
\end{eqnarray}

All in all, one finds
\begin{eqnarray}
\sigma^{\gamma^*}_{T} \Big|_{q\bar{q}g;\, \textrm{UV finite terms}}
&=& 4 N_c \alpha_{em}  \left(\frac{\alpha_s C_F}{\pi}\right)
\sum_{f}  e_f^2
\widetilde{\sum_{{\cal P.S.}\, q_0 \bar{q}_1g_2; 4}}\;
{\textrm{Re}}\left[1-{\cal S}^{(3)}_{012}\right]\;
\nonumber\\
&&  \hspace{1cm}
\times
\frac{Q^2}{X_{012}^2}\; \Big[\textrm{K}_1\!\left(Q X_{012}\right)\Big]^2\;
\bigg\{ \Upsilon^{(a')}_{\textrm{inst.}}+ \Upsilon^{(b')}_{\textrm{inst.}} +\Upsilon^{(ab)}_{ \textrm{ interf.}} \bigg\}
\label{sigma_T_qqbarg_finite_piece}
\end{eqnarray}
where
\begin{eqnarray}
\Upsilon^{(a')}_{\textrm{inst.}}
&=& \frac{(k_{0}^+)^2 (k_{1}^+)^2 (k_{2}^+)^2}{(q^+)^4(k_{0}^+\!+\!k_{2}^+)^2}
 -\frac{2(k_{0}^+)^2 (k_{1}^+)^3 k_{2}^+}{(q^+)^5(k_{0}^+\!+\!k_{2}^+)}\;
 \frac{\left(\x_{0+2;1}\!\cdot\! \x_{20}\right)}{\x_{20}^2}
+\frac{2(k_{0}^+)^2 k_{1}^+(k_{1}^+\!+\!k_{2}^+)^2 k_{2}^+}{(q^+)^5(k_{0}^+\!+\!k_{2}^+)}\;
 \frac{\left(\x_{0;1+2}\!\cdot\! \x_{21}\right)}{\x_{21}^2}
\label{upsilon_a_prime}
\\
\Upsilon^{(b')}_{\textrm{inst.}}
&=&
 \frac{(k_{0}^+)^2 (k_{1}^+)^2 (k_{2}^+)^2}{(q^+)^4(k_{1}^+\!+\!k_{2}^+)^2}
 -\frac{2 k_{0}^+ (k_{1}^+)^2 (k_{0}^+\!+\!k_{2}^+)^2 k_{2}^+}{(q^+)^5(k_{1}^+\!+\!k_{2}^+)}\;
 \frac{\left(\x_{0+2;1}\!\cdot\! \x_{20}\right)}{\x_{20}^2}
+\frac{2(k_{0}^+)^3 (k_{1}^+)^2 k_{2}^+}{(q^+)^5(k_{1}^+\!+\!k_{2}^+)}\;
 \frac{\left(\x_{0;1+2}\!\cdot\! \x_{21}\right)}{\x_{21}^2}
\label{upsilon_b_prime}
\\
\Upsilon^{(ab)}_{ \textrm{ interf.}}
&=&
-\frac{2 k_{0}^+ k_{1}^+}{(q^+)^6}\, \Big[k_{1}^+(k_{0}^+\!+\!k_{2}^+)+k_{0}^+(k_{1}^+\!+\!k_{2}^+)\Big]
\Big[k_{0}^+(k_{0}^+\!+\!k_{2}^+)+k_{1}^+(k_{1}^+\!+\!k_{2}^+)\Big]
 \frac{\left(\x_{20}\!\cdot\! \x_{21}\right)}{\x_{20}^2\, \x_{21}^2}\,
 \left(\x_{0+2;1}\!\cdot\! \x_{0;1+2}\right)
\nonumber\\
&&
+\frac{2 k_{0}^+ k_{1}^+ (k_{2}^+)^2 (k_{0}^+\!-\!k_{1}^+)^2}{(q^+)^4(k_{0}^+\!+\!k_{2}^+)(k_{1}^+\!+\!k_{2}^+)}\:
 \frac{\left(\x_{20}\wedge \x_{21}\right)^2}{\x_{20}^2\, \x_{21}^2}
\label{upsilon_ab_interf}
 \, .
\end{eqnarray}


\subsection{Quark-antiquark-gluon contribution: $|a|^2$ and $|b|^2$ terms}

In generic dimension $D$, the $q\bar{q}g$ contribution to the transverse photon cross section is given by
\begin{eqnarray}
\sigma^{\gamma^*}_{T} \Big|_{q\bar{q}g}
&=& N_c C_F
\sum_{f}
\widetilde{\sum_{{\cal P.S.}\, q_0 \bar{q}_1g_2; D}}\;
\frac{(2\pi)}{2 (2k^+_0)(2k^+_1)}\;
{\textrm{Re}}\left[1-{\cal S}^{(3)}_{012}\right]\;
\frac{1}{(D\!-\!2)}\;
\sum_{T\textrm{ pol. }\lambda, \lambda_2}\;\;  \sum_{h_0, h_1 = \pm 1/2}
\left|\widetilde{\psi}_{\gamma_{\lambda}^*\rightarrow q_0 \bar{q_1} g_2}\right|^2\;
\, ,
\end{eqnarray}
and the contribution of the diagram $(a)$ to the LFWF is
\begin{eqnarray}
&& \hspace{-0.7cm} \widetilde{\psi}_{\gamma_{\lambda}^{*}\rightarrow q_0\bar{q}_1g_2}^{(a)}
=
-\frac{ e\, e_f\, g\,
\varepsilon_{\lambda}^{i} \varepsilon_{\lambda_2}^{j *}}{(k_{0}^+\!+\!k_{2}^+)q^+}\;
   {\cal I}^{lm}\!\left(a\right)\,
  \overline{u_G}(0) \gamma^+\!\! \left[(2k_{0}^+\!+\!k_{2}^+) \delta^{jm}
   + \frac{k_{2}^+}{2} [\gamma^j,\gamma^m]\right]\!
   \left[(2k_{1}^+\!-\!q^+) \delta^{il}
   + \frac{q^+}{2} [\gamma^i,\gamma^l]\right]\!  v_G(1)
\, .
\end{eqnarray}
Hence, one has
\begin{eqnarray}
&&\sum_{T\textrm{ pol. }\lambda, \lambda_2}\;\;  \sum_{h_0, h_1 = \pm 1/2}
\left|\widetilde{\psi}_{\gamma_{\lambda}^*\rightarrow q_0 \bar{q_1} g_2}^{(a)} \right|^2
=
\frac{e^2\, e_f^2\, g^2}{(q^+)^2 (k_{0}^+\!+\!k_{2}^+)^2}\, \delta^{i'i}\, \delta^{j'j}\,
{\cal I}^{l'm'}\!\left(a\right)^*\, {\cal I}^{lm}\!\left(a\right)
\nonumber\\
& & \hspace{1.2cm}
\times
 \!\!\!\sum_{h_0, h_1 = \pm 1/2} \!\!\!
\overline{v_G}(1) \gamma^+
\!\left[(2k_{1}^+\!-\!q^+) \delta^{i'l'} -\frac{q^+}{2} [\gamma^{i'},\gamma^{l'}]\right]\! \left[(2k_{0}^+\!+\!k_{2}^+) \delta^{j'm'} -\frac{k_{2}^+}{2} [\gamma^{j'},\gamma^{m'}]\right]\!
u_G(0)
\nonumber\\
& & \hspace{2.5cm}
\times\,
\overline{u_G}(0) \gamma^+\!\!
\left[(2k_{0}^+\!+\!k_{2}^+) \delta^{jm} +\frac{k_{2}^+}{2} [\gamma^j,\gamma^m]\right]\!
   \left[(2k_{1}^+\!-\!q^+) \delta^{il} +\frac{q^+}{2} [\gamma^i,\gamma^l]\right]\!  v_G(1)
\nonumber\\
&=& \frac{(2k_0^+)(2k_1^+)\, e^2\, e_f^2\, g^2}{(q^+)^2 (k_{0}^+\!+\!k_{2}^+)^2}\,
{\cal I}^{l'm'}\!\left(a\right)^*\, {\cal I}^{lm}\!\left(a\right)\;
\textrm{Tr}\Bigg\{{\cal P}_{G}
\bigg[
4(k_{0}^+\!+\!k_{2}^+)^2\, \delta^{m'm}
+\Big(4k_{2}^+(k_{0}^+\!+\!k_{2}^+)-(D\!-\!2)(k_{2}^+)^2\Big)\gamma^{m'}\gamma^{m}
\bigg]
\nonumber\\
& &
\hspace{2cm} \times\:
\bigg[
4(q^+\!-\!k_{1}^+)^2\, \delta^{l'l}
+\Big(4q^+(q^+\!-\!k_{1}^+)-(D\!-\!2)(q^+)^2\Big)\gamma^{l}\gamma^{l'}
\bigg]
\Bigg\}
\nonumber\\
&=&
 2 (4\pi)^2\,\alpha_{em}\, e_f^2\, \alpha_s\,  \frac{(2k_0^+)(2k_1^+)}{(q^+)^2 (k_{0}^+\!+\!k_{2}^+)^2}\,
\bigg[4k_{0}^+(k_{0}^+\!+\!k_{2}^+)+(D\!-\!2)(k_{2}^+)^2\bigg]\!
\bigg[(D\!-\!2)(q^+)^2\!-\!4k_{1}^+(q^+\!-\!k_{1}^+)\bigg]
\left|{\cal I}^{lm}\!\left(a\right)\right|^2\;
\, .
\end{eqnarray}
In the last step of that calculation, one uses the fact that ${\cal I}^{lm}\!\left(a\right)\propto \x_{0+2;1}^l \,\x_{20}^m$ (see Eq.~\eqref{int_Ilm_result_1}) and thus ${\cal I}^{l'm'}\!\left(a\right)^* \propto \x_{0+2;1}^{l'} \,\x_{20}^{m'}$, so that only the piece symmetric both in $l',l$ and in $m',m$ in the Dirac trace survives.

Thus, the $|a|^2$ contribution to the transverse photon cross section writes
\begin{eqnarray}
 \sigma^{\gamma^*}_{T} \Big|_{q\bar{q}g, |a|^2}
&=& 4 N_c \alpha_{em}  \left(\frac{\alpha_s C_F}{\pi}\right)
\sum_{f}  e_f^2
\widetilde{\sum_{{\cal P.S.}\, q_0 \bar{q}_1g_2; D}}\;
{\textrm{Re}}\left[1-{\cal S}^{(3)}_{012}\right]\; \frac{(2\pi)^4}{2}\;
 \left|{\cal I}^{lm}\!\left(a\right)\right|^2\, 
\nonumber\\
&& \hspace{1cm}
\times
\frac{\big[(D\!-\!2)(q^+)^2\!-\!4k_{1}^+(q^+\!-\!k_{1}^+)\big]}{(D\!-\!2)\, (q^+)^2}\;
\frac{\big[4k_{0}^+(k_{0}^+\!+\!k_{2}^+)+(D\!-\!2)(k_{2}^+)^2\big]}{(k_{0}^+\!+\!k_{2}^+)^2}\;
\, .
\label{sigma_T_qqbarg_a2}
\end{eqnarray}

The calculation of the $|b|^2$ contribution is analog, and gives
\begin{eqnarray}
\sigma^{\gamma^*}_{T} \Big|_{q\bar{q}g, |b|^2}
&=& 4 N_c \alpha_{em}  \left(\frac{\alpha_s C_F}{\pi}\right)
\sum_{f}  e_f^2
\widetilde{\sum_{{\cal P.S.}\, q_0 \bar{q}_1g_2; D}}\;
{\textrm{Re}}\left[1-{\cal S}^{(3)}_{012}\right]\;
\frac{(2\pi)^4}{2}\;
 \left|{\cal I}^{lm}\!\left(b\right)\right|^2
\nonumber\\
&& \hspace{1cm}
\times
\frac{\big[(D\!-\!2)(q^+)^2\!-\!4k_{0}^+(q^+\!-\!k_{0}^+)\big]}{(D\!-\!2)\, (q^+)^2}\;
\frac{\big[4k_{1}^+(k_{1}^+\!+\!k_{2}^+)+(D\!-\!2)(k_{2}^+)^2\big]}{(k_{1}^+\!+\!k_{2}^+)^2}\;
 \, .
 \label{sigma_T_qqbarg_b2}
\end{eqnarray}


\subsection{UV subtraction}

The UV divergences of the $|a|^2$ and $|b|^2$ contributions to the transverse photon cross section at NLO can be dealt with in the same way as in the longitudinal photon case. Thanks to the UV behavior
\begin{eqnarray}
{\cal I}^{lm}\!\left(a\right)
= {\cal I}^{lm}\!\left(\x_{0+2;1},\x_{20};\overline{Q}_{(a)}^2,{\cal C}_{(a)}\right)
\sim {\cal I}^{lm}_{\textrm{UV}}\!\left(\x_{01},\x_{20};\overline{Q}_{(a)}^2\right) \quad \textrm{for} \quad \x_{20}\rightarrow 0\, ,
\end{eqnarray}
of the Fourier integral, with ${\cal I}^{lm}_{\textrm{UV}}$ defined in Eq.~\eqref{Ilm_UV_approx},
one can construct the UV subtraction term for the $|a|^2$ contribution as
\begin{eqnarray}
\sigma^{\gamma^*}_{T} \Big|_{UV, |a|^2}
&=& 4 N_c \alpha_{em}  \left(\frac{\alpha_s C_F}{\pi}\right)
\sum_{f}  e_f^2
\widetilde{\sum_{{\cal P.S.}\, q_0 \bar{q}_1g_2; D}}\;
{\textrm{Re}}\left[1-{\cal S}_{01}\right]\; \frac{(2\pi)^4}{2}\,
\nonumber\\
&& \times
\frac{\big[(D\!-\!2)(q^+)^2\!-\!4k_{1}^+(q^+\!-\!k_{1}^+)\big]}{(D\!-\!2)\, (q^+)^2}\;
\frac{\big[4k_{0}^+(k_{0}^+\!+\!k_{2}^+)+(D\!-\!2)(k_{2}^+)^2\big]}{(k_{0}^+\!+\!k_{2}^+)^2}\;
\nonumber\\
&& 
\times
\bigg[
\Big|{\cal I}^{lm}_{\textrm{UV}}\!\left(\x_{01},\x_{20};\overline{Q}_{(a)}^2\right)\Big|^2
-{\textrm{Re}}\bigg({\cal I}^{lm}_{\textrm{UV}}\!\left(\x_{01},\x_{20};\overline{Q}_{(a)}^2\right)^*\;
{\cal I}^{lm}_{\textrm{UV}}\!\left(\x_{01},\x_{21};\overline{Q}_{(a)}^2\right)
\bigg)
\bigg]
\, ,
\label{UV_subtr_term_a2_T}
\end{eqnarray}
in order to extract the UV divergence at $\x_{2}\rightarrow \x_{0}$ without producing any collinear divergence in the regime $\x_{20}^2\simeq \x_{21}^2\gg \x_{01}^2$. The corresponding UV subtraction term for the $|b|^2$ contribution is
\begin{eqnarray}
\sigma^{\gamma^*}_{T} \Big|_{UV, |b|^2}
&=& 4 N_c \alpha_{em}  \left(\frac{\alpha_s C_F}{\pi}\right)
\sum_{f}  e_f^2
\widetilde{\sum_{{\cal P.S.}\, q_0 \bar{q}_1g_2; D}}\;
{\textrm{Re}}\left[1-{\cal S}_{01}\right]\; \frac{(2\pi)^4}{2}\,
\nonumber\\
&& \times
\frac{\big[(D\!-\!2)(q^+)^2\!-\!4k_{0}^+(q^+\!-\!k_{0}^+)\big]}{(D\!-\!2)\, (q^+)^2}\;
\frac{\big[4k_{1}^+(k_{1}^+\!+\!k_{2}^+)+(D\!-\!2)(k_{2}^+)^2\big]}{(k_{1}^+\!+\!k_{2}^+)^2}
\nonumber\\
&& 
\times\,
\bigg[
\Big|{\cal I}^{lm}_{\textrm{UV}}\!\left(\x_{01},\x_{21};\overline{Q}_{(b)}^2\right)\Big|^2
-{\textrm{Re}}\bigg({\cal I}^{lm}_{\textrm{UV}}\!\left(\x_{01},\x_{21};\overline{Q}_{(b)}^2\right)^*\;
{\cal I}^{lm}_{\textrm{UV}}\!\left(\x_{01},\x_{20};\overline{Q}_{(b)}^2\right)
\bigg)
\bigg]
\, .
\label{UV_subtr_term_b2_T}
\end{eqnarray}

These UV subtraction terms can be calculated in the same way as in the longitudinal photon case. Then, using the minimal cut-off prescription (for $k_2^+$ only), one finds
\begin{eqnarray}
&&\sigma^{\gamma^*}_{T} \Big|_{UV, |a|^2}
= 4 N_c \alpha_{em} \sum_{f} e_f^2\;
\int_{k^+_{\min}}^{+\infty}\!\!\!\!\!\! dk^+_0\; \int_{0}^{+\infty}\!\!\!\!\!\! dk^+_1\;
\frac{\delta(k_0^+\!+\!k_1^+\!-\!q^+)}{q^+}\;
\frac{\big[(D\!-\!2)(q^+)^2\!-\!4k_{1}^+(q^+\!-\!k_{1}^+)\big]}{(D\!-\!2)\, (q^+)^2}\;
\int \frac{d^{D-2} \x_0}{2\pi} \int \frac{d^{D-2} \x_1}{2\pi}\;
\nonumber\\
& & \hspace{0.5cm}
\times
 {\textrm{Re}}\left[1-{\cal S}_{01}\right]\;
\Big[(2\pi)^2\, \mu^2\, \x_{01}^2\Big]^{\frac{D}{2}-2}
\Big(\overline{Q}^2\Big)^{\frac{D}{2}-1}
\left[\textrm{K}_{\frac{D}{2}-1}\Big(|\x_{01}|\, \overline{Q}\Big)\right]^2\,
\left(\frac{\alpha_s\, C_F}{\pi}\right)
\widetilde{{\cal V}}_{UV, |a|^2}\;
\, ,
\end{eqnarray}
with the same $\widetilde{{\cal V}}_{UV, |a|^2}$ as in the longitudinal photon case, given in Eq.~\eqref{V_UV_subtr_term_a2}. The results for $|b|^2$ are the same, up to the exchange of $(k_0^+,\x_0)$ and $(k_1^+,\x_1)$. Combining the UV subtraction terms with the $q\bar{q}$ contribution and then dropping the unnecessary $k^+_{\min}$ cut-off like in the longitudinal photon case, one gets
\begin{eqnarray}
\sigma^{\gamma^*}_{T} \Big|_{\textrm{dipole}}
&=&\left[\sigma^{\gamma^*}_{T} \Big|_{q\bar{q}} +\sigma^{\gamma^*}_{T} \Big|_{UV, |a|^2}
+\sigma^{\gamma^*}_{T} \Big|_{UV, |b|^2}\right]
\nonumber\\
&=& 4 N_c \alpha_{em} \sum_{f} e_f^2\;
\int_{0}^{+\infty}\!\!\!\!\!\! dk^+_0\; \int_{0}^{+\infty}\!\!\!\!\!\! dk^+_1\;
\frac{\delta(k_0^+\!+\!k_1^+\!-\!q^+)}{q^+}\;
\frac{k_0^+\, k_1^+}{(q^+)^4}
\Big[(k_{0}^+)^2+(k_{1}^+)^2\Big]\,
\nonumber\\
& & \hspace{0.5cm}
\times
\int \frac{d^2 \x_0}{2\pi} \int \frac{d^2 \x_1}{2\pi}\;
{\textrm{Re}}\left[1-{\cal S}_{01}\right]\;
Q^2
\left[\textrm{K}_{1}\Big(|\x_{01}|\, \overline{Q}\Big)\right]^2
\left[1+\left(\frac{\alpha_s\, C_F}{\pi}\right)
\widetilde{{\cal V}}_{\textrm{reg.}}\right]\;
\label{sigma_T_dipole}
\end{eqnarray}
with $\widetilde{{\cal V}}_{\textrm{reg.}}$ defined in Eq.~\eqref{V_reg}.

Moreover, combining the $|a|^2$ contribution \eqref{sigma_T_qqbarg_a2} and its subtraction term \eqref{UV_subtr_term_a2_T} at integrand level, one can take the $D\rightarrow 4$ limit, and find
\begin{eqnarray}
&&\left[\sigma^{\gamma^*}_{T} \Big|_{q\bar{q}g, |a|^2} -\sigma^{\gamma^*}_{T} \Big|_{UV, |a|^2}\right]
= 4 N_c \alpha_{em} \left(\frac{\alpha_s\, C_F}{\pi}\right)\sum_{f} e_f^2\;
\widetilde{\sum_{{\cal P.S.}\, q_0 \bar{q}_1g_2; 4}}\;
\frac{k_1^+\, (q^+\!-\!k_{1}^+)}{(q^+)^4}\,
\Big[(q^+\!-\!k_{1}^+)^2+(k_{1}^+)^2\Big]\,
\nonumber\\
&&  \hspace{1.5cm}
\times
\frac{\big[2k_{0}^+(k_{0}^+\!+\!k_{2}^+)+(k_{2}^+)^2\big]}{(k_{0}^+\!+\!k_{2}^+)^2}\;
 \Bigg\{
\frac{1}{\x_{20}^2}\,
\left[\frac{k_1^+ (q^+\!-\!k_{1}^+)\, (\x_{0+2;1})^2}{(q^+)^2\, X_{012}^2}\right]\,
Q^2 \Big[\textrm{K}_1\!\left(Q X_{012}\right)\Big]^2\;
{\textrm{Re}}\left[1-{\cal S}^{(3)}_{012}\right]
\nonumber\\
&&  \hspace{7cm}
-\, \frac{\x_{20}}{\x_{20}^2} \!\cdot\! \left(\frac{\x_{20}}{\x_{20}^2}\!-\!\frac{\x_{21}}{\x_{21}^2}\right)\,
Q^2 \Big[\textrm{K}_{1}\left(\overline{Q}_{(a)}\, |\x_{01}|\right)\Big]^2\;
{\textrm{Re}}\left[1-{\cal S}_{01}\right]
\Bigg\}
\, .
\label{sigma_T_a2_subtracted}
\end{eqnarray}
Similarly, one gets for the $|b|^2$ contribution
\begin{eqnarray}
&&\left[\sigma^{\gamma^*}_{T} \Big|_{q\bar{q}g, |b|^2} -\sigma^{\gamma^*}_{T} \Big|_{UV, |b|^2}\right]
= 4 N_c \alpha_{em} \left(\frac{\alpha_s\, C_F}{\pi}\right)\sum_{f} e_f^2\;
\widetilde{\sum_{{\cal P.S.}\, q_0 \bar{q}_1g_2; 4}}\;
\frac{k_0^+\, (q^+\!-\!k_{0}^+)}{(q^+)^4}\,
\Big[(k_{0}^+)^2+(q^+\!-\!k_{0}^+)^2\Big]\,
\nonumber\\
&&  \hspace{1.5cm}
\times
\frac{[2k_{1}^+(k_{1}^+\!+\!k_{2}^+)+(k_{2}^+)^2]}{(k_{1}^+\!+\!k_{2}^+)^2}\;
\Bigg\{
\frac{1}{\x_{21}^2}\,
\left[\frac{k_0^+ (q^+\!-\!k_{0}^+)\, (\x_{0;1+2})^2}{(q^+)^2\, X_{012}^2}\right]\,
Q^2 \Big[\textrm{K}_1\!\left(Q X_{012}\right)\Big]^2\;
{\textrm{Re}}\left[1-{\cal S}^{(3)}_{012}\right]
\nonumber\\
&&  \hspace{7cm}
-\, \frac{\x_{21}}{\x_{21}^2} \!\cdot\! \left(\frac{\x_{21}}{\x_{21}^2}\!-\!\frac{\x_{20}}{\x_{20}^2}\right)\,
Q^2 \Big[\textrm{K}_{1}\left(\overline{Q}_{(b)}\, |\x_{01}|\right)\Big]^2\;
{\textrm{Re}}\left[1-{\cal S}_{01}\right]
\Bigg\}
\, .
\label{sigma_T_b2_subtracted}
\end{eqnarray}


\subsection{Rearranging the NLO cross-section for transverse photon}

At this stage, the transverse photon cross section at NLO is given by the sum of the expressions
\eqref{sigma_T_qqbarg_finite_piece}, \eqref{sigma_T_dipole}, \eqref{sigma_T_a2_subtracted} and \eqref{sigma_T_b2_subtracted}. Each of these terms is UV finite.
One can check from these expressions that the $q\bar{q}g$ piece is consistent with the results of Ref.~\cite{Beuf:2011xd} (see Eqs.~(43) and (51) there).
 The aim of this section is to show how to simplify these expressions, in order to get the transverse photon cross section at NLO in a form analog to Eq.~\eqref{sigma_L_NLO_fixed}. In order to do so, one has to eliminate the transverse vectors $\x_{0+2;1}$ and $\x_{0;1+2}$ in favor of simpler ones. According to the definition \eqref{def_pre-split_parent_dipole}, they write
\begin{eqnarray}
\x_{0+2;1}&=&   - \left(\frac{k^+_0}{k^+_0\!+\!k^+_2}\right) \x_{20} + \x_{21} =\x_{01} + \left(\frac{k^+_2}{k^+_0\!+\!k^+_2}\right) \x_{20}
\label{pre-split_parent_dipole_0} \\
\x_{0;1+2}&=& -\x_{20} + \left(\frac{k^+_1}{k^+_1\!+\!k^+_2}\right) \x_{21}
= \x_{01} - \left(\frac{k^+_2}{k^+_1\!+\!k^+_2}\right) \x_{21}
\label{pre-split_parent_dipole_1}
\, .
\end{eqnarray}
Then, their squares can calculated as
\begin{eqnarray}
(\x_{0+2;1})^2 &=& \frac{(q^+)^2}{k^+_1(k^+_0\!+\!k^+_2)}\, X_{012}^2
- \frac{q^+\, k^+_0\, k^+_2}{k^+_1(k^+_0\!+\!k^+_2)^2}\, \x_{20}^2  \\
(\x_{0;1+2})^2 &=& \frac{(q^+)^2}{k^+_0(k^+_1\!+\!k^+_2)}\, X_{012}^2
- \frac{q^+\, k^+_1\, k^+_2}{k^+_0(k^+_1\!+\!k^+_2)^2}\, \x_{21}^2
\, .
\end{eqnarray}
Thanks to these relations, the expressions \eqref{sigma_T_a2_subtracted} and \eqref{sigma_T_b2_subtracted}
become
\begin{eqnarray}
&&\left[\sigma^{\gamma^*}_{T} \Big|_{q\bar{q}g, |a|^2} -\sigma^{\gamma^*}_{T} \Big|_{UV, |a|^2}\right]
= 4 N_c \alpha_{em} \left(\frac{\alpha_s\, C_F}{\pi}\right)\sum_{f} e_f^2\;
\widetilde{\sum_{{\cal P.S.}\, q_0 \bar{q}_1g_2; 4}}\;
\frac{k_1^+\, (q^+\!-\!k_{1}^+)}{(q^+)^4}\,
\Big[(q^+\!-\!k_{1}^+)^2+(k_{1}^+)^2\Big]\,
\nonumber\\
&&  \hspace{1.5cm}
\times
\frac{\big[2k_{0}^+(k_{0}^+\!+\!k_{2}^+)+(k_{2}^+)^2\big]}{(k_{0}^+\!+\!k_{2}^+)^2}\;
\Bigg\{
\frac{\x_{20}}{\x_{20}^2} \!\cdot\! \left(\frac{\x_{20}}{\x_{20}^2}\!-\!\frac{\x_{21}}{\x_{21}^2}\right)\,
\bigg[Q^2 \Big(\textrm{K}_1\!\left(Q X_{012}\right)\Big)^2\;
{\textrm{Re}}\left(1-{\cal S}^{(3)}_{012}\right)-\Big(\x_{2}\rightarrow \x_{0}\Big)\bigg]
\nonumber\\
&&  \hspace{5.5cm}
+\left[\frac{\x_{20}\!\cdot\!\x_{21}}{\x_{20}^2 \x_{21}^2}
  -\frac{k_{0}^+ k_{2}^+}{q^+ (k_{0}^+\!+\!k_{2}^+) X_{012}^2}\right]\,
Q^2 \Big(\textrm{K}_1\!\left(Q X_{012}\right)\Big)^2\;
{\textrm{Re}}\left(1-{\cal S}^{(3)}_{012}\right)
\Bigg\}
\label{sigma_T_a2_subtracted_2}
\end{eqnarray}
and
\begin{eqnarray}
&&\left[\sigma^{\gamma^*}_{T} \Big|_{q\bar{q}g, |b|^2} -\sigma^{\gamma^*}_{T} \Big|_{UV, |b|^2}\right]
= 4 N_c \alpha_{em} \left(\frac{\alpha_s\, C_F}{\pi}\right)\sum_{f} e_f^2\;
\widetilde{\sum_{{\cal P.S.}\, q_0 \bar{q}_1g_2; 4}}\;
\frac{k_0^+\, (q^+\!-\!k_{0}^+)}{(q^+)^4}\,
\Big[(k_{0}^+)^2+(q^+\!-\!k_{0}^+)^2\Big]\,
\nonumber\\
&&  \hspace{1.5cm}
\times
\frac{\big[2k_{1}^+(k_{1}^+\!+\!k_{2}^+)+(k_{2}^+)^2\big]}{(k_{1}^+\!+\!k_{2}^+)^2}\;
\Bigg\{
\frac{\x_{21}}{\x_{21}^2} \!\cdot\! \left(\frac{\x_{21}}{\x_{21}^2}\!-\!\frac{\x_{20}}{\x_{20}^2}\right)\,
\bigg[Q^2 \Big(\textrm{K}_1\!\left(Q X_{012}\right)\Big)^2\;
{\textrm{Re}}\left(1-{\cal S}^{(3)}_{012}\right)-\Big(\x_{2}\rightarrow \x_{1}\Big)\bigg]
\nonumber\\
&&  \hspace{5.5cm}
+\left[\frac{\x_{20}\!\cdot\!\x_{21}}{\x_{20}^2 \x_{21}^2}
  -\frac{k_{1}^+ k_{2}^+}{q^+ (k_{1}^+\!+\!k_{2}^+) X_{012}^2}\right]\,
Q^2 \Big(\textrm{K}_1\!\left(Q X_{012}\right)\Big)^2\;
{\textrm{Re}}\left(1-{\cal S}^{(3)}_{012}\right)
\Bigg\}
\label{sigma_T_b2_subtracted_2}
\, ,
\end{eqnarray}
which make the cancellation of UV divergences more obvious.

From Eqs.~\eqref{pre-split_parent_dipole_0} and \eqref{pre-split_parent_dipole_1} one can also find
\begin{eqnarray}
\frac{(\x_{0+2;1}\!\cdot\!\x_{20})}{\x_{20}^2}
&=& \frac{(\x_{20}\!\cdot\!\x_{21})}{\x_{20}^2} - \left(\frac{k^+_0}{k^+_0\!+\!k^+_2}\right)
\\
\frac{(\x_{0;1+2}\!\cdot\!\x_{21})}{\x_{21}^2}
&=& - \frac{(\x_{20}\!\cdot\!\x_{21})}{\x_{21}^2} + \left(\frac{k^+_1}{k^+_1\!+\!k^+_2}\right)
\\
(\x_{0+2;1}\!\cdot\!\x_{0;1+2}) &=& \frac{(q^+)^2\, X_{012}^2}{(k^+_0\!+\!k^+_2)(k^+_1\!+\!k^+_2)}
- \frac{q^+\, k^+_2\, (\x_{20}\!\cdot\!\x_{21})}{(k^+_0\!+\!k^+_2)(k^+_1\!+\!k^+_2)}
\, .
\end{eqnarray}
Moreover, one has
\begin{eqnarray}
 \frac{\left(\x_{20}\wedge \x_{21}\right)^2}{\x_{20}^2\, \x_{21}^2}
 &=& 1- \frac{\left(\x_{20}\!\cdot\! \x_{21}\right)^2}{\x_{20}^2\, \x_{21}^2}
\, ,
\end{eqnarray}
since the left-hand side is the square of the sine of the angle between $\x_{20}$ and $\x_{21}$.
With all these identities, the expressions \eqref{upsilon_a_prime}, \eqref{upsilon_b_prime} and \eqref{upsilon_ab_interf} can be rewritten as
\begin{eqnarray}
\Upsilon^{(a')}_{\textrm{inst.}}
&=&  -\frac{2(k_{0}^+)^2 (k_{1}^+)^3 k_{2}^+}{(q^+)^5(k_{0}^+\!+\!k_{2}^+)}\;
 \frac{\left(\x_{20}\!\cdot\! \x_{21}\right)}{\x_{20}^2}
-\frac{2(k_{0}^+)^2 k_{1}^+(k_{1}^+\!+\!k_{2}^+)^2 k_{2}^+}{(q^+)^5(k_{0}^+\!+\!k_{2}^+)}\;
 \frac{\left(\x_{20}\!\cdot\! \x_{21}\right)}{\x_{21}^2}
\nonumber\\
&&
+\frac{(k_{0}^+)^2 (k_{1}^+)^2 k_{2}^+}{(q^+)^5(k_{0}^+\!+\!k_{2}^+)^2}\,
\Big[k_{0}^+ k_{1}^+ + 3(k_{0}^+\!+\!k_{2}^+)(k_{1}^+\!+\!k_{2}^+)\Big]
\\
\Upsilon^{(b')}_{\textrm{inst.}}
&=&
-\frac{2k_{0}^+ (k_{1}^+)^2(k_{0}^+\!+\!k_{2}^+)^2 k_{2}^+}{(q^+)^5(k_{1}^+\!+\!k_{2}^+)}\;
 \frac{\left(\x_{20}\!\cdot\! \x_{21}\right)}{\x_{20}^2}
-\frac{2(k_{0}^+)^3 (k_{1}^+)^2 k_{2}^+}{(q^+)^5(k_{1}^+\!+\!k_{2}^+)}\;
 \frac{\left(\x_{20}\!\cdot\! \x_{21}\right)}{\x_{21}^2}
\nonumber\\
&&
+\frac{(k_{0}^+)^2 (k_{1}^+)^2 k_{2}^+}{(q^+)^5(k_{1}^+\!+\!k_{2}^+)^2}\,
\Big[k_{0}^+ k_{1}^+ + 3(k_{0}^+\!+\!k_{2}^+)(k_{1}^+\!+\!k_{2}^+)\Big]
\end{eqnarray}
and
\begin{eqnarray}
\Upsilon^{(ab)}_{ \textrm{ interf.}}
&=&
-\frac{2 k_{0}^+ k_{1}^+\Big[k_{1}^+(k_{0}^+\!+\!k_{2}^+)+k_{0}^+(k_{1}^+\!+\!k_{2}^+)\Big]
\Big[k_{0}^+(k_{0}^+\!+\!k_{2}^+)+k_{1}^+(k_{1}^+\!+\!k_{2}^+)\Big]}{(q^+)^4(k^+_0\!+\!k^+_2)(k^+_1\!+\!k^+_2)}\,
 \frac{\left(\x_{20}\!\cdot\! \x_{21}\right)}{\x_{20}^2\, \x_{21}^2}\, X_{012}^2
\nonumber\\
&&
+\frac{4 (k_{0}^+)^2  (k_{1}^+)^2 k_{2}^+}{(q^+)^5(k_{0}^+\!+\!k_{2}^+)(k_{1}^+\!+\!k_{2}^+)}
\Big[(k_{0}^+\!+\!k_{2}^+)^2 +(k_{1}^+\!+\!k_{2}^+)^2\Big]
\frac{\left(\x_{20}\!\cdot\! \x_{21}\right)^2}{\x_{20}^2\, \x_{21}^2}
+\frac{2 k_{0}^+ k_{1}^+ (k_{2}^+)^2 (k_{0}^+\!-\!k_{1}^+)^2}{(q^+)^4(k_{0}^+\!+\!k_{2}^+)(k_{1}^+\!+\!k_{2}^+)}
\\
&=&
-\frac{2 k_{0}^+ k_{1}^+}{(q^+)^4}\,
\bigg\{ \Big[(q^+\!-\!k_{1}^+)^2\!+\!(k_{1}^+)^2\Big]
+\Big[(k_{0}^+)^2\!+\!(q^+\!-\!k_{0}^+)^2\Big] \bigg\}
 \frac{\left(\x_{20}\!\cdot\! \x_{21}\right)}{\x_{20}^2\, \x_{21}^2}\, X_{012}^2
\nonumber\\
&&
+\frac{2 k_{0}^+  k_{1}^+ k_{2}^+}{(q^+)^5}
\Big[(k_{0}^+\!+\!k_{2}^+)^2 +(k_{1}^+\!+\!k_{2}^+)^2\Big]
\bigg\{ \left(\frac{k^+_1}{k^+_1\!+\!k^+_2}\right) \frac{(\x_{20}\!\cdot\!\x_{21})}{\x_{20}^2}
+ \left(\frac{k^+_0}{k^+_0\!+\!k^+_2}\right) \frac{(\x_{20}\!\cdot\!\x_{21})}{\x_{21}^2}
\bigg\}
\nonumber\\
&&+\frac{2 k_{0}^+ k_{1}^+ (k_{2}^+)^2 (k_{0}^+\!-\!k_{1}^+)^2}{(q^+)^4(k_{0}^+\!+\!k_{2}^+)(k_{1}^+\!+\!k_{2}^+)}
\, ,
\end{eqnarray}
so that their sum gives
\begin{eqnarray}
\Upsilon^{(ab)}_{ \textrm{ interf.}}+\Upsilon^{(a')}_{\textrm{inst.}}+\Upsilon^{(b')}_{\textrm{inst.}}
&=&
-\frac{2 k_{0}^+ k_{1}^+}{(q^+)^4}\,
\bigg\{ \Big[(q^+\!-\!k_{1}^+)^2\!+\!(k_{1}^+)^2\Big]
+\Big[(k_{0}^+)^2\!+\!(q^+\!-\!k_{0}^+)^2\Big] \bigg\}
 \frac{\left(\x_{20}\!\cdot\! \x_{21}\right)}{\x_{20}^2\, \x_{21}^2}\, X_{012}^2
\nonumber\\
&&
+\frac{2k_{0}^+ (k_{1}^+)^2 (k_{2}^+)^2}{(q^+)^4(k_{0}^+\!+\!k_{2}^+)}\;
 \frac{\left(\x_{20}\!\cdot\! \x_{21}\right)}{\x_{20}^2}
+\frac{2(k_{0}^+)^2 k_{1}^+ (k_{2}^+)^2}{(q^+)^4(k_{1}^+\!+\!k_{2}^+)}\;
 \frac{\left(\x_{20}\!\cdot\! \x_{21}\right)}{\x_{21}^2}
\nonumber\\
&&+\frac{k_{0}^+ k_{1}^+ k_{2}^+}{(q^+)^5}\;
\bigg\{
(k_{0}^+\!+\!k_{2}^+)^2+(k_{1}^+\!+\!k_{2}^+)^2+2(k_{0}^+)^2+2(k_{1}^+)^2
\nonumber\\
&&\hspace{1cm}
+\frac{(k_{0}^+)^2 (k_{1}^+)^2}{(k_{0}^+\!+\!k_{2}^+)^2}+\frac{(k_{0}^+)^2 (k_{1}^+)^2}{(k_{1}^+\!+\!k_{2}^+)^2}-q^+ k_{2}^+\bigg[\left(\frac{k_{1}^+\!+\!k_{2}^+}{k_{0}^+\!+\!k_{2}^+}\right) +\left(\frac{k_{0}^+\!+\!k_{2}^+}{k_{1}^+\!+\!k_{2}^+}\right)\bigg]\bigg\}
\, .
\end{eqnarray}
There are important cancellations between these terms and the contributions \eqref{sigma_T_a2_subtracted_2}
and \eqref{sigma_T_b2_subtracted_2}. One can then write
\begin{eqnarray}
\sigma^{\gamma^*}_{T} \Big|_{q\rightarrow g}+\sigma^{\gamma^*}_{T} \Big|_{\bar{q}\rightarrow g}
&=&\left[\sigma^{\gamma^*}_{T} \Big|_{q\bar{q}g, |a|^2} -\sigma^{\gamma^*}_{T} \Big|_{UV, |a|^2}\right]
+ \left[\sigma^{\gamma^*}_{T} \Big|_{q\bar{q}g, |b|^2} -\sigma^{\gamma^*}_{T} \Big|_{UV, |b|^2}\right]
+\sigma^{\gamma^*}_{T} \Big|_{q\bar{q}g;\, \textrm{UV finite terms}}
\, ,
\end{eqnarray}
where
\begin{eqnarray}
\sigma^{\gamma^*}_{T} \Big|_{q\rightarrow g}
&=& 4 N_c \alpha_{em} \left(\frac{\alpha_s\, C_F}{\pi}\right)\sum_{f} e_f^2\;
\widetilde{\sum_{{\cal P.S.}\, q_0 \bar{q}_1g_2; 4}}\;
\Bigg\{
\frac{k_1^+\, (q^+\!-\!k_{1}^+)}{(q^+)^4}\,
\Big[(q^+\!-\!k_{1}^+)^2+(k_{1}^+)^2\Big]\,
\frac{\big[2k_{0}^+(k_{0}^+\!+\!k_{2}^+)+(k_{2}^+)^2\big]}{(k_{0}^+\!+\!k_{2}^+)^2}\;
\nonumber\\
&&  \hspace{2cm}
\times\;
\frac{\x_{20}}{\x_{20}^2} \!\cdot\! \left(\frac{\x_{20}}{\x_{20}^2}\!-\!\frac{\x_{21}}{\x_{21}^2}\right)\,
\bigg[Q^2 \Big(\textrm{K}_1\!\left(Q X_{012}\right)\Big)^2\;
{\textrm{Re}}\left(1-{\cal S}^{(3)}_{012}\right)-\Big(\x_{2}\rightarrow \x_{0}\Big)\bigg]
\nonumber\\
&&  \hspace{0.5cm}
+\,
\left[
\Big[(q^+\!-\!k_{1}^+)^2+(k_{1}^+)^2\Big]\,
\frac{(\x_{20}\!\cdot\!\x_{21})}{\x_{20}^2 \x_{21}^2}
+2k_{0}^+ k_{1}^+\;
 \frac{\left(\x_{20}\!\cdot\! \x_{21}\right)}{\x_{20}^2 X_{012}^2}
  -\frac{k_{0}^+(k_{1}^+\!+\!k_{2}^+)}{ X_{012}^2}\right]\,
\nonumber\\
&& \hspace{2cm}
\times\; \frac{k_1^+\, (k_2^+)^2}{(q^+)^4 (k_{0}^+\!+\!k_{2}^+)}\,
Q^2 \Big(\textrm{K}_1\!\left(Q X_{012}\right)\Big)^2\;
{\textrm{Re}}\left(1-{\cal S}^{(3)}_{012}\right)
\Bigg\}
\label{sigma_T_q_to_g_plus}
\end{eqnarray}
and
\begin{eqnarray}
\sigma^{\gamma^*}_{T} \Big|_{\bar{q}\rightarrow g}
&=& 4 N_c \alpha_{em} \left(\frac{\alpha_s\, C_F}{\pi}\right)\sum_{f} e_f^2\;
\widetilde{\sum_{{\cal P.S.}\, q_0 \bar{q}_1g_2; 4}}\;
\Bigg\{
\frac{k_0^+\, (q^+\!-\!k_{0}^+)}{(q^+)^4}\,
\Big[(k_{0}^+)^2+(q^+\!-\!k_{0}^+)^2\Big]\,
\frac{\big[2k_{1}^+(k_{1}^+\!+\!k_{2}^+)+(k_{2}^+)^2\big]}{(k_{1}^+\!+\!k_{2}^+)^2}\;
\nonumber\\
&&  \hspace{2cm}
\times\;
\frac{\x_{21}}{\x_{21}^2} \!\cdot\! \left(\frac{\x_{21}}{\x_{21}^2}\!-\!\frac{\x_{20}}{\x_{20}^2}\right)\,
\bigg[Q^2 \Big(\textrm{K}_1\!\left(Q X_{012}\right)\Big)^2\;
{\textrm{Re}}\left(1-{\cal S}^{(3)}_{012}\right)-\Big(\x_{2}\rightarrow \x_{1}\Big)\bigg]
\nonumber\\
&&  \hspace{0.5cm}
+\,
\left[
\Big[(q^+\!-\!k_{0}^+)^2+(k_{0}^+)^2\Big]\,
\frac{(\x_{20}\!\cdot\!\x_{21})}{\x_{20}^2 \x_{21}^2}
+2k_{0}^+ k_{1}^+\;
 \frac{\left(\x_{20}\!\cdot\! \x_{21}\right)}{\x_{21}^2 X_{012}^2}
  -\frac{k_{1}^+(k_{0}^+\!+\!k_{2}^+)}{ X_{012}^2}\right]\,
\nonumber\\
&& \hspace{2cm}
\times\;
\frac{k_0^+\, (k_2^+)^2}{(q^+)^4 (k_{1}^+\!+\!k_{2}^+)}\,
Q^2 \Big(\textrm{K}_1\!\left(Q X_{012}\right)\Big)^2\;
{\textrm{Re}}\left(1-{\cal S}^{(3)}_{012}\right)
\Bigg\}
\label{sigma_T_qbar_to_g_plus}
\, .
\end{eqnarray}
Note that only the first part of the expressions \eqref{sigma_T_q_to_g_plus} and \eqref{sigma_T_qbar_to_g_plus}, including the UV subtraction term, has a potential divergence at low $k_2^+$ and thus contribute to high-energy leading logarithms, whereas the second part of the expressions \eqref{sigma_T_q_to_g_plus} and \eqref{sigma_T_qbar_to_g_plus} is regular at low $k_2^+$ thanks to the explicit $(k_2^+)^2$ factor compensating the $1/k_2^+$ factor included in  the phase-space integration \eqref{phase_space_qqbarg}.

Like in the longitudinal photon case, the integrands in Eqs.~ \eqref{sigma_T_q_to_g_plus} and \eqref{sigma_T_qbar_to_g_plus} are related by exchange of the quark and antiquark, $(k_{0}^+,\x_0)\leftrightarrow (k_{1}^+,\x_1)$. Hence, the corresponding integrals over the phase-space are equal:
\begin{eqnarray}
\sigma^{\gamma^*}_{T} \Big|_{\bar{q}\rightarrow g} = \sigma^{\gamma^*}_{T} \Big|_{q\rightarrow g}
\, .
\end{eqnarray}

All in all, the transverse photon cross section at NLO can be written as
\begin{eqnarray}
\sigma^{\gamma^*}_{T}&=&
\sigma^{\gamma^*}_{T} \Big|_{\textrm{dipole}} + \sigma^{\gamma^*}_{T} \Big|_{q\rightarrow g}
+\sigma^{\gamma^*}_{T} \Big|_{\bar{q}\rightarrow g}
=\sigma^{\gamma^*}_{T} \Big|_{\textrm{dipole}} +2\, \sigma^{\gamma^*}_{T} \Big|_{q\rightarrow g}
\label{sigma_T_NLO_fixed}
\, ,
\end{eqnarray}
with the various terms given in Eqs.~\eqref{sigma_T_dipole}, \eqref{sigma_T_q_to_g_plus} and \eqref{sigma_T_qbar_to_g_plus}.

\subsection{Final result for the transverse photon cross section at NLO before high-energy resummation\label{sec:final_res_fixed_NLO_T}}

Like in the longitudinal photon case, it is convenient to switch to light-cone momentum fraction variables.
Using the same change of variables as in sec.~\ref{sec:final_res_fixed_NLO_L} for each of the three terms in the transverse photon cross section at NLO given in Eq.~\eqref{sigma_T_NLO_fixed}, one gets
\begin{eqnarray}
\sigma^{\gamma^*}_{T} \Big|_{\textrm{dipole}}
&=& 4 N_c \alpha_{em} \sum_{f} e_f^2\;
\theta(1\!-\!2 z_c)
\int_{z_c}^{1-z_c}\!\! dz\;  z(1\!-\!z) \big[z^2+(1\!-\!z)^2\big]
\int \frac{d^2 \x_0}{2\pi} \int \frac{d^2 \x_1}{2\pi}\;
Q^2 \bigg( \textrm{K}_{1}\Big(Q \sqrt{z(1\!-\!z)}|\x_{01}|\Big)\bigg)^2\;
\nonumber\\
& &\hspace{2.5cm}
\times\; {\textrm{Re}}\big(1-{\cal S}_{01}\big)\;
\Bigg\{1+\left(\frac{\alpha_s\, C_F}{\pi}\right)\;
\bigg[\frac{1}{2} \left(\log\left(\frac{z}{1\!-\!z}\right)\right)^2
-\frac{\pi^2}{6}
 +\frac{5}{2}
\bigg]\Bigg\}\;\;
\label{sigma_T_dipole_frac}
\end{eqnarray}
for the dipole-like term,
\begin{eqnarray}
&&\sigma^{\gamma^*}_{T} \Big|_{q\rightarrow g}
= 4 N_c \alpha_{em} \left(\frac{\alpha_s\, C_F}{\pi}\right)\sum_{f} e_f^2\;
\theta\!\left(1\!-\! \frac{k^+_{\min}}{q^+}\!-\! z_c\right)
\int_{\frac{k^+_{\min}}{q^+}}^{1- z_c}\! dz\; z(1\!-\!z)
\int_{\frac{k^+_{\min}}{z q^+}}^{1}
d \zeta\;
\int \frac{d^2 \x_0}{2\pi} \int \frac{d^2 \x_1}{2\pi}\int \frac{d^2 \x_2}{2\pi}\;
\nonumber\\
&&\hspace{0.5cm}
\times \Bigg\{
\Big[z^2+(1\!-\!z)^2\Big]\,
\left[\frac{1+(1\!-\!\zeta)^2}{\zeta}\right]\;
\frac{\x_{20}}{\x_{20}^2} \!\cdot\! \left(\frac{\x_{20}}{\x_{20}^2}\!-\!\frac{\x_{21}}{\x_{21}^2}\right)\,
\bigg[Q^2 \Big(\textrm{K}_1\!\left(Q X_{012}\right)\Big)^2\;
{\textrm{Re}}\left(1-{\cal S}^{(3)}_{012}\right)-\Big(\x_{2}\rightarrow \x_{0}\Big)\bigg]
\nonumber\\
&&  \hspace{1cm}
+\zeta\,
\left[
\Big[z^2+(1\!-\!z)^2\Big]\,
\frac{(\x_{20}\!\cdot\!\x_{21})}{\x_{20}^2 \x_{21}^2}
+2z(1\!-\!z)(1\!-\!\zeta)\;
 \frac{\left(\x_{20}\!\cdot\! \x_{21}\right)}{\x_{20}^2 X_{012}^2}
  -\frac{z(1\!-\!\zeta)(1\!-\!z\!+\!\zeta z)}{ X_{012}^2}\right]\,
\nonumber\\
&& \hspace{3cm}
\times\;
Q^2 \Big(\textrm{K}_1\!\left(Q X_{012}\right)\Big)^2\;
{\textrm{Re}}\left(1-{\cal S}^{(3)}_{012}\right)
\Bigg\}
\label{sigma_T_q_to_g_frac}
\end{eqnarray}
for the $q\rightarrow g$ contribution (with $X_{012}$ given by Eq.~\eqref{def_X012_q2g}), and
\begin{eqnarray}
\sigma^{\gamma^*}_{T} \Big|_{\bar{q}\rightarrow g}
&=& 4 N_c \alpha_{em} \left(\frac{\alpha_s\, C_F}{\pi}\right)\sum_{f} e_f^2\;
\theta\!\left(1\!-\! \frac{k^+_{\min}}{q^+}\!-\! z_c\right)
\int_{z_c}^{1- \frac{k^+_{\min}}{q^+}}\! dz\; z(1\!-\!z)
\int_{\frac{k^+_{\min}}{(1\!-\!z) q^+}}^{1}
d \zeta\;
\int \frac{d^2 \x_0}{2\pi} \int \frac{d^2 \x_1}{2\pi}\int \frac{d^2 \x_2}{2\pi}\;
\nonumber\\
&&
\times \Bigg\{
\Big[z^2+(1\!-\!z)^2\Big]\,
\left[\frac{1+(1\!-\!\zeta)^2}{\zeta}\right]\;
\frac{\x_{21}}{\x_{21}^2} \!\cdot\! \left(\frac{\x_{21}}{\x_{21}^2}\!-\!\frac{\x_{20}}{\x_{20}^2}\right)\,
\bigg[Q^2 \Big(\textrm{K}_1\!\left(Q X_{012}\right)\Big)^2\;
{\textrm{Re}}\left(1-{\cal S}^{(3)}_{012}\right)-\Big(\x_{2}\rightarrow \x_{0}\Big)\bigg]
\nonumber\\
&&  \hspace{0.5cm}
+\zeta\,
\left[
\Big[z^2+(1\!-\!z)^2\Big]\,
\frac{(\x_{20}\!\cdot\!\x_{21})}{\x_{20}^2 \x_{21}^2}
+2z(1\!-\!z)(1\!-\!\zeta)\;
 \frac{\left(\x_{20}\!\cdot\! \x_{21}\right)}{\x_{21}^2 X_{012}^2}
  -\frac{(1\!-\!z)(1\!-\!\zeta)(z\!+\!\zeta (1\!-\!z))}{ X_{012}^2}\right]\,
\nonumber\\
&& \hspace{3cm}
\times\;
Q^2 \Big(\textrm{K}_1\!\left(Q X_{012}\right)\Big)^2\;
{\textrm{Re}}\left(1-{\cal S}^{(3)}_{012}\right)
\Bigg\}
\label{sigma_T_qbar_to_g_frac}
\end{eqnarray}
for the $\bar{q}\rightarrow g$ contribution (with now $X_{012}$ given by Eq.~\eqref{def_X012_qbar2g}).

Like in sec.~\ref{sec:final_res_fixed_NLO_L}, the parameter $z_c$ has been introduced in order to account for the $k^+$ regularization scheme dependence: $z_c=0$ for the minimal cut-off prescription, whereas $z_c=k^+_{\min}/q^+$ for the non-minimal cut-off prescription of sec.~\ref{sec:non_mini_cut}.


\section{High-energy LL resummation\label{sec:LL_resum}}

The transverse or longitudinal photon cross sections at NLO calculated so far correspond to the scattering of the photon on a classical gluon field shockwave. In order to obtain the corresponding photon-target total cross sections encountered in the neutral current DIS process, Eq.~\eqref{DIS_xsect_one_photon}, it remains to take the functional average over the classical gluon field shockwave configurations. Since this classical field appears only via the Wilson lines contained in the dipole and tripole amplitude, ${\cal S}_{01}$ and ${\cal S}^{(3)}_{012}$, the photon-target total cross sections are obtained by making the replacements ${\cal S}_{01}\mapsto \langle{\cal S}_{01}\rangle_{0}$ and ${\cal S}^{(3)}_{012}\mapsto \langle{\cal S}^{(3)}_{012}\rangle_{0}$
in the results of the previous sections. The index $0$ for this statistical average indicates that only the long lived partons in the target are included as sources for the semi-classical gluon field, in order to avoid double counting.

Indeed, the short-lived partons in the target are already taken into account via higher Fock states in the dressed photon. By consistency the minimal lifetime along $x^-$ for the sources included in the target average should be set by $1/k^+_{\min}$, since $k^+_{\min}$ represent the minimal $k^+$ for the partons (or at least gluons) included in the dressed photon projectile. It is convenient to choose $k^+_{\min}$ as small as possible, in order to leave the valence partons and the non-perturbative fluctuations inside the target, and to include most of the perturbative radiation into the projectile. As discussed in sec.~II.B of Ref.~\cite{Beuf:2014uia}, this amounts to choose $k^+_{\min}$ as the typical $k^+$ scale associated with the non-perturbative partons in the target before any evolution,\footnote{With an exact calculation taking into account the target mass $M_T$ in the last step of Eq.~\eqref{def_k_plus_min}, the $x_{Bj}$ would be replaced by $x_N\equiv 2\, x_{Bj}/\left[1+\sqrt{1+\frac{4 x_{Bj}^2 M_{T}^2}{Q^2}}\right]\, ,$  which is the Nachtmann variable.  However, in the regime of interest, $x_{Bj}<10^{-2}$ and $Q> 1$ GeV, the difference between $x_N$ and $x_{Bj}$ seems completely negligible.}
\begin{eqnarray}
k^+_{\min} \equiv \frac{Q_0^2}{2 x_0\, P^-} \simeq  \frac{Q_0^2\, x_{Bj}}{x_0\, Q^2}\, q^+
\label{def_k_plus_min}
\, ,
\end{eqnarray}
where $P^-$ is the momentum of the target, and $x_0$ and $Q_0$ are the typical momentum fraction and transverse mass (or virtuality) of the partons in the target at the initial condition for the high-energy evolution. Hence, $Q_0^2/x_0$ will be a parameter of the fit to the data, together with the shape of the initial condition. It should depend on the type of target, but not on the observable. One can take for example $x_0=10^{-2}$ and expect $Q_0$ to be a non-perturbative scale, presumably related to the initial saturation scale of the target.


In the $q\rightarrow g$ contributions \eqref{sigma_L_q_to_g_frac} and  \eqref{sigma_T_q_to_g_frac} to the longitudinal and transverse photon cross sections, the integration over the gluon momentum fraction $\zeta$ leads to a large logarithm $\log(z q^+/k^+_{\min})$. Such large logarithms can be subtracted from the NLO correction, and combined (up to higher order corrections) with the dipole amplitude $\langle{\cal S}_{01}\rangle_{0}$ appearing inside the LO contribution to the cross section, effectively replacing the unevolved $\langle{\cal S}_{01}\rangle_{0}$ by $\langle{\cal S}_{01}\rangle_{Y_f^+}$, evolved from the cut-off $k^+_{\min}$ up to a factorization scale $k_f^+\lesssim z q^+$ (where $Y_f^+=\log(k_f^+/k^+_{\min})$) with a high-energy LL evolution equation like BK, JIMWLK or BFKL, depending on the other approximations made.

Moreover, yet higher order corrections to the cross section will contain contributions with one such large logarithm per power of $\alpha_s$, at most.
At each order, the large logs can be subtracted and resummed into the evolution of multipole amplitudes appearing in lower order terms.
Hence, the unevolved amplitudes $\langle{\cal S}_{01}\rangle_{0}$
and $\langle{\cal S}^{(3)}_{012}\rangle_{0}$
appearing in the NLO contribution should be replaced by evolved ones when performing the resummation of LL pieces contained in the terms of order NNLO and beyond in the cross section.
However, it is not clear a priori what the evolution range should be in that case, since the higher order corrections beyond NLO are not yet known.

Nevertheless, there are several arguments in favor of an evolution up to the scale provided by the extra gluon in the NLO contribution. First, as found in Ref.~\cite{Beuf:2014uia}, this is necessary in order to have a smooth interplay between the high-energy regime and the collinear regime, so that the Regge and the Bjorken limits can commute.
Second, this is the choice of evolution range which allows to extend to arbitrary order the picture of LL resummation in which the LL contributions can be subtracted and resummed step by step, from the highest to lowest order, in a consistent way. A similar observation has also been made in Ref.~\cite{Iancu:2016vyg}.

Then, there are two main approaches to the high-energy LL resummation for the photon-target cross sections at NLO.
\begin{itemize}
  \item The first approach, advertised in Ref.~\cite{Iancu:2016vyg}, is simply to promote the dipole and tripole amplitude appearing in the $q\rightarrow g$ and $\bar{q}\rightarrow g$ contributions to evolved ones, over the range $Y_2^+=\log(k_2^+/k^+_{\min})$, leaving the dipole amplitude in the LO contribution unevolved. In the following, this will be called the unsubtracted form for the LL resummation.

  \item The second (and more traditional) approach to high-energy LL resummation is to go all way, subtracting as well the LL arising from the integration over $\zeta$ in the $q\rightarrow g$ and $\bar{q}\rightarrow g$ contributions, and evolving the dipole amplitude in the LO contribution accordingly. In the following, this will be called the subtracted form for the LL resummation.
\end{itemize}


\subsection{Unsubtracted form for the high-energy LL resummation\label{sec:unsub_LL}}

All in all, the photon-target cross sections at NLO with high-energy LL resummation in the unsubtracted form can be written as
\begin{eqnarray}
\sigma^{\gamma^*}_{T,L} \Big|_{\textrm{NLO+LL; unsub.}} &=&
\sigma^{\gamma^*}_{T,L} \Big|_{\textrm{dipole; I.C.}}
+2\; \sigma^{\gamma^*}_{T,L} \Big|_{q\rightarrow g;\, \textrm{LL}}
\label{sigma_TL_NLO_LL_unsub}
\, ,
\end{eqnarray}
with
\begin{eqnarray}
\sigma^{\gamma^*}_{L} \Big|_{\textrm{dipole; I.C.}}
&=& 4 N_c \alpha_{em} \sum_{f} e_f^2\; \theta(1\!-\!2 z_c)
\int_{z_c}^{1-z_c}\!\! dz\; 4 z^2(1\!-\!z)^2
\int \frac{d^2 \x_0}{2\pi} \int \frac{d^2 \x_1}{2\pi}\;
{\textrm{Re}}\Big[1-\langle{\cal S}_{01}\rangle_{0}\Big]\;
\nonumber\\
& &\hspace{1cm}
\times\;
Q^2 \bigg( \textrm{K}_{0}\Big(Q \sqrt{z(1\!-\!z)}|\x_{01}|\Big)\bigg)^2\;
\Bigg\{1+\left(\frac{\alpha_s\, C_F}{\pi}\right)\;
\bigg[\frac{1}{2} \left(\log\left(\frac{z}{1\!-\!z}\right)\right)^2
-\frac{\pi^2}{6}
 +\frac{5}{2}
\bigg]\Bigg\}\;\;
\label{sigma_L_dipole_LL_unsub}
\end{eqnarray}
and
\begin{eqnarray}
\sigma^{\gamma^*}_{L} \Big|_{q\rightarrow g;\, \textrm{LL}}
&=& 4 N_c \alpha_{em} \left(\frac{\alpha_s\, C_F}{\pi}\right)\sum_{f} e_f^2\;
\theta\!\left(1\!-\! z_{\min}\!-\! z_c\right)
\int_{z_{\min}}^{1- z_c}\! dz\; 4 z^2(1\!-\!z)^2
\int_{\frac{z_{\min}}{z}}^{1}
d \zeta\;
\int \frac{d^2 \x_0}{2\pi} \int \frac{d^2 \x_1}{2\pi}\int \frac{d^2 \x_2}{2\pi}\;
\nonumber\\
&&\hspace{0.5cm}
\times \Bigg\{
\left[\frac{1+(1\!-\!\zeta)^2}{\zeta}\right]\;
\frac{\x_{20}}{\x_{20}^2} \!\cdot\! \left(\frac{\x_{20}}{\x_{20}^2}\!-\!\frac{\x_{21}}{\x_{21}^2}\right)\,
\bigg[Q^2 \Big(\textrm{K}_0\!\left(Q X_{012}\right)\Big)^2\;
{\textrm{Re}}\left[1-\langle{\cal S}^{(3)}_{012}\rangle_{Y_2^+}\right]-\Big(\x_{2}\rightarrow \x_{0}\Big)\bigg]
\nonumber\\
&&  \hspace{3cm}
+\zeta\, \frac{(\x_{20}\!\cdot\!\x_{21})}{\x_{20}^2 \x_{21}^2}
\,
Q^2 \Big(\textrm{K}_0\!\left(Q X_{012}\right)\Big)^2\;
{\textrm{Re}}\left[1-\langle{\cal S}^{(3)}_{012}\rangle_{Y_2^+}\right]
\Bigg\}
\label{sigma_L_q_to_g_LL_unsub}
\end{eqnarray}
in the longitudinal photon case,
and
\begin{eqnarray}
\sigma^{\gamma^*}_{T} \Big|_{\textrm{dipole; I.C.}}
&=& 4 N_c \alpha_{em} \sum_{f} e_f^2\;
\theta(1\!-\!2 z_c)
\int_{z_c}^{1-z_c}\!\! dz\;  z(1\!-\!z) \big[z^2+(1\!-\!z)^2\big]
\int \frac{d^2 \x_0}{2\pi} \int \frac{d^2 \x_1}{2\pi}\;
{\textrm{Re}}\Big[1-\langle{\cal S}_{01}\rangle_{0}\Big]\;
\nonumber\\
& &\hspace{1cm}
\times\;
Q^2 \bigg( \textrm{K}_{1}\Big(Q \sqrt{z(1\!-\!z)}|\x_{01}|\Big)\bigg)^2\;
\Bigg\{1+\left(\frac{\alpha_s\, C_F}{\pi}\right)\;
\bigg[\frac{1}{2} \left(\log\left(\frac{z}{1\!-\!z}\right)\right)^2
-\frac{\pi^2}{6}
 +\frac{5}{2}
\bigg]\Bigg\}\;\;
\label{sigma_T_dipole_LL_unsub}
\end{eqnarray}
and
\begin{eqnarray}
&&\sigma^{\gamma^*}_{T} \Big|_{q\rightarrow g;\, \textrm{LL}}
= 4 N_c \alpha_{em} \left(\frac{\alpha_s\, C_F}{\pi}\right)\sum_{f} e_f^2\;
\theta\!\left(1\!-\! z_{\min}\!-\! z_c\right)
\int_{z_{\min}}^{1- z_c}\! dz\; z(1\!-\!z)
\int_{\frac{z_{\min}}{z}}^{1}
d \zeta\;
\int \frac{d^2 \x_0}{2\pi} \int \frac{d^2 \x_1}{2\pi}\int \frac{d^2 \x_2}{2\pi}\;
\nonumber\\
&&\hspace{0.5cm}
\times \Bigg\{
\Big[z^2+(1\!-\!z)^2\Big]\,
\left[\frac{1+(1\!-\!\zeta)^2}{\zeta}\right]\;
\frac{\x_{20}}{\x_{20}^2} \!\cdot\! \left(\frac{\x_{20}}{\x_{20}^2}\!-\!\frac{\x_{21}}{\x_{21}^2}\right)\,
\bigg[Q^2 \Big(\textrm{K}_1\!\left(Q X_{012}\right)\Big)^2\;
{\textrm{Re}}\left[1-\langle{\cal S}^{(3)}_{012}\rangle_{Y_2^+}\right]-\Big(\x_{2}\rightarrow \x_{0}\Big)\bigg]
\nonumber\\
&&  \hspace{1cm}
+\zeta\,
\left[
\Big[z^2+(1\!-\!z)^2\Big]\,
\frac{(\x_{20}\!\cdot\!\x_{21})}{\x_{20}^2 \x_{21}^2}
+2z(1\!-\!z)(1\!-\!\zeta)\;
 \frac{\left(\x_{20}\!\cdot\! \x_{21}\right)}{\x_{20}^2 X_{012}^2}
  -\frac{z(1\!-\!\zeta)(1\!-\!z\!+\!\zeta z)}{ X_{012}^2}\right]\,
\nonumber\\
&& \hspace{3cm}
\times\;
Q^2 \Big(\textrm{K}_1\!\left(Q X_{012}\right)\Big)^2\;
{\textrm{Re}}\left[1-\langle{\cal S}^{(3)}_{012}\rangle_{Y_2^+}\right]
\Bigg\}
\label{sigma_T_q_to_g_LL_unsub}
\end{eqnarray}
in the transverse photon case. In Eqs.~\eqref{sigma_L_q_to_g_LL_unsub} and \eqref{sigma_T_q_to_g_LL_unsub},
$X_{012}$ is as given by Eq.~\eqref{def_X012_q2g} and $Y_2^+$ is defined by
\begin{eqnarray}
Y_2^+ &\equiv & \log\left(\frac{k_2^+}{k^+_{\min}}\right) = \log\left(\zeta\, z\,
\frac{Q^2\, x_0}{Q_0^2\, x_{Bj}}\right)
\label{Y_2_plus}
\, .
\end{eqnarray}
Moreover,
\begin{eqnarray}
z_{\min} &\equiv & \frac{k^+_{\min}}{q^+} =\frac{Q_0^2\, x_{Bj}}{Q^2\, x_0}
\label{z_min}
\, ,
\end{eqnarray}
and $z_c=0$ if the minimal cut-off prescription is chosen for the $k^+$ regularization, or $z_c=z_{\min}$ if the non-minimal cut-off prescription of sec.~\ref{sec:non_mini_cut} is chosen.

In order to obtain the evolution of the tripole and dipole amplitudes from $0$ to $Y_2^+$, one can use either the B-JIMWLK evolution, or the BK equation, after rewriting the tripole in terms of dipoles in the usual way, or the BFKL equation, after taking the two gluons exchange approximation. Moreover, it should be consistent to use these evolution equations either at naive LL accuracy or with the kinematical improvement presented in Refs.~\cite{Beuf:2014uia} or \cite{Iancu:2015vea}. Indeed, the calculation of the photon-target cross sections at NLO have been performed using a cut-off in $k^+$, so that it should be used in conjunction with
high-energy evolution equations where $k^+$ (or its $\log$) plays the role of evolution variable, in order to have a consistent factorization scheme for high-energy logs. The various versions of kinematically consistent BK equation provided in Refs.~\cite{Beuf:2014uia}, \cite{Iancu:2015vea} are only available as evolution equations in $k^+$, and would have a different expression in other factorization schemes. By contrast, at naive LL accuracy, the kernel of the high-energy evolution equations is factorization-scheme independent.

As a final remark, note that it is not clear a priori if the dipole amplitude present in the NLO part of the dipole-like contributions \eqref{sigma_L_dipole_LL_unsub} and \eqref{sigma_T_dipole_LL_unsub} should be taken at the initial condition or evolved over some range. All the available results and arguments constrain only the scale of the dipole amplitude in the LO contribution, and the one of the dipole and tripole coming with a $d\zeta/\zeta$ enhancement. By contrast, a calculation of the cross sections at NNLO seems necessary in order to obtain a clear guidance in the choice of scale for the dipole amplitude in the NLO part of Eqs.~\eqref{sigma_L_dipole_LL_unsub} and \eqref{sigma_T_dipole_LL_unsub}. For simplicity, the dipole has been chosen to stay unevolved there, but other prescriptions can be used.


\subsection{Subtracted form for the high-energy LL resummation\label{sec:sub_LL}}

In the subtracted form for the high-energy LL resummation, the photon-target cross sections at NLO become
\begin{eqnarray}
\sigma^{\gamma^*}_{T,L} \Big|_{\textrm{NLO+LL; sub.}} &=&
\sigma^{\gamma^*}_{T,L} \Big|_{\textrm{dipole; }Y_f^+}
+2\; \sigma^{\gamma^*}_{T,L} \Big|_{q\rightarrow g;\, \textrm{LL}} + \sigma^{\gamma^*}_{T,L} \Big|_{\textrm{LL sub. term}}
\label{sigma_TL_NLO_LL_sub}
\, ,
\end{eqnarray}
with
\begin{eqnarray}
\sigma^{\gamma^*}_{L} \Big|_{\textrm{dipole; }Y_f^+}
&=& 4 N_c \alpha_{em} \sum_{f} e_f^2\; \theta(1\!-\!2 z_c)
\int_{z_c}^{1-z_c}\!\! dz\; 4 z^2(1\!-\!z)^2
\int \frac{d^2 \x_0}{2\pi} \int \frac{d^2 \x_1}{2\pi}\;
{\textrm{Re}}\Big[1-\langle{\cal S}_{01}\rangle_{Y_f^+}\Big]\;
\nonumber\\
& &\hspace{1cm}
\times\;
Q^2 \bigg( \textrm{K}_{0}\Big(Q \sqrt{z(1\!-\!z)}|\x_{01}|\Big)\bigg)^2\;
\Bigg\{1+\left(\frac{\alpha_s\, C_F}{\pi}\right)\;
\bigg[\frac{1}{2} \left(\log\left(\frac{z}{1\!-\!z}\right)\right)^2
-\frac{\pi^2}{6}
 +\frac{5}{2}
\bigg]\Bigg\}
\label{sigma_L_dipole_LL_sub}
\end{eqnarray}
and
\begin{eqnarray}
\sigma^{\gamma^*}_{T} \Big|_{\textrm{dipole; }Y_f^+}
&=& 4 N_c \alpha_{em} \sum_{f} e_f^2\;
\theta(1\!-\!2 z_c)
\int_{z_c}^{1-z_c}\!\! dz\;  z(1\!-\!z) \big[z^2+(1\!-\!z)^2\big]
\int \frac{d^2 \x_0}{2\pi} \int \frac{d^2 \x_1}{2\pi}\;
{\textrm{Re}}\Big[1-\langle{\cal S}_{01}\rangle_{Y_f^+}\Big]\;
\nonumber\\
& &\hspace{1cm}
\times\;
Q^2 \bigg( \textrm{K}_{1}\Big(Q \sqrt{z(1\!-\!z)}|\x_{01}|\Big)\bigg)^2\;
\Bigg\{1+\left(\frac{\alpha_s\, C_F}{\pi}\right)\;
\bigg[\frac{1}{2} \left(\log\left(\frac{z}{1\!-\!z}\right)\right)^2
-\frac{\pi^2}{6}
 +\frac{5}{2}
\bigg]\Bigg\}
\label{sigma_T_dipole_LL_sub}
\end{eqnarray}
with $Y_f^+\equiv \log(k_f^+/k^+_{\min})$, whereas the expressions \eqref{sigma_L_q_to_g_LL_unsub} and \eqref{sigma_T_q_to_g_LL_unsub} are kept. A LL subtraction term is included in Eq.~\eqref{sigma_TL_NLO_LL_sub}, in order to account for the mismatch between the evolved dipole amplitude in Eqs.~\eqref{sigma_L_dipole_LL_sub} and \eqref{sigma_T_dipole_LL_sub} on the one hand, and the unevolved dipole amplitude in Eqs.~\eqref{sigma_L_dipole_LL_unsub} and \eqref{sigma_T_dipole_LL_unsub} on the other hand. It writes
\begin{eqnarray}
\sigma^{\gamma^*}_{L} \Big|_{\textrm{LL sub. term}}
&=& 4 N_c \alpha_{em} \sum_{f} e_f^2\; \theta(1\!-\!2 z_c)
\int_{z_c}^{1-z_c}\!\! dz\; 4 z^2(1\!-\!z)^2
\int \frac{d^2 \x_0}{2\pi} \int \frac{d^2 \x_1}{2\pi}\;
\nonumber\\
& &\hspace{1.5cm}
\times\;
Q^2 \bigg( \textrm{K}_{0}\Big(Q \sqrt{z(1\!-\!z)}|\x_{01}|\Big)\bigg)^2\;\;
{\textrm{Re}}\int_{0}^{Y_f^+} \!\!\!\!dY^+\; \partial_{Y^+} \langle{\cal S}_{01}\rangle_{Y^+}
\label{sigma_L_LL_sub_term}
\end{eqnarray}
in the longitudinal photon case, and
\begin{eqnarray}
\sigma^{\gamma^*}_{T} \Big|_{\textrm{LL sub. term}}
&=& 4 N_c \alpha_{em} \sum_{f} e_f^2\;
\theta(1\!-\!2 z_c)
\int_{z_c}^{1-z_c}\!\! dz\;  z(1\!-\!z) \big[z^2+(1\!-\!z)^2\big]
\int \frac{d^2 \x_0}{2\pi} \int \frac{d^2 \x_1}{2\pi}\;
\nonumber\\
& &\hspace{1.5cm}
\times\;
Q^2 \bigg( \textrm{K}_{1}\Big(Q \sqrt{z(1\!-\!z)}|\x_{01}|\Big)\bigg)^2\;\;
 {\textrm{Re}}\int_{0}^{Y_f^+} \!\!\!\!dY^+\; \partial_{Y^+} \langle{\cal S}_{01}\rangle_{Y^+}
\label{sigma_T_LL_sub_term}
\end{eqnarray}
in the transverse photon case.
Combining these subtraction terms with the corresponding $q\rightarrow g$ terms \eqref{sigma_L_q_to_g_LL_unsub} and \eqref{sigma_T_q_to_g_LL_unsub}, one should get a cancellation of the $d\zeta/\zeta$ terms\footnote{Note that it is straightforward to perform this combination of subtraction terms and $q\rightarrow g$ terms if the non-minimal cut-off prescription in $k^+$ is chosen, since the $z$-integral has the same bounds in both cases, for $z_{\min}=z_c$. By contrast, in the minimal cut-off prescription, the combination would require to neglect the difference in the integration bounds, which is indeed a power-suppressed effect for $x_{Bj}\rightarrow 0$ or for $Q^2\rightarrow +\infty$. This observation provides a motivation to favor the non-minimal over the minimal cut-off prescription in $k^+$.}, meaning that the high-energy LL contribution have been indeed resummed into the $\langle{\cal S}_{01}\rangle_{Y_f^+}$ appearing in the LO contribution in Eqs.~\eqref{sigma_L_dipole_LL_sub} and \eqref{sigma_T_dipole_LL_sub}.
Obviously, in order to have consistent results, the same high-energy evolution equation should be used in order to evaluate the various terms in Eq.~\eqref{sigma_TL_NLO_LL_sub}.

Moreover, in order to avoid an oversubtraction of LL, the factorization scale $k_f^+$ should obey $k_f^+ \lesssim z q^+$ and $k_f^+ \lesssim (1\!-\!z) q^+$. In the case of the minimal cut-off prescription (with $z_c=0$), it seems complicated to construct a factorization scale $k_f^+$ obeying these inequalities and simultaneously guarantying the necessary positivity of $Y_f^+$ over the whole phase space. On the other hand, in the case of the non-minimal cut-off prescription (with $z_c=z_{\min}$), the choice $k_f^+ \equiv \min\{z, (1\!-\!z)\}\, q^+$ seems to satisfy all of the requirements.

For simplicity, the dipole amplitude in the NLO part of the dipole-like contributions \eqref{sigma_L_dipole_LL_sub} and \eqref{sigma_T_dipole_LL_sub} is taken to be evolved over the same range as the dipole amplitude in the LO term, instead of staying unevolved like in the unsubtracted form \eqref{sigma_TL_NLO_LL_unsub} for the LL resummation. This change leads to the only quantitative difference between the unsubtracted and subtracted forms of the resummation (as presented in the present section), which are otherwise identical up to a reorganization of the terms. This mismatch between the two forms is a sum of terms of order $\alpha_s (\alpha_s\, Y_f^+)^n$ relative to the LO term, with $n\geq 1$. Hence, even though the two forms of the LL resummation should lead to different quantitative results, this difference is relevant only at NNLO and/or at NLL accuracy. Moreover, it is definitely possible to construct variants of the unsubtracted and subtracted forms for the LL resummation which would agree identically with each other.


\subsubsection{Naive LL resummation\label{sec:naiveLL_resum}}

The standard B-JIMWLK evolution of the dipole amplitude at LL accuracy is given by
\begin{eqnarray}
 \partial_{Y^+} \langle{\cal S}_{01}\rangle_{Y^+}
&=& \frac{2 \alpha_s C_F}{\pi}\, \int \frac{d^2 \x_2}{2\pi}\;
\frac{\x_{10}^2 }{\x_{20}^2 \x_{21}^2}
\bigg[\langle{\cal S}^{(3)}_{012}\rangle_{Y^+}
- \langle{\cal S}_{01}\rangle_{Y^+}
\bigg]
\label{BK_dipole}
\, .
\end{eqnarray}
Hence, if it is chosen in order to make the high-energy LL resummation for the photon-target cross section, one should substitute
\begin{eqnarray}
\int_{0}^{Y_f^+} \!\!\!\!dY^+\;  \partial_{Y^+} \langle{\cal S}_{01}\rangle_{Y^+}
&=& \frac{2 \alpha_s C_F}{\pi}\, \int_{0}^{Y_f^+} \!\!\!\!dY^+\;  \int \frac{d^2 \x_2}{2\pi}\;
\frac{\x_{10}^2 }{\x_{20}^2 \x_{21}^2}
\bigg[\langle{\cal S}^{(3)}_{012}\rangle_{Y^+}
- \langle{\cal S}_{01}\rangle_{Y^+}
\bigg]
\nonumber\\
&=& \frac{2 \alpha_s C_F}{\pi}\,
\int_{\frac{z_{\min}}{z}}^{\frac{k_f^+}{z q^+}}
\frac{d \zeta}{\zeta}\;  \int \frac{d^2 \x_2}{2\pi}\;
\frac{\x_{10}^2 }{\x_{20}^2 \x_{21}^2}
\bigg[\langle{\cal S}^{(3)}_{012}\rangle_{Y_2^+}
- \langle{\cal S}_{01}\rangle_{Y_2^+}
\bigg]
\label{BK_dipole_int}
\end{eqnarray}
in Eqs.~\eqref{sigma_L_LL_sub_term} and \eqref{sigma_T_LL_sub_term}, with $Y_2^+$ and $z_{\min}$ defined by Eqs.~\eqref{Y_2_plus} and \eqref{z_min} respectively.

Subtraction terms \eqref{sigma_L_LL_sub_term} and \eqref{sigma_T_LL_sub_term} constructed from the strict LL evolution \eqref{BK_dipole} are enough to cancel the logarithmic sensitivity in the cut-off $k^+_{\min}$ from the $q\rightarrow g$ terms \eqref{sigma_L_q_to_g_LL_unsub} and \eqref{sigma_T_q_to_g_LL_unsub} (and the corresponding $\bar{q}\rightarrow g$ terms). However, as shown in Ref.~\cite{Beuf:2014uia}, they do not provide a fully accurate subtraction of high-energy LLs. Indeed, such subtraction terms overestimate the phase-space giving rise to LLs in the $q\rightarrow g$ contributions, in particular because they neglect the non-trivial interplay of longitudinal momentum fractions and transverse positions arising through the variable $X_{012}$. Then, one is left after such naive LL subtraction with large transverse logarithms due to the inaccurate treatment of the region $\x_{10}^2 \lesssim \zeta\, \x_{20}^2 \simeq \zeta\, \x_{21}^2$ for small $\zeta$.


\subsubsection{Kinematically consistent LL resummation\label{sec:kcLL_resum}}

In order to perform a fully accurate subtraction of high-energy LLs from the NLO correction to an observable, one should use a kinematically improved (a.k.a. kinematically consistent) LL evolution equation \cite{Ciafaloni:1987ur,Andersson:1995ju,Kwiecinski:1996td,Salam:1998tj,Motyka:2009gi,Beuf:2014uia,Iancu:2015vea}. Such equations effectively resum partons cascades which are strictly ordered both in $k^+$ and in $k^-$ simultaneously, by contrast to naive LL evolution equations which correspond to an ordering along the evolution variable ($k^+$ in our case) only. The strict double ordering is necessary and sufficient to reproduce the non-trivial interplay between longitudinal and transverse variables encountered in higher order corrections to high-energy observables.

At the non-linear level, the easiest is to use the kinematically consistent LL evolution equation
\begin{eqnarray}
 \partial_{Y^+} \langle{\cal S}_{01}\rangle_{Y^+}
&=& \frac{2 \alpha_s C_F}{\pi}\, \int \frac{d^2 \x_2}{2\pi}\;
\frac{\x_{10}^2 }{\x_{20}^2 \x_{21}^2}\, \theta\left(Y^+\!-\! \Delta_{012}\right)
\bigg[\langle{\cal S}^{(3)}_{012}\rangle_{Y^+\!-\! \Delta_{012}}
- \langle{\cal S}_{01}\rangle_{Y^+}
\bigg]
\label{kcBK_dipole}
\end{eqnarray}
constructed in Ref.~\cite{Beuf:2014uia}. In Eq.~\eqref{kcBK_dipole}, the shift $\Delta_{012}$ allows to mix the transverse and longitudinal variables in the appropriate way in order to emulate the $k^-$ ordering of the parton cascade.
Various definitions for $\Delta_{012}$ can be chosen, for example
\begin{eqnarray}
\Delta_{012} &=& \max\left\{0, \log\left(\frac{\min \{\x_{20}^2,\x_{21}^2\}}{\x_{10}^2}\right) \right\}
\label{def_Delta012_1}
\end{eqnarray}
or
\begin{eqnarray}
\Delta_{012} &=&  \max\left\{0, \log\left(\frac{|\x_{20}\cdot \x_{21}|}{\x_{10}^2}\right) \right\}
\label{def_Delta012_2}
\, .
\end{eqnarray}

Then, one should substitute
\begin{eqnarray}
\int_{0}^{Y_f^+} \!\!\!\!dY^+\;  \partial_{Y^+} \langle{\cal S}_{01}\rangle_{Y^+}
&=& \frac{2 \alpha_s C_F}{\pi}\, \int_{0}^{Y_f^+} \!\!\!\!dY^+\;  \int \frac{d^2 \x_2}{2\pi}\;
\frac{\x_{10}^2 }{\x_{20}^2 \x_{21}^2}\, \theta\left(Y^+\!-\! \Delta_{012}\right)
\bigg[\langle{\cal S}^{(3)}_{012}\rangle_{Y^+\!-\! \Delta_{012}}
- \langle{\cal S}_{01}\rangle_{Y^+}
\bigg]
\label{kcBK_dipole_int}
\end{eqnarray}
in Eqs.~\eqref{sigma_L_LL_sub_term} and \eqref{sigma_T_LL_sub_term}. As shown in Ref.~\cite{Beuf:2014uia}, based on the earlier partial results for DIS at NLO from Ref.~\cite{Beuf:2011xd}, this allows to accurately subtract the high-energy LLs from the NLO contribution over the whole phase-space.

In Ref.~\cite{Iancu:2015vea}, another non-linear kinematically improved LL evolution equation has been constructed, as an alternative to Eq.~\eqref{kcBK_dipole}. Both equations are in principle equivalent. However, the equation from Ref.~\cite{Iancu:2015vea} is local in $Y^+$, whereas Eq.~\eqref{kcBK_dipole}, features memory effects in $Y^+$-space.\footnote{Note that the memory effects are not really a problem in practice, since numerical simulations of Eq.~\eqref{kcBK_dipole} have been performed successfully, see Ref.~\cite{Albacete:2015xza}.} The price to pay in order to obtain a local evolution equation is that the solution of the equation from Ref.~\cite{Iancu:2015vea} corresponds to the physical dipole amplitude only in part of the phase-space, not all of it, and is obtained starting from a modified initial condition, not from the physical one. Because of this, it is not consistent to plug directly the kinematically improved equation of Ref.~\cite{Iancu:2015vea} in the formulas \eqref{sigma_L_LL_sub_term} and \eqref{sigma_T_LL_sub_term}. Most likely, it should be possible to use that kinematically improved evolution equation if an extra term is added to Eq.~\eqref{sigma_TL_NLO_LL_sub}, in order to account for the difference between the physical initial condition for the dipole amplitude, appearing in the LO contribution before LL resummation, and the modified initial condition required in order to write the LL subtraction terms as Eqs.~\eqref{sigma_L_LL_sub_term} and \eqref{sigma_T_LL_sub_term}. A complete study of this issue is however beyond the scope of the present paper.


\subsection{Running coupling effects\label{sec:RC}}

In principle, the methods given so far in this section provide the complete high-energy LL resummation for the NLO photon-target cross sections. However, it is well known both from numerical simulations \cite{GolecBiernat:2001if,Rummukainen:2003ns,Albacete:2004gw} and from theoretical arguments \cite{Gribov:1984tu,Mueller:2002zm,Munier:2003sj} that the running coupling effects, which appear formally beyond LL accuracy, have a dramatic impact on the solutions of non-linear LL evolution equations, such as the BK or JIMWLK equations. Hence, running coupling versions of these equations have to be used in the LL resummation procedure. Then, by consistency, the $\alpha_s$ appearing in the NLO corrections to the cross sections should also be taken as running. Since the NNLO corrections to the cross sections are unknown, it is not known for sure at what scale this running $\alpha_s$ should be taken. Still, in order to maintain the self-consistency of the LL resummation, the prescription for the running $\alpha_s$ in the $q\rightarrow g$ (and $\bar{q}\rightarrow g$) contributions \eqref{sigma_L_q_to_g_LL_unsub} and \eqref{sigma_T_q_to_g_LL_unsub} to the cross sections has to be compatible with the prescription  for the running $\alpha_s$ in the LL evolution equations.

The most widely preferred running coupling prescription for the BK equation in the literature is Balitsky's prescription, from Ref.~\cite{Balitsky:2006wa}. In that case, the running coupling related piece from the NLL BK kernel is identified following the BLM prescription \cite{Brodsky:1982gc}, and then resummed according to renormalon based arguments, in order to obtain a running coupling prescription.
Unfortunately, Balitsky's prescription leads to a quite complicated expression of the coupling, entangled with the BK kernel.\footnote{Moreover, the BK kernel with Balitsky's running coupling prescription has some puzzling and unwanted features: it is not always positive, thus ruining its probabilistic interpretation as dipole splitting probability density, and it has a sizable sensitivity on the infrared regularization procedure necessary to deal with the Landau pole.} For that reason, it is both tricky and ambiguous to construct a running coupling prescription for NLO cross sections, which would be compatible with Balitsky's prescription for the BK equation. This issue has already been encountered in Ref.~\cite{Ducloue:2017mpb}, in the case of the NLO corrections to the single inclusive forward hadron production in pA collisions.

Instead, it is possible to use a simpler running coupling prescription, such as the parent dipole prescription or the smallest dipole prescription, which can be implemented straightforwardly both in the BK equation and in the NLO corrections to the cross section. However, such prescriptions are less accurate, since they are not based on higher order calculations.

Yet another running coupling prescription can be constructed, which combine the positive aspects of the previous ones, while eliminating their negative aspects, by following the BLM method \cite{Brodsky:1982gc} all the way. The first step of the BLM method consists in identifying the running coupling related terms in the NLO correction to the evolution equation by focusing on the quark loop contributions, as done in Ref.~\cite{Balitsky:2006wa}. The second step is simply to find the value for the UV renormalization scale which cancels these running coupling related terms, and take the running coupling $\alpha_s$ at that scale. In that way, one finds the prescription $\alpha_s \mapsto \alpha_s\left(Q_{012}\right)$ for the BK equation, with the momentum scale $Q_{012}$ defined as
\begin{eqnarray}
Q_{012}^2 \equiv \frac{4\, C^2}{\x_{10}^2}\;\; \left(\frac{\x_{20}^2}{\x_{21}^2}\right)^{\left(\frac{\x_{20}^2-\x_{21}^2}{\x_{10}^2}\right)}
\label{def_BLM_RC_scale}
\end{eqnarray}
where $C^2$ is a constant. Its best motivated values are $C^2=e^{-2 \gamma_E -5/3}$ or
$C^2=e^{-2 \gamma_E}$. In each UV or collinear limit, the scale $Q_{012}$ is determined by the smallest dipole size only. This running coupling prescription is identical, at the perturbative level, to the one proposed in Ref.~\cite{Iancu:2015joa} (see Eq.~(12) there), called fac (for \emph{fastest apparent convergence}) there. However, writing the running coupling prescription as a single coupling $\alpha_s$ taken at a single scale $Q_{012}$ related to the smallest dipole size allows to limit the sensitivity on the details of the IR regularization procedure for the Landau pole, for example a freezing of the coupling. Hence, the variant of the BLM prescription using the single scale $Q_{012}$ should be less IR sensitive than both the variant of the BLM prescription presented in Ref.~\cite{Iancu:2015joa} and Balitsky's prescription \cite{Balitsky:2006wa}, even though these three prescriptions contain the same NLO information and are thus as valid formally.

It is straightforward to use the BLM running coupling prescription based on Eq.~\eqref{def_BLM_RC_scale} both in the BK equation and in the $q\rightarrow g$ contributions \eqref{sigma_L_q_to_g_LL_unsub} and \eqref{sigma_T_q_to_g_LL_unsub} to the photon-target cross sections. There is also an $\alpha_s$ present in the dipole-like contribution to the cross-sections. However, the integration over $\x_2$ has already been performed there. The most natural is then to use the parent dipole prescription in that term.


\subsection{NLL resummation\label{sec:NLL}}

Up to now, the focus in this section has been on the high-energy LL resummation of the NLO photon-target
cross sections, in order to obtain results at NLO+LL accuracy. Since the high-energy evolution equations are now available at NLL accuracy \cite{Fadin:1998py,Ciafaloni:1998gs,Balitsky:2008zz,Balitsky:2009xg,Balitsky:2013fea,Lublinsky:2016meo}, one should also think about the NLL resummation procedure, in order to obtain results at NLO+NLL accuracy for the photon-target cross sections.

The first attempt for a NLL resummation would be simply to replace the LL evolution equations by their NLL generalizations everywhere, both in the unsubtracted expression \eqref{sigma_TL_NLO_LL_unsub} and in the subtracted expression \eqref{sigma_TL_NLO_LL_sub}. By consistency, the subtraction terms
\eqref{sigma_L_LL_sub_term} and \eqref{sigma_T_LL_sub_term} should include the NLL part of the evolution equations. Such contributions to the subtraction terms correspond both to genuine NLL contributions for the cross sections, and to NNLO corrections which should in principle be discarded. Hence, it seems that there might be a conflict between including all terms of NLL accuracy and dropping terms of NNLO accuracy.

The NLL evolutions equations are derived in ideal setups approximating generic high-energy observables, in principle sharing the same large logarithms. Hence, it is expected that the cross sections should contain such terms of order NNLO and NLL before any high-energy resummation, which will be canceled by the
subtraction terms \eqref{sigma_L_LL_sub_term} and \eqref{sigma_T_LL_sub_term} if the NLL evolution is included. Thus, when performing the NLL resummation in the subtracted form \eqref{sigma_TL_NLO_LL_sub}, the NLL contributions present in the NNLO corrections are canceled by combination with the subtraction term constructed from the NLL evolution equation. It is then safe to neglect the leftover NNLO corrections, in which no large high-energy log is left. Therefore, it should be consistent to use Eq.~\eqref{sigma_TL_NLO_LL_sub} together with a NLL evolution equation, in order to get results at NLO+NLL accuracy.

By contrast, in the unsubtracted form of the high-energy resummation, given in Eq.~\eqref{sigma_TL_NLO_LL_unsub}, the NLL contribution present in the NNLO correction is left there. Hence, when truncating the series to drop the NNLO corrections, one is necessarily also dropping NLL contributions to the cross section, in that case. For that reason, the unsubtracted form \eqref{sigma_TL_NLO_LL_unsub} of the LL resummation cannot be upgraded to a consistent NLL resummation by using a NLL evolution equation instead of a LL one, unless some appropriate extra term\footnote{This issue has been mentioned in the latest version of Ref.~\cite{Iancu:2016vyg}.} is added in Eq.~\eqref{sigma_TL_NLO_LL_unsub}.


\subsection{Summary and discussion about the high-energy resummation\label{sec:discussion}}

In summary, it is possible to perform the high-energy resummation (at LL accuracy or beyond) on top of the fixed-order NLO results for the photon-target cross sections obtained in the sections \ref{sec:sigma_L_NLO} and \ref{sec:sigma_T_NLO}. However, various prescriptions and choices are necessary in order to fully specify the resummation.
\begin{description}
  \item[Factorization scheme] Various factorization schemes are in principle possible for the high-energy resummation, associated in particular to the choice of the evolution variable for the evolution equation. Since the results for the NLO cross sections have been obtained assuming a cut-off in $k^+$ independent of the transverse variables, only a factorization scheme in $k^+$ can be used consistently for the moment.
  \item[Implementation of the $k^+$ cut-off] Both a minimal and a non-minimal (see sec.~\ref{sec:non_mini_cut}) prescriptions have been presented for the cut-off in $k^+$. They should give results differing by power-suppressed terms at high energy.
  \item[Subtracted or unsubtracted form] The LL resummation has been presented both in an unsubtracted form \eqref{sigma_TL_NLO_LL_unsub}, following Ref.~\cite{Iancu:2016vyg}, and in a subtracted form \eqref{sigma_TL_NLO_LL_sub}. At NLO+LL accuracy, both forms should give equivalent results.
  \item[Evolution range for the dipole in the NLO dipole-like term] There is no real theoretical guidance for the range over which that dipole should be evolved. For simplicity, it has been taken evolved over the same range as the dipole in the LO term: $0$ in the unsubtracted case and $Y_f^+$ in the subtracted case. But other choices are definitely possible.
  \item[Evolution equation] Of course, the evolution equation used for the resummation needs to be specified: JIMWLK, BK or BFKL? With what running coupling prescription? At LL or NLL accuracy? With or without kinematical improvement?
\end{description}

Since essentially all of these choices are allowed by the theory, it would be useful to explore numerically the impact of each prescription, both in order to check if they are all consistent, and in order to get a handle on the associated theoretical uncertainty.

Still, one can make the following remarks, before numerical studies:
\begin{itemize}
  \item The non-minimal prescription of sec.~\ref{sec:non_mini_cut} for the $k^+$ cut-off seems slightly preferred, since it does not require to drop any further power-suppressed term when performing the high-energy resummation in the subtracted form.
  \item In the $k^+$ factorization scheme used thoroughly, the results for the photon-target cross sections obtained with a naive LL resummation will not be consistent with the DGLAP evolution of the target in the DIS regime, but only with the DGLAP evolution of the photon in the resolved photoproduction limit, as discussed in Ref.~\cite{Beuf:2014uia} for example, whereas the reverse would be true in a factorization scheme effectively using $k^-$ as evolution variable. By construction, using a kinematically consistent LL evolution equation \cite{Beuf:2014uia,Iancu:2015vea} instead of a naive one allows to obtain correct results in both DGLAP limits simultaneously. Hence, in the $k^+$ factorization scheme, a dramatic improvement is expected for the photon-target cross sections
      when using a kinematically consistent LL equation instead of a naive one.
  \item Both the unsubtracted and the subtracted forms \eqref{sigma_TL_NLO_LL_unsub} and \eqref{sigma_TL_NLO_LL_sub} of the high-energy resummation can be used in order to get results at NLO+LL accuracy (with or without kinematical improvement). However, as discussed in sec.~\ref{sec:NLL}, only the subtracted form \eqref{sigma_TL_NLO_LL_sub} can be promoted to NLL accuracy without further modifications.
\end{itemize}


\section{Conclusion\label{sec:conclusion}}

In the present study, the complete calculation of the NLO corrections to the transverse and longitudinal photon-target cross sections at low $x_{Bj}$ has been finished, following Ref.~\cite{Beuf:2016wdz} in which intermediate results have been obtained. These photon-target cross sections, equivalent to the structure functions, determine the neutral current inclusive DIS cross section in the one-photon exchange approximation, see Eq.~\eqref{DIS_xsect_one_photon}. The calculation has been done in the eikonal approximation, and thus provides the leading power in the high-energy (or equivalently low $x_{Bj}$) limit, and the results have been presented in a generalized version of the dipole factorization formula, including full gluon saturation effects.

At LO, only the $q\bar{q}$ Fock sector inside the virtual photon contributes to the photon-target cross sections, hence the dipole factorization formula. By contrast, at NLO there are contributions both from the
$q\bar{q}$ Fock sector at one-loop and from the $q\bar{q}g$ Fock sector at tree level. For the first time, both types of contributions have been explicitly calculated, whereas in the previous studies \cite{Balitsky:2010ze} and \cite{Beuf:2011xd} only the $q\bar{q}g$ contribution was calculated directly, and the $q\bar{q}$ contribution at NLO was instead reconstructed indirectly.

Both the $q\bar{q}$ and the $q\bar{q}g$ contributions to the photon-target cross sections at NLO are UV divergent, and these UV divergences cannot be eliminated via renormalization at this order in perturbation theory. By using CDR as UV regulator all the way in these calculations, it has been possible to show the expected cancellation of the UV divergences between the $q\bar{q}$ and the $q\bar{q}g$ contributions, as well as of the regularization scheme dependent artifacts, such as the extra rational terms induced by the numerator algebra (about $\gamma^{\mu}$s and polarization vectors) performed in arbitrary dimension $D$ instead of $D=4$.

The high-energy LL resummation of the obtained NLO results has been discussed in depth. Several prescriptions for that resummation have been provided. The final results for the photon-target cross section at NLO+LL accuracy are presented in Eq.~\eqref{sigma_TL_NLO_LL_unsub}, using the unsubtracted form of the LL resummation, or in  Eq.~\eqref{sigma_TL_NLO_LL_sub}, using the subtracted form of the LL resummation.
It has also been argued that the results at NLO+LL in the subtracted form can be promoted to NLO+NLL accuracy simply by using NLL evolution equations instead of LL ones.

Hence, the present article opens the way for future numerical and phenomenological studies in DIS at low $x_{Bj}$ with gluon saturation, including new fits to HERA data, first at NLO+LL accuracy (either without or with kinematical improvement of the BK equation \cite{Beuf:2014uia,Iancu:2015vea}), and then at NLO+NLL accuracy.
However, the results presented here include only the case of massless quarks. But massive quarks provide a non-negligible contribution to DIS in the kinematical range of HERA, and quark masses seem important in practice in the earlier fits at LO+LL accuracy \cite{Albacete:2009fh,Albacete:2010sy,Lappi:2013zma}.
Therefore, quark mass effects are expected to be important as well at NLO+(N)LL accuracy. A natural extension of the calculation presented in Ref.~\cite{Beuf:2016wdz} and in the present article is then to calculate the NLO corrections to the contribution of massive quarks to DIS in the dipole factorization.

In general, future calculations of other high-energy observables at NLO will be facilitated by the present calculation. On the one hand, several technical issues relevant for any such NLO calculation have been elucidated, such as the proper implementation of CDR in LFPT,\footnote{Note also Ref.~\cite{Lappi:2016oup}, where the use of the four-dimensional helicity scheme (FDH) as an alternative to CDR for the UV regularization in LFPT is promoted.} or the construction of UV subtraction terms making explicit the cancellation of UV divergences between contributions from different Fock sectors. On the other hand, the obtained expressions for the LFWFs for the $q\bar{q}$ and $q\bar{q}g$ Fock states inside a dressed photon are also important building blocks in the calculation of other DIS observables. For example, a cross check of the expression for the $q\bar{q}g$ contribution to DIS obtained in Ref.~\cite{Beuf:2011xd} and in the present study has been provided by the calculation of trijet production in DIS in Ref.~\cite{Boussarie:2014lxa}, and a cross-check of the one-loop $q\bar{q}$ LFWFs in a photon derived in Ref.~\cite{Beuf:2016wdz} has been provided by the calculation of dijet production in DIS at NLO in Ref.~\cite{Boussarie:2016ogo}.

Finally, it would be advisable to check the consistency of the results presented here for NLO DIS at low $x_{Bj}$ with the results of Ref.~\cite{Balitsky:2010ze}. However, these results are presented in two different forms, with different integrals already performed or left to do. Therefore, there does not seem to be a simple way to compare the two results in full generality. Still, a possibility would be to linearize the present results, in the two gluons approximation, and go back to full momentum space. In this way, it should be possible to obtain a result for the NLO photon impact factor, which could then be compared with the expressions obtained in Ref.~\cite{Balitsky:2012bs} or in Refs.~\cite{Bartels:2000gt,Bartels:2001mv,Bartels:2002uz,Bartels:2004bi}. This would be a rather long and technical calculation, which is left for further studies.


\begin{acknowledgments}
I thank Bertrand Duclou\'e, Henri H\"anninen, Tuomas Lappi, Risto Paatelainen and Yan Zhu for communication on their ongoing projects before publication.	
\end{acknowledgments}


\appendix


\section{Fourier transform integrals\label{sec:FTs}}

When performing the transverse Fourier transform from full momentum space to mixed space for the $q\bar{q}g$ LFWF inside a dressed transverse or longitudinal photon, one encounters the integrals
\begin{eqnarray}
{\cal I}\!\left(\r,\r';\overline{Q}^2,{\cal C}\right)
&\equiv & (\mu^2)^{2-\frac{D}{2}}\,
\int \frac{d^{D-2} \P}{(2\pi)^{D-2}}\; \int \frac{d^{D-2} \K}{(2\pi)^{D-2}}\;
\frac{e^{i\K\cdot \r'}\:
e^{i\P\cdot \r}\,}{\big\{\K^2 + {\cal C} \big[\P^2+\overline{Q}^2\big]\big\}}
\label{def_int_I}
\\
{\cal I}^{lm}\!\left(\r,\r';\overline{Q}^2,{\cal C}\right)
&\equiv & (\mu^2)^{2-\frac{D}{2}}\,
\int \frac{d^{D-2} \P}{(2\pi)^{D-2}}\; \int \frac{d^{D-2} \K}{(2\pi)^{D-2}}\;
\frac{\P^l\: \K^m\;  e^{i\K\cdot \r'}\:
e^{i\P\cdot \r}\,}{ \big[\P^2+\overline{Q}^2\big] \big\{\K^2 + {\cal C} \big[\P^2+\overline{Q}^2\big]\big\}}
\label{def_int_Ilm}
\\
{\cal I}^{m}\!\left(\r,\r';\overline{Q}^2,{\cal C}\right)
&\equiv & (\mu^2)^{2-\frac{D}{2}}\,
\int \frac{d^{D-2} \P}{(2\pi)^{D-2}}\; \int \frac{d^{D-2} \K}{(2\pi)^{D-2}}\;
\frac{\K^m\:   e^{i\K\cdot \r'}\:
e^{i\P\cdot \r}\,}{ \big[\P^2+\overline{Q}^2\big] \big\{\K^2 + {\cal C} \big[\P^2+\overline{Q}^2\big]\big\}}
\label{def_int_Im}
\, .
\end{eqnarray}
The standard way to start their calculation is to introduce a Schwinger parametrisation for each denominator, and then perform the transverse momentum integrals, which are then Gaussian.

In the case of the integral \eqref{def_int_I}, one is left with an integral over a single Schwinger parameter, which can be performed for generic dimension $D$, and gives
\begin{eqnarray}
{\cal I}\!\left(\r,\r';\overline{Q}^2,{\cal C}\right)
&= &  (2\pi)^{2-D}\, \left(\frac{\mu^2}{{\cal C}}\right)^{2-\frac{D}{2}}\,
\left[\frac{\overline{Q}}{\sqrt{\r^2+{\cal C}\, {\r'}^2}}\right]^{D-3}\,
 \textrm{K}_{D-3}\Big(\overline{Q}\, \sqrt{\r^2+{\cal C}\, {\r'}^2} \Big)\;
\label{int_I_result}
\, ,
\end{eqnarray}
where $\textrm{K}_{\beta}(x)$ is the modified Bessel function of the second kind.

By contrast, in the case of the integrals \eqref{def_int_Ilm} and \eqref{def_int_Im}, one is left with integrals over two Schwinger parameters, which read
\begin{eqnarray}
{\cal I}^{lm}\!\left(\r,\r';\overline{Q}^2,{\cal C}\right)
&= & -\frac{1}{4}\, (2\pi)^{2-D}\, \left(\mu^2\right)^{2-\frac{D}{2}}\,
\left({\r'}^2\right)^{1-\frac{D}{2}}\,
 {\r'}^m\, {\r}^l\;
 \int_{0}^{+\infty} \!\!\!\! d \sigma \; \sigma^{-\frac{D}{2}}\;
 e^{-\sigma \overline{Q}^2}\,  e^{-\frac{\r^2}{4\sigma}}
 \int_{\frac{{\r'}^2 {\cal C}}{4\sigma}}^{+\infty}\! d u\; u^{\frac{D}{2}-2}\; e^{-u}
\label{int_Ilm_result_1}
\nonumber\\
{\cal I}^{m}\!\left(\r,\r';\overline{Q}^2,{\cal C}\right)
&= & \frac{i}{2}\, (2\pi)^{2-D}\, \left(\mu^2\, {\r'}^2\right)^{2-\frac{D}{2}}\,
   \left( \frac{{\r'}^m}{{\r'}^2}\right)\;
 \int_{0}^{+\infty} \!\!\!\! d \sigma \; \sigma^{1-\frac{D}{2}}\;
 e^{-\sigma \overline{Q}^2}\,  e^{-\frac{\r^2}{4\sigma}}
 \int_{\frac{{\r'}^2 {\cal C}}{4\sigma}}^{+\infty}\! d u\; u^{\frac{D}{2}-2}\; e^{-u}
\label{int_Im_result_1}
\, .
\end{eqnarray}
For generic dimension $D$, the integral over $u$ would then give a $\sigma$-dependent incomplete Gamma function, preventing us to express the final result in terms of elementary functions and familiar special functions.

In the particular case of $D=4$, the integral over $u$ simply gives an exponential, making the integral over $\sigma$ tractable. Then, one finds
\begin{eqnarray}
{\cal I}^{lm}\!\left(\r,\r';\overline{Q}^2,{\cal C}\right)
&= & -\frac{{\r}^l}{(2\pi)^2}\, \left(\frac{{\r'}^m}{{\r'}^2}\right)\,
 \; \left[\frac{\overline{Q}}{\sqrt{\r^2+{\cal C}\, {\r'}^2}}\right]\,
 \textrm{K}_{1}\Big(\overline{Q}\, \sqrt{\r^2+{\cal C}\, {\r'}^2} \Big)\;
\label{int_Ilm_result_4D}
\\
{\cal I}^{m}\!\left(\r,\r';\overline{Q}^2,{\cal C}\right)
&= & \frac{i}{(2\pi)^2}\, \left(\frac{{\r'}^m}{{\r'}^2}\right)\;
 \textrm{K}_{0}\Big(\overline{Q}\, \sqrt{\r^2+{\cal C}\, {\r'}^2} \Big)\;
\label{int_Im_result_4D}
\, .
\end{eqnarray}

Moreover, in generic dimension $D$, it is straightforward to show that the integrals ${\cal I}^{lm}$ and ${\cal I}^{m}$ behave as
\begin{eqnarray}
{\cal I}^{lm}\!\left(\r,\r';\overline{Q}^2,{\cal C}\right)
&\sim & {\cal I}^{lm}_{\textrm{UV}}\!\left(\r,\r';\overline{Q}^2\right)
\\
{\cal I}^{m}\!\left(\r,\r';\overline{Q}^2,{\cal C}\right)
&\sim & {\cal I}^{m}_{\textrm{UV}}\!\left(\r,\r';\overline{Q}^2\right)
\end{eqnarray}
in the $\r'\rightarrow 0$ limit, where
\begin{eqnarray}
{\cal I}^{m}_{\textrm{UV}}\!\left(\r,\r';\overline{Q}^2\right)
&\equiv & {\r'}^m\,  \left({\r'}^2\right)^{1-\frac{D}{2}}\,
\frac{i}{(2\pi)^2}\; \Gamma\!\left(\frac{D}{2}\!-\!1\right)\,
\left[\frac{2 \overline{Q}}{(2\pi)^2\mu^2 |\r|}\right]^{\frac{D}{2}-2}\,
 \textrm{K}_{\frac{D}{2}-2}\Big(\overline{Q}\, |\r|\Big)
\label{Im_UV_approx}
\\
{\cal I}^{lm}_{\textrm{UV}}\!\left(\r,\r';\overline{Q}^2\right)
&\equiv & {\r'}^m\,  \left({\r'}^2\right)^{1-\frac{D}{2}}\,
\frac{(-1)}{2(2\pi)^2}\; \Gamma\!\left(\frac{D}{2}\!-\!1\right)\,
\left[(2\pi)^2\mu^2\right]^{2-\frac{D}{2}}\,  \r^l\,
\left[\frac{2 \overline{Q}}{|\r|}\right]^{\frac{D}{2}-1}\,
 \textrm{K}_{\frac{D}{2}-1}\Big(\overline{Q}\, |\r|\Big)
 \label{Ilm_UV_approx}
 \, .
\end{eqnarray}




\bibliography{MaBiblioHEQCD}

\begin{thebibliography}{83}%
\makeatletter
\providecommand \@ifxundefined [1]{%
 \@ifx{#1\undefined}
}%
\providecommand \@ifnum [1]{%
 \ifnum #1\expandafter \@firstoftwo
 \else \expandafter \@secondoftwo
 \fi
}%
\providecommand \@ifx [1]{%
 \ifx #1\expandafter \@firstoftwo
 \else \expandafter \@secondoftwo
 \fi
}%
\providecommand \natexlab [1]{#1}%
\providecommand \enquote  [1]{``#1''}%
\providecommand \bibnamefont  [1]{#1}%
\providecommand \bibfnamefont [1]{#1}%
\providecommand \citenamefont [1]{#1}%
\providecommand \href@noop [0]{\@secondoftwo}%
\providecommand \href [0]{\begingroup \@sanitize@url \@href}%
\providecommand \@href[1]{\@@startlink{#1}\@@href}%
\providecommand \@@href[1]{\endgroup#1\@@endlink}%
\providecommand \@sanitize@url [0]{\catcode `\\12\catcode `\$12\catcode
  `\&12\catcode `\#12\catcode `\^12\catcode `\_12\catcode `\%12\relax}%
\providecommand \@@startlink[1]{}%
\providecommand \@@endlink[0]{}%
\providecommand \url  [0]{\begingroup\@sanitize@url \@url }%
\providecommand \@url [1]{\endgroup\@href {#1}{\urlprefix }}%
\providecommand \urlprefix  [0]{URL }%
\providecommand \Eprint [0]{\href }%
\providecommand \doibase [0]{http://dx.doi.org/}%
\providecommand \selectlanguage [0]{\@gobble}%
\providecommand \bibinfo  [0]{\@secondoftwo}%
\providecommand \bibfield  [0]{\@secondoftwo}%
\providecommand \translation [1]{[#1]}%
\providecommand \BibitemOpen [0]{}%
\providecommand \bibitemStop [0]{}%
\providecommand \bibitemNoStop [0]{.\EOS\space}%
\providecommand \EOS [0]{\spacefactor3000\relax}%
\providecommand \BibitemShut  [1]{\csname bibitem#1\endcsname}%
\let\auto@bib@innerbib\@empty
\bibitem [{\citenamefont {Beuf}(2016)}]{Beuf:2016wdz}%
  \BibitemOpen
  \bibfield  {author} {\bibinfo {author} {\bibfnamefont {G.}~\bibnamefont
  {Beuf}},\ }\href {\doibase 10.1103/PhysRevD.94.054016} {\bibfield  {journal}
  {\bibinfo  {journal} {Phys. Rev.}\ }\textbf {\bibinfo {volume} {D94}},\
  \bibinfo {pages} {054016} (\bibinfo {year} {2016})},\ \Eprint
  {http://arxiv.org/abs/1606.00777} {arXiv:1606.00777 [hep-ph]} \BibitemShut
  {NoStop}%
\bibitem [{\citenamefont {Lipatov}(1976)}]{Lipatov:1976zz}%
  \BibitemOpen
  \bibfield  {author} {\bibinfo {author} {\bibfnamefont {L.~N.}\ \bibnamefont
  {Lipatov}},\ }\href@noop {} {\bibfield  {journal} {\bibinfo  {journal} {Sov.
  J. Nucl. Phys.}\ }\textbf {\bibinfo {volume} {23}},\ \bibinfo {pages} {338}
  (\bibinfo {year} {1976})}\BibitemShut {NoStop}%
\bibitem [{\citenamefont {Kuraev}\ \emph {et~al.}(1977)\citenamefont {Kuraev},
  \citenamefont {Lipatov},\ and\ \citenamefont {Fadin}}]{Kuraev:1977fs}%
  \BibitemOpen
  \bibfield  {author} {\bibinfo {author} {\bibfnamefont {E.~A.}\ \bibnamefont
  {Kuraev}}, \bibinfo {author} {\bibfnamefont {L.~N.}\ \bibnamefont {Lipatov}},
  \ and\ \bibinfo {author} {\bibfnamefont {V.~S.}\ \bibnamefont {Fadin}},\
  }\href@noop {} {\bibfield  {journal} {\bibinfo  {journal} {Sov. Phys. JETP}\
  }\textbf {\bibinfo {volume} {45}},\ \bibinfo {pages} {199} (\bibinfo {year}
  {1977})}\BibitemShut {NoStop}%
\bibitem [{\citenamefont {Balitsky}\ and\ \citenamefont
  {Lipatov}(1978)}]{Balitsky:1978ic}%
  \BibitemOpen
  \bibfield  {author} {\bibinfo {author} {\bibfnamefont {I.~I.}\ \bibnamefont
  {Balitsky}}\ and\ \bibinfo {author} {\bibfnamefont {L.~N.}\ \bibnamefont
  {Lipatov}},\ }\href@noop {} {\bibfield  {journal} {\bibinfo  {journal} {Sov.
  J. Nucl. Phys.}\ }\textbf {\bibinfo {volume} {28}},\ \bibinfo {pages} {822}
  (\bibinfo {year} {1978})}\BibitemShut {NoStop}%
\bibitem [{\citenamefont {Gribov}\ \emph {et~al.}(1983)\citenamefont {Gribov},
  \citenamefont {Levin},\ and\ \citenamefont {Ryskin}}]{Gribov:1984tu}%
  \BibitemOpen
  \bibfield  {author} {\bibinfo {author} {\bibfnamefont {L.~V.}\ \bibnamefont
  {Gribov}}, \bibinfo {author} {\bibfnamefont {E.~M.}\ \bibnamefont {Levin}}, \
  and\ \bibinfo {author} {\bibfnamefont {M.~G.}\ \bibnamefont {Ryskin}},\
  }\href {\doibase 10.1016/0370-1573(83)90022-4} {\bibfield  {journal}
  {\bibinfo  {journal} {Phys. Rept.}\ }\textbf {\bibinfo {volume} {100}},\
  \bibinfo {pages} {1} (\bibinfo {year} {1983})}\BibitemShut {NoStop}%
\bibitem [{\citenamefont {Mueller}\ and\ \citenamefont
  {Qiu}(1986)}]{Mueller:1985wy}%
  \BibitemOpen
  \bibfield  {author} {\bibinfo {author} {\bibfnamefont {A.~H.}\ \bibnamefont
  {Mueller}}\ and\ \bibinfo {author} {\bibfnamefont {J.-w.}\ \bibnamefont
  {Qiu}},\ }\href {\doibase 10.1016/0550-3213(86)90164-1} {\bibfield  {journal}
  {\bibinfo  {journal} {Nucl. Phys.}\ }\textbf {\bibinfo {volume} {B268}},\
  \bibinfo {pages} {427} (\bibinfo {year} {1986})}\BibitemShut {NoStop}%
\bibitem [{\citenamefont {McLerran}\ and\ \citenamefont
  {Venugopalan}(1994{\natexlab{a}})}]{McLerran:1993ni}%
  \BibitemOpen
  \bibfield  {author} {\bibinfo {author} {\bibfnamefont {L.~D.}\ \bibnamefont
  {McLerran}}\ and\ \bibinfo {author} {\bibfnamefont {R.}~\bibnamefont
  {Venugopalan}},\ }\href {\doibase 10.1103/PhysRevD.49.2233} {\bibfield
  {journal} {\bibinfo  {journal} {Phys. Rev.}\ }\textbf {\bibinfo {volume}
  {D49}},\ \bibinfo {pages} {2233} (\bibinfo {year} {1994}{\natexlab{a}})},\
  \Eprint {http://arxiv.org/abs/hep-ph/9309289} {arXiv:hep-ph/9309289}
  \BibitemShut {NoStop}%
\bibitem [{\citenamefont {McLerran}\ and\ \citenamefont
  {Venugopalan}(1994{\natexlab{b}})}]{McLerran:1993ka}%
  \BibitemOpen
  \bibfield  {author} {\bibinfo {author} {\bibfnamefont {L.~D.}\ \bibnamefont
  {McLerran}}\ and\ \bibinfo {author} {\bibfnamefont {R.}~\bibnamefont
  {Venugopalan}},\ }\href {\doibase 10.1103/PhysRevD.49.3352} {\bibfield
  {journal} {\bibinfo  {journal} {Phys. Rev.}\ }\textbf {\bibinfo {volume}
  {D49}},\ \bibinfo {pages} {3352} (\bibinfo {year} {1994}{\natexlab{b}})},\
  \Eprint {http://arxiv.org/abs/hep-ph/9311205} {arXiv:hep-ph/9311205}
  \BibitemShut {NoStop}%
\bibitem [{\citenamefont {McLerran}\ and\ \citenamefont
  {Venugopalan}(1994{\natexlab{c}})}]{McLerran:1994vd}%
  \BibitemOpen
  \bibfield  {author} {\bibinfo {author} {\bibfnamefont {L.~D.}\ \bibnamefont
  {McLerran}}\ and\ \bibinfo {author} {\bibfnamefont {R.}~\bibnamefont
  {Venugopalan}},\ }\href {\doibase 10.1103/PhysRevD.50.2225} {\bibfield
  {journal} {\bibinfo  {journal} {Phys. Rev.}\ }\textbf {\bibinfo {volume}
  {D50}},\ \bibinfo {pages} {2225} (\bibinfo {year} {1994}{\natexlab{c}})},\
  \Eprint {http://arxiv.org/abs/hep-ph/9402335} {arXiv:hep-ph/9402335}
  \BibitemShut {NoStop}%
\bibitem [{\citenamefont {Balitsky}(1996)}]{Balitsky:1995ub}%
  \BibitemOpen
  \bibfield  {author} {\bibinfo {author} {\bibfnamefont {I.}~\bibnamefont
  {Balitsky}},\ }\href@noop {} {\bibfield  {journal} {\bibinfo  {journal}
  {Nucl. Phys.}\ }\textbf {\bibinfo {volume} {B463}},\ \bibinfo {pages} {99}
  (\bibinfo {year} {1996})},\ \Eprint {http://arxiv.org/abs/hep-ph/9509348}
  {hep-ph/9509348} \BibitemShut {NoStop}%
\bibitem [{\citenamefont {Jalilian-Marian}\ \emph {et~al.}(1997)\citenamefont
  {Jalilian-Marian}, \citenamefont {Kovner}, \citenamefont {Leonidov},\ and\
  \citenamefont {Weigert}}]{Jalilian-Marian:1997jx}%
  \BibitemOpen
  \bibfield  {author} {\bibinfo {author} {\bibfnamefont {J.}~\bibnamefont
  {Jalilian-Marian}}, \bibinfo {author} {\bibfnamefont {A.}~\bibnamefont
  {Kovner}}, \bibinfo {author} {\bibfnamefont {A.}~\bibnamefont {Leonidov}}, \
  and\ \bibinfo {author} {\bibfnamefont {H.}~\bibnamefont {Weigert}},\
  }\href@noop {} {\bibfield  {journal} {\bibinfo  {journal} {Nucl. Phys.}\
  }\textbf {\bibinfo {volume} {B504}},\ \bibinfo {pages} {415} (\bibinfo {year}
  {1997})},\ \Eprint {http://arxiv.org/abs/hep-ph/9701284} {hep-ph/9701284}
  \BibitemShut {NoStop}%
\bibitem [{\citenamefont {Jalilian-Marian}\ \emph
  {et~al.}(1998{\natexlab{a}})\citenamefont {Jalilian-Marian}, \citenamefont
  {Kovner}, \citenamefont {Leonidov},\ and\ \citenamefont
  {Weigert}}]{Jalilian-Marian:1997gr}%
  \BibitemOpen
  \bibfield  {author} {\bibinfo {author} {\bibfnamefont {J.}~\bibnamefont
  {Jalilian-Marian}}, \bibinfo {author} {\bibfnamefont {A.}~\bibnamefont
  {Kovner}}, \bibinfo {author} {\bibfnamefont {A.}~\bibnamefont {Leonidov}}, \
  and\ \bibinfo {author} {\bibfnamefont {H.}~\bibnamefont {Weigert}},\
  }\href@noop {} {\bibfield  {journal} {\bibinfo  {journal} {Phys. Rev.}\
  }\textbf {\bibinfo {volume} {D59}},\ \bibinfo {pages} {014014} (\bibinfo
  {year} {1998}{\natexlab{a}})},\ \Eprint {http://arxiv.org/abs/hep-ph/9706377}
  {hep-ph/9706377} \BibitemShut {NoStop}%
\bibitem [{\citenamefont {Jalilian-Marian}\ \emph
  {et~al.}(1998{\natexlab{b}})\citenamefont {Jalilian-Marian}, \citenamefont
  {Kovner},\ and\ \citenamefont {Weigert}}]{Jalilian-Marian:1997dw}%
  \BibitemOpen
  \bibfield  {author} {\bibinfo {author} {\bibfnamefont {J.}~\bibnamefont
  {Jalilian-Marian}}, \bibinfo {author} {\bibfnamefont {A.}~\bibnamefont
  {Kovner}}, \ and\ \bibinfo {author} {\bibfnamefont {H.}~\bibnamefont
  {Weigert}},\ }\href@noop {} {\bibfield  {journal} {\bibinfo  {journal} {Phys.
  Rev.}\ }\textbf {\bibinfo {volume} {D59}},\ \bibinfo {pages} {014015}
  (\bibinfo {year} {1998}{\natexlab{b}})},\ \Eprint
  {http://arxiv.org/abs/hep-ph/9709432} {hep-ph/9709432} \BibitemShut {NoStop}%
\bibitem [{\citenamefont {Kovner}\ \emph {et~al.}(2000)\citenamefont {Kovner},
  \citenamefont {Milhano},\ and\ \citenamefont {Weigert}}]{Kovner:2000pt}%
  \BibitemOpen
  \bibfield  {author} {\bibinfo {author} {\bibfnamefont {A.}~\bibnamefont
  {Kovner}}, \bibinfo {author} {\bibfnamefont {J.~G.}\ \bibnamefont {Milhano}},
  \ and\ \bibinfo {author} {\bibfnamefont {H.}~\bibnamefont {Weigert}},\
  }\href@noop {} {\bibfield  {journal} {\bibinfo  {journal} {Phys. Rev.}\
  }\textbf {\bibinfo {volume} {D62}},\ \bibinfo {pages} {114005} (\bibinfo
  {year} {2000})},\ \Eprint {http://arxiv.org/abs/hep-ph/0004014}
  {hep-ph/0004014} \BibitemShut {NoStop}%
\bibitem [{\citenamefont {Weigert}(2002)}]{Weigert:2000gi}%
  \BibitemOpen
  \bibfield  {author} {\bibinfo {author} {\bibfnamefont {H.}~\bibnamefont
  {Weigert}},\ }\href@noop {} {\bibfield  {journal} {\bibinfo  {journal} {Nucl.
  Phys.}\ }\textbf {\bibinfo {volume} {A703}},\ \bibinfo {pages} {823}
  (\bibinfo {year} {2002})},\ \Eprint {http://arxiv.org/abs/hep-ph/0004044}
  {hep-ph/0004044} \BibitemShut {NoStop}%
\bibitem [{\citenamefont {Iancu}\ \emph
  {et~al.}(2001{\natexlab{a}})\citenamefont {Iancu}, \citenamefont {Leonidov},\
  and\ \citenamefont {McLerran}}]{Iancu:2000hn}%
  \BibitemOpen
  \bibfield  {author} {\bibinfo {author} {\bibfnamefont {E.}~\bibnamefont
  {Iancu}}, \bibinfo {author} {\bibfnamefont {A.}~\bibnamefont {Leonidov}}, \
  and\ \bibinfo {author} {\bibfnamefont {L.~D.}\ \bibnamefont {McLerran}},\
  }\href@noop {} {\bibfield  {journal} {\bibinfo  {journal} {Nucl. Phys.}\
  }\textbf {\bibinfo {volume} {A692}},\ \bibinfo {pages} {583} (\bibinfo {year}
  {2001}{\natexlab{a}})},\ \Eprint {http://arxiv.org/abs/hep-ph/0011241}
  {hep-ph/0011241} \BibitemShut {NoStop}%
\bibitem [{\citenamefont {Iancu}\ \emph
  {et~al.}(2001{\natexlab{b}})\citenamefont {Iancu}, \citenamefont {Leonidov},\
  and\ \citenamefont {McLerran}}]{Iancu:2001ad}%
  \BibitemOpen
  \bibfield  {author} {\bibinfo {author} {\bibfnamefont {E.}~\bibnamefont
  {Iancu}}, \bibinfo {author} {\bibfnamefont {A.}~\bibnamefont {Leonidov}}, \
  and\ \bibinfo {author} {\bibfnamefont {L.~D.}\ \bibnamefont {McLerran}},\
  }\href@noop {} {\bibfield  {journal} {\bibinfo  {journal} {Phys. Lett.}\
  }\textbf {\bibinfo {volume} {B510}},\ \bibinfo {pages} {133} (\bibinfo {year}
  {2001}{\natexlab{b}})},\ \Eprint {http://arxiv.org/abs/hep-ph/0102009}
  {hep-ph/0102009} \BibitemShut {NoStop}%
\bibitem [{\citenamefont {Ferreiro}\ \emph {et~al.}(2002)\citenamefont
  {Ferreiro}, \citenamefont {Iancu}, \citenamefont {Leonidov},\ and\
  \citenamefont {McLerran}}]{Ferreiro:2001qy}%
  \BibitemOpen
  \bibfield  {author} {\bibinfo {author} {\bibfnamefont {E.}~\bibnamefont
  {Ferreiro}}, \bibinfo {author} {\bibfnamefont {E.}~\bibnamefont {Iancu}},
  \bibinfo {author} {\bibfnamefont {A.}~\bibnamefont {Leonidov}}, \ and\
  \bibinfo {author} {\bibfnamefont {L.}~\bibnamefont {McLerran}},\ }\href@noop
  {} {\bibfield  {journal} {\bibinfo  {journal} {Nucl. Phys.}\ }\textbf
  {\bibinfo {volume} {A703}},\ \bibinfo {pages} {489} (\bibinfo {year}
  {2002})},\ \Eprint {http://arxiv.org/abs/hep-ph/0109115} {hep-ph/0109115}
  \BibitemShut {NoStop}%
\bibitem [{\citenamefont {Kovchegov}(1999)}]{Kovchegov:1999yj}%
  \BibitemOpen
  \bibfield  {author} {\bibinfo {author} {\bibfnamefont {Y.~V.}\ \bibnamefont
  {Kovchegov}},\ }\href@noop {} {\bibfield  {journal} {\bibinfo  {journal}
  {Phys. Rev.}\ }\textbf {\bibinfo {volume} {D60}},\ \bibinfo {pages} {034008}
  (\bibinfo {year} {1999})},\ \Eprint {http://arxiv.org/abs/hep-ph/9901281}
  {hep-ph/9901281} \BibitemShut {NoStop}%
\bibitem [{\citenamefont {Kovchegov}(2000)}]{Kovchegov:1999ua}%
  \BibitemOpen
  \bibfield  {author} {\bibinfo {author} {\bibfnamefont {Y.~V.}\ \bibnamefont
  {Kovchegov}},\ }\href@noop {} {\bibfield  {journal} {\bibinfo  {journal}
  {Phys. Rev.}\ }\textbf {\bibinfo {volume} {D61}},\ \bibinfo {pages} {074018}
  (\bibinfo {year} {2000})},\ \Eprint {http://arxiv.org/abs/hep-ph/9905214}
  {hep-ph/9905214} \BibitemShut {NoStop}%
\bibitem [{\citenamefont {Bjorken}\ \emph {et~al.}(1971)\citenamefont
  {Bjorken}, \citenamefont {Kogut},\ and\ \citenamefont
  {Soper}}]{Bjorken:1970ah}%
  \BibitemOpen
  \bibfield  {author} {\bibinfo {author} {\bibfnamefont {J.}~\bibnamefont
  {Bjorken}}, \bibinfo {author} {\bibfnamefont {J.~B.}\ \bibnamefont {Kogut}},
  \ and\ \bibinfo {author} {\bibfnamefont {D.~E.}\ \bibnamefont {Soper}},\
  }\href {\doibase 10.1103/PhysRevD.3.1382} {\bibfield  {journal} {\bibinfo
  {journal} {Phys.Rev.}\ }\textbf {\bibinfo {volume} {D3}},\ \bibinfo {pages}
  {1382} (\bibinfo {year} {1971})}\BibitemShut {NoStop}%
\bibitem [{\citenamefont {Nikolaev}\ and\ \citenamefont
  {Zakharov}(1991)}]{Nikolaev:1990ja}%
  \BibitemOpen
  \bibfield  {author} {\bibinfo {author} {\bibfnamefont {N.~N.}\ \bibnamefont
  {Nikolaev}}\ and\ \bibinfo {author} {\bibfnamefont {B.~G.}\ \bibnamefont
  {Zakharov}},\ }\href {\doibase 10.1007/BF01483577} {\bibfield  {journal}
  {\bibinfo  {journal} {Z. Phys.}\ }\textbf {\bibinfo {volume} {C49}},\
  \bibinfo {pages} {607} (\bibinfo {year} {1991})}\BibitemShut {NoStop}%
\bibitem [{\citenamefont {Nikolaev}\ and\ \citenamefont
  {Zakharov}(1992)}]{Nikolaev:1991et}%
  \BibitemOpen
  \bibfield  {author} {\bibinfo {author} {\bibfnamefont {N.}~\bibnamefont
  {Nikolaev}}\ and\ \bibinfo {author} {\bibfnamefont {B.~G.}\ \bibnamefont
  {Zakharov}},\ }\href {\doibase 10.1007/BF01597573} {\bibfield  {journal}
  {\bibinfo  {journal} {Z.Phys.}\ }\textbf {\bibinfo {volume} {C53}},\ \bibinfo
  {pages} {331} (\bibinfo {year} {1992})}\BibitemShut {NoStop}%
\bibitem [{\citenamefont {Kopeliovich}\ and\ \citenamefont
  {Zakharov}(1991)}]{Kopeliovich:1991pu}%
  \BibitemOpen
  \bibfield  {author} {\bibinfo {author} {\bibfnamefont {B.}~\bibnamefont
  {Kopeliovich}}\ and\ \bibinfo {author} {\bibfnamefont {B.}~\bibnamefont
  {Zakharov}},\ }\href {\doibase 10.1103/PhysRevD.44.3466} {\bibfield
  {journal} {\bibinfo  {journal} {Phys.Rev.}\ }\textbf {\bibinfo {volume}
  {D44}},\ \bibinfo {pages} {3466} (\bibinfo {year} {1991})}\BibitemShut
  {NoStop}%
\bibitem [{\citenamefont {Mueller}(1994)}]{Mueller:1993rr}%
  \BibitemOpen
  \bibfield  {author} {\bibinfo {author} {\bibfnamefont {A.~H.}\ \bibnamefont
  {Mueller}},\ }\href {\doibase 10.1016/0550-3213(94)90116-3} {\bibfield
  {journal} {\bibinfo  {journal} {Nucl. Phys.}\ }\textbf {\bibinfo {volume}
  {B415}},\ \bibinfo {pages} {373} (\bibinfo {year} {1994})}\BibitemShut
  {NoStop}%
\bibitem [{\citenamefont {Mueller}\ and\ \citenamefont
  {Patel}(1994)}]{Mueller:1994jq}%
  \BibitemOpen
  \bibfield  {author} {\bibinfo {author} {\bibfnamefont {A.~H.}\ \bibnamefont
  {Mueller}}\ and\ \bibinfo {author} {\bibfnamefont {B.}~\bibnamefont
  {Patel}},\ }\href {\doibase 10.1016/0550-3213(94)90284-4} {\bibfield
  {journal} {\bibinfo  {journal} {Nucl. Phys.}\ }\textbf {\bibinfo {volume}
  {B425}},\ \bibinfo {pages} {471} (\bibinfo {year} {1994})},\ \Eprint
  {http://arxiv.org/abs/hep-ph/9403256} {arXiv:hep-ph/9403256} \BibitemShut
  {NoStop}%
\bibitem [{\citenamefont {Mueller}(1995)}]{Mueller:1994gb}%
  \BibitemOpen
  \bibfield  {author} {\bibinfo {author} {\bibfnamefont {A.~H.}\ \bibnamefont
  {Mueller}},\ }\href {\doibase 10.1016/0550-3213(94)00480-3} {\bibfield
  {journal} {\bibinfo  {journal} {Nucl. Phys.}\ }\textbf {\bibinfo {volume}
  {B437}},\ \bibinfo {pages} {107} (\bibinfo {year} {1995})},\ \Eprint
  {http://arxiv.org/abs/hep-ph/9408245} {arXiv:hep-ph/9408245} \BibitemShut
  {NoStop}%
\bibitem [{\citenamefont {Albacete}\ \emph {et~al.}(2009)\citenamefont
  {Albacete}, \citenamefont {Armesto}, \citenamefont {Milhano},\ and\
  \citenamefont {Salgado}}]{Albacete:2009fh}%
  \BibitemOpen
  \bibfield  {author} {\bibinfo {author} {\bibfnamefont {J.~L.}\ \bibnamefont
  {Albacete}}, \bibinfo {author} {\bibfnamefont {N.}~\bibnamefont {Armesto}},
  \bibinfo {author} {\bibfnamefont {J.~G.}\ \bibnamefont {Milhano}}, \ and\
  \bibinfo {author} {\bibfnamefont {C.~A.}\ \bibnamefont {Salgado}},\ }\href
  {\doibase 10.1103/PhysRevD.80.034031} {\bibfield  {journal} {\bibinfo
  {journal} {Phys. Rev.}\ }\textbf {\bibinfo {volume} {D80}},\ \bibinfo {pages}
  {034031} (\bibinfo {year} {2009})},\ \Eprint {http://arxiv.org/abs/0902.1112}
  {arXiv:0902.1112 [hep-ph]} \BibitemShut {NoStop}%
\bibitem [{\citenamefont {Albacete}\ \emph {et~al.}(2011)\citenamefont
  {Albacete}, \citenamefont {Armesto}, \citenamefont {Milhano}, \citenamefont
  {Quiroga-Arias},\ and\ \citenamefont {Salgado}}]{Albacete:2010sy}%
  \BibitemOpen
  \bibfield  {author} {\bibinfo {author} {\bibfnamefont {J.~L.}\ \bibnamefont
  {Albacete}}, \bibinfo {author} {\bibfnamefont {N.}~\bibnamefont {Armesto}},
  \bibinfo {author} {\bibfnamefont {J.~G.}\ \bibnamefont {Milhano}}, \bibinfo
  {author} {\bibfnamefont {P.}~\bibnamefont {Quiroga-Arias}}, \ and\ \bibinfo
  {author} {\bibfnamefont {C.~A.}\ \bibnamefont {Salgado}},\ }\href {\doibase
  10.1140/epjc/s10052-011-1705-3} {\bibfield  {journal} {\bibinfo  {journal}
  {Eur.Phys.J.}\ }\textbf {\bibinfo {volume} {C71}},\ \bibinfo {pages} {1705}
  (\bibinfo {year} {2011})},\ \Eprint {http://arxiv.org/abs/1012.4408}
  {arXiv:1012.4408 [hep-ph]} \BibitemShut {NoStop}%
\bibitem [{\citenamefont {Lappi}\ and\ \citenamefont
  {Mäntysaari}(2013)}]{Lappi:2013zma}%
  \BibitemOpen
  \bibfield  {author} {\bibinfo {author} {\bibfnamefont {T.}~\bibnamefont
  {Lappi}}\ and\ \bibinfo {author} {\bibfnamefont {H.}~\bibnamefont
  {Mäntysaari}},\ }\href {\doibase 10.1103/PhysRevD.88.114020} {\bibfield
  {journal} {\bibinfo  {journal} {Phys. Rev.}\ }\textbf {\bibinfo {volume}
  {D88}},\ \bibinfo {pages} {114020} (\bibinfo {year} {2013})},\ \Eprint
  {http://arxiv.org/abs/1309.6963} {arXiv:1309.6963 [hep-ph]} \BibitemShut
  {NoStop}%
\bibitem [{\citenamefont {Balitsky}(2007)}]{Balitsky:2006wa}%
  \BibitemOpen
  \bibfield  {author} {\bibinfo {author} {\bibfnamefont {I.}~\bibnamefont
  {Balitsky}},\ }\href@noop {} {\bibfield  {journal} {\bibinfo  {journal}
  {Phys. Rev.}\ }\textbf {\bibinfo {volume} {D75}},\ \bibinfo {pages} {014001}
  (\bibinfo {year} {2007})},\ \Eprint {http://arxiv.org/abs/hep-ph/0609105}
  {hep-ph/0609105} \BibitemShut {NoStop}%
\bibitem [{\citenamefont {Kovchegov}\ and\ \citenamefont
  {Weigert}(2007)}]{Kovchegov:2006vj}%
  \BibitemOpen
  \bibfield  {author} {\bibinfo {author} {\bibfnamefont {Y.~V.}\ \bibnamefont
  {Kovchegov}}\ and\ \bibinfo {author} {\bibfnamefont {H.}~\bibnamefont
  {Weigert}},\ }\href@noop {} {\bibfield  {journal} {\bibinfo  {journal} {Nucl.
  Phys.}\ }\textbf {\bibinfo {volume} {A784}},\ \bibinfo {pages} {188}
  (\bibinfo {year} {2007})},\ \Eprint {http://arxiv.org/abs/hep-ph/0609090}
  {hep-ph/0609090} \BibitemShut {NoStop}%
\bibitem [{\citenamefont {Balitsky}\ and\ \citenamefont
  {Chirilli}(2011)}]{Balitsky:2010ze}%
  \BibitemOpen
  \bibfield  {author} {\bibinfo {author} {\bibfnamefont {I.}~\bibnamefont
  {Balitsky}}\ and\ \bibinfo {author} {\bibfnamefont {G.~A.}\ \bibnamefont
  {Chirilli}},\ }\href {\doibase 10.1103/PhysRevD.83.031502} {\bibfield
  {journal} {\bibinfo  {journal} {Phys.Rev.}\ }\textbf {\bibinfo {volume}
  {D83}},\ \bibinfo {pages} {031502} (\bibinfo {year} {2011})},\ \Eprint
  {http://arxiv.org/abs/1009.4729} {arXiv:1009.4729 [hep-ph]} \BibitemShut
  {NoStop}%
\bibitem [{\citenamefont {Bartels}\ \emph
  {et~al.}(2001{\natexlab{a}})\citenamefont {Bartels}, \citenamefont
  {Gieseke},\ and\ \citenamefont {Qiao}}]{Bartels:2000gt}%
  \BibitemOpen
  \bibfield  {author} {\bibinfo {author} {\bibfnamefont {J.}~\bibnamefont
  {Bartels}}, \bibinfo {author} {\bibfnamefont {S.}~\bibnamefont {Gieseke}}, \
  and\ \bibinfo {author} {\bibfnamefont {C.~F.}\ \bibnamefont {Qiao}},\ }\href
  {\doibase 10.1103/PhysRevD.63.056014, 10.1103/PhysRevD.65.079902} {\bibfield
  {journal} {\bibinfo  {journal} {Phys. Rev.}\ }\textbf {\bibinfo {volume}
  {D63}},\ \bibinfo {pages} {056014} (\bibinfo {year} {2001}{\natexlab{a}})},\
  \bibinfo {note} {[Erratum: Phys. Rev.D65,079902(2002)]},\ \Eprint
  {http://arxiv.org/abs/hep-ph/0009102} {arXiv:hep-ph/0009102 [hep-ph]}
  \BibitemShut {NoStop}%
\bibitem [{\citenamefont {Bartels}\ \emph
  {et~al.}(2001{\natexlab{b}})\citenamefont {Bartels}, \citenamefont
  {Gieseke},\ and\ \citenamefont {Kyrieleis}}]{Bartels:2001mv}%
  \BibitemOpen
  \bibfield  {author} {\bibinfo {author} {\bibfnamefont {J.}~\bibnamefont
  {Bartels}}, \bibinfo {author} {\bibfnamefont {S.}~\bibnamefont {Gieseke}}, \
  and\ \bibinfo {author} {\bibfnamefont {A.}~\bibnamefont {Kyrieleis}},\ }\href
  {\doibase 10.1103/PhysRevD.65.014006} {\bibfield  {journal} {\bibinfo
  {journal} {Phys.Rev.}\ }\textbf {\bibinfo {volume} {D65}},\ \bibinfo {pages}
  {014006} (\bibinfo {year} {2001}{\natexlab{b}})},\ \Eprint
  {http://arxiv.org/abs/hep-ph/0107152} {arXiv:hep-ph/0107152 [hep-ph]}
  \BibitemShut {NoStop}%
\bibitem [{\citenamefont {Bartels}\ \emph {et~al.}(2002)\citenamefont
  {Bartels}, \citenamefont {Colferai}, \citenamefont {Gieseke},\ and\
  \citenamefont {Kyrieleis}}]{Bartels:2002uz}%
  \BibitemOpen
  \bibfield  {author} {\bibinfo {author} {\bibfnamefont {J.}~\bibnamefont
  {Bartels}}, \bibinfo {author} {\bibfnamefont {D.}~\bibnamefont {Colferai}},
  \bibinfo {author} {\bibfnamefont {S.}~\bibnamefont {Gieseke}}, \ and\
  \bibinfo {author} {\bibfnamefont {A.}~\bibnamefont {Kyrieleis}},\ }\href
  {\doibase 10.1103/PhysRevD.66.094017} {\bibfield  {journal} {\bibinfo
  {journal} {Phys.Rev.}\ }\textbf {\bibinfo {volume} {D66}},\ \bibinfo {pages}
  {094017} (\bibinfo {year} {2002})},\ \Eprint
  {http://arxiv.org/abs/hep-ph/0208130} {arXiv:hep-ph/0208130 [hep-ph]}
  \BibitemShut {NoStop}%
\bibitem [{\citenamefont {Bartels}\ and\ \citenamefont
  {Kyrieleis}(2004)}]{Bartels:2004bi}%
  \BibitemOpen
  \bibfield  {author} {\bibinfo {author} {\bibfnamefont {J.}~\bibnamefont
  {Bartels}}\ and\ \bibinfo {author} {\bibfnamefont {A.}~\bibnamefont
  {Kyrieleis}},\ }\href {\doibase 10.1103/PhysRevD.70.114003} {\bibfield
  {journal} {\bibinfo  {journal} {Phys.Rev.}\ }\textbf {\bibinfo {volume}
  {D70}},\ \bibinfo {pages} {114003} (\bibinfo {year} {2004})},\ \Eprint
  {http://arxiv.org/abs/hep-ph/0407051} {arXiv:hep-ph/0407051 [hep-ph]}
  \BibitemShut {NoStop}%
\bibitem [{\citenamefont {Balitsky}\ and\ \citenamefont
  {Chirilli}(2013{\natexlab{a}})}]{Balitsky:2012bs}%
  \BibitemOpen
  \bibfield  {author} {\bibinfo {author} {\bibfnamefont {I.}~\bibnamefont
  {Balitsky}}\ and\ \bibinfo {author} {\bibfnamefont {G.~A.}\ \bibnamefont
  {Chirilli}},\ }\href {\doibase 10.1103/PhysRevD.87.014013} {\bibfield
  {journal} {\bibinfo  {journal} {Phys. Rev.}\ }\textbf {\bibinfo {volume}
  {D87}},\ \bibinfo {pages} {014013} (\bibinfo {year} {2013}{\natexlab{a}})},\
  \Eprint {http://arxiv.org/abs/1207.3844} {arXiv:1207.3844 [hep-ph]}
  \BibitemShut {NoStop}%
\bibitem [{\citenamefont {Beuf}(2012)}]{Beuf:2011xd}%
  \BibitemOpen
  \bibfield  {author} {\bibinfo {author} {\bibfnamefont {G.}~\bibnamefont
  {Beuf}},\ }\href {\doibase 10.1103/PhysRevD.85.034039} {\bibfield  {journal}
  {\bibinfo  {journal} {Phys. Rev.}\ }\textbf {\bibinfo {volume} {D85}},\
  \bibinfo {pages} {034039} (\bibinfo {year} {2012})},\ \Eprint
  {http://arxiv.org/abs/1112.4501} {arXiv:1112.4501 [hep-ph]} \BibitemShut
  {NoStop}%
\bibitem [{\citenamefont {Boussarie}\ \emph {et~al.}(2014)\citenamefont
  {Boussarie}, \citenamefont {Grabovsky}, \citenamefont {Szymanowski},\ and\
  \citenamefont {Wallon}}]{Boussarie:2014lxa}%
  \BibitemOpen
  \bibfield  {author} {\bibinfo {author} {\bibfnamefont {R.}~\bibnamefont
  {Boussarie}}, \bibinfo {author} {\bibfnamefont {A.~V.}\ \bibnamefont
  {Grabovsky}}, \bibinfo {author} {\bibfnamefont {L.}~\bibnamefont
  {Szymanowski}}, \ and\ \bibinfo {author} {\bibfnamefont {S.}~\bibnamefont
  {Wallon}},\ }\href {\doibase 10.1007/JHEP09(2014)026} {\bibfield  {journal}
  {\bibinfo  {journal} {JHEP}\ }\textbf {\bibinfo {volume} {09}},\ \bibinfo
  {pages} {026} (\bibinfo {year} {2014})},\ \Eprint
  {http://arxiv.org/abs/1405.7676} {arXiv:1405.7676 [hep-ph]} \BibitemShut
  {NoStop}%
\bibitem [{\citenamefont {Boussarie}\ \emph
  {et~al.}(2016{\natexlab{a}})\citenamefont {Boussarie}, \citenamefont
  {Grabovsky}, \citenamefont {Szymanowski},\ and\ \citenamefont
  {Wallon}}]{Boussarie:2016ogo}%
  \BibitemOpen
  \bibfield  {author} {\bibinfo {author} {\bibfnamefont {R.}~\bibnamefont
  {Boussarie}}, \bibinfo {author} {\bibfnamefont {A.~V.}\ \bibnamefont
  {Grabovsky}}, \bibinfo {author} {\bibfnamefont {L.}~\bibnamefont
  {Szymanowski}}, \ and\ \bibinfo {author} {\bibfnamefont {S.}~\bibnamefont
  {Wallon}},\ }\href {\doibase 10.1007/JHEP11(2016)149} {\bibfield  {journal}
  {\bibinfo  {journal} {JHEP}\ }\textbf {\bibinfo {volume} {11}},\ \bibinfo
  {pages} {149} (\bibinfo {year} {2016}{\natexlab{a}})},\ \Eprint
  {http://arxiv.org/abs/1606.00419} {arXiv:1606.00419 [hep-ph]} \BibitemShut
  {NoStop}%
\bibitem [{\citenamefont {Boussarie}\ \emph
  {et~al.}(2016{\natexlab{b}})\citenamefont {Boussarie}, \citenamefont
  {Grabovsky}, \citenamefont {Ivanov}, \citenamefont {Szymanowski},\ and\
  \citenamefont {Wallon}}]{Boussarie:2016bkq}%
  \BibitemOpen
  \bibfield  {author} {\bibinfo {author} {\bibfnamefont {R.}~\bibnamefont
  {Boussarie}}, \bibinfo {author} {\bibfnamefont {A.~V.}\ \bibnamefont
  {Grabovsky}}, \bibinfo {author} {\bibfnamefont {D.~{\relax Yu}.}\
  \bibnamefont {Ivanov}}, \bibinfo {author} {\bibfnamefont {L.}~\bibnamefont
  {Szymanowski}}, \ and\ \bibinfo {author} {\bibfnamefont {S.}~\bibnamefont
  {Wallon}},\ }\href@noop {} {\  (\bibinfo {year} {2016}{\natexlab{b}})},\
  \Eprint {http://arxiv.org/abs/1612.08026} {arXiv:1612.08026 [hep-ph]}
  \BibitemShut {NoStop}%
\bibitem [{\citenamefont {Chirilli}\ \emph
  {et~al.}(2012{\natexlab{a}})\citenamefont {Chirilli}, \citenamefont {Xiao},\
  and\ \citenamefont {Yuan}}]{Chirilli:2011km}%
  \BibitemOpen
  \bibfield  {author} {\bibinfo {author} {\bibfnamefont {G.~A.}\ \bibnamefont
  {Chirilli}}, \bibinfo {author} {\bibfnamefont {B.-W.}\ \bibnamefont {Xiao}},
  \ and\ \bibinfo {author} {\bibfnamefont {F.}~\bibnamefont {Yuan}},\ }\href
  {\doibase 10.1103/PhysRevLett.108.122301} {\bibfield  {journal} {\bibinfo
  {journal} {Phys. Rev. Lett.}\ }\textbf {\bibinfo {volume} {108}},\ \bibinfo
  {pages} {122301} (\bibinfo {year} {2012}{\natexlab{a}})},\ \Eprint
  {http://arxiv.org/abs/1112.1061} {arXiv:1112.1061 [hep-ph]} \BibitemShut
  {NoStop}%
\bibitem [{\citenamefont {Chirilli}\ \emph
  {et~al.}(2012{\natexlab{b}})\citenamefont {Chirilli}, \citenamefont {Xiao},\
  and\ \citenamefont {Yuan}}]{Chirilli:2012jd}%
  \BibitemOpen
  \bibfield  {author} {\bibinfo {author} {\bibfnamefont {G.~A.}\ \bibnamefont
  {Chirilli}}, \bibinfo {author} {\bibfnamefont {B.-W.}\ \bibnamefont {Xiao}},
  \ and\ \bibinfo {author} {\bibfnamefont {F.}~\bibnamefont {Yuan}},\ }\href
  {\doibase 10.1103/PhysRevD.86.054005} {\bibfield  {journal} {\bibinfo
  {journal} {Phys.Rev.}\ }\textbf {\bibinfo {volume} {D86}},\ \bibinfo {pages}
  {054005} (\bibinfo {year} {2012}{\natexlab{b}})},\ \Eprint
  {http://arxiv.org/abs/1203.6139} {arXiv:1203.6139 [hep-ph]} \BibitemShut
  {NoStop}%
\bibitem [{\citenamefont {Altinoluk}\ \emph {et~al.}(2016)\citenamefont
  {Altinoluk}, \citenamefont {Armesto}, \citenamefont {Beuf}, \citenamefont
  {Kovner},\ and\ \citenamefont {Lublinsky}}]{Altinoluk:2015vax}%
  \BibitemOpen
  \bibfield  {author} {\bibinfo {author} {\bibfnamefont {T.}~\bibnamefont
  {Altinoluk}}, \bibinfo {author} {\bibfnamefont {N.}~\bibnamefont {Armesto}},
  \bibinfo {author} {\bibfnamefont {G.}~\bibnamefont {Beuf}}, \bibinfo {author}
  {\bibfnamefont {A.}~\bibnamefont {Kovner}}, \ and\ \bibinfo {author}
  {\bibfnamefont {M.}~\bibnamefont {Lublinsky}},\ }\href {\doibase
  10.1103/PhysRevD.93.054049} {\bibfield  {journal} {\bibinfo  {journal} {Phys.
  Rev.}\ }\textbf {\bibinfo {volume} {D93}},\ \bibinfo {pages} {054049}
  (\bibinfo {year} {2016})},\ \Eprint {http://arxiv.org/abs/1511.09415}
  {arXiv:1511.09415 [hep-ph]} \BibitemShut {NoStop}%
\bibitem [{\citenamefont {Benic}\ and\ \citenamefont
  {Fukushima}(2017)}]{Benic:2016yqt}%
  \BibitemOpen
  \bibfield  {author} {\bibinfo {author} {\bibfnamefont {S.}~\bibnamefont
  {Benic}}\ and\ \bibinfo {author} {\bibfnamefont {K.}~\bibnamefont
  {Fukushima}},\ }\href {\doibase 10.1016/j.nuclphysa.2016.11.003} {\bibfield
  {journal} {\bibinfo  {journal} {Nucl. Phys.}\ }\textbf {\bibinfo {volume}
  {A958}},\ \bibinfo {pages} {1} (\bibinfo {year} {2017})},\ \Eprint
  {http://arxiv.org/abs/1602.01989} {arXiv:1602.01989 [hep-ph]} \BibitemShut
  {NoStop}%
\bibitem [{\citenamefont {Benic}\ \emph {et~al.}(2017)\citenamefont {Benic},
  \citenamefont {Fukushima}, \citenamefont {Garcia-Montero},\ and\
  \citenamefont {Venugopalan}}]{Benic:2016uku}%
  \BibitemOpen
  \bibfield  {author} {\bibinfo {author} {\bibfnamefont {S.}~\bibnamefont
  {Benic}}, \bibinfo {author} {\bibfnamefont {K.}~\bibnamefont {Fukushima}},
  \bibinfo {author} {\bibfnamefont {O.}~\bibnamefont {Garcia-Montero}}, \ and\
  \bibinfo {author} {\bibfnamefont {R.}~\bibnamefont {Venugopalan}},\ }\href
  {\doibase 10.1007/JHEP01(2017)115} {\bibfield  {journal} {\bibinfo  {journal}
  {JHEP}\ }\textbf {\bibinfo {volume} {01}},\ \bibinfo {pages} {115} (\bibinfo
  {year} {2017})},\ \Eprint {http://arxiv.org/abs/1609.09424} {arXiv:1609.09424
  [hep-ph]} \BibitemShut {NoStop}%
\bibitem [{\citenamefont {Fadin}\ and\ \citenamefont
  {Lipatov}(1998)}]{Fadin:1998py}%
  \BibitemOpen
  \bibfield  {author} {\bibinfo {author} {\bibfnamefont {V.~S.}\ \bibnamefont
  {Fadin}}\ and\ \bibinfo {author} {\bibfnamefont {L.~N.}\ \bibnamefont
  {Lipatov}},\ }\href {\doibase 10.1016/S0370-2693(98)00473-0} {\bibfield
  {journal} {\bibinfo  {journal} {Phys. Lett.}\ }\textbf {\bibinfo {volume}
  {B429}},\ \bibinfo {pages} {127} (\bibinfo {year} {1998})},\ \Eprint
  {http://arxiv.org/abs/hep-ph/9802290} {arXiv:hep-ph/9802290} \BibitemShut
  {NoStop}%
\bibitem [{\citenamefont {Ciafaloni}\ and\ \citenamefont
  {Camici}(1998)}]{Ciafaloni:1998gs}%
  \BibitemOpen
  \bibfield  {author} {\bibinfo {author} {\bibfnamefont {M.}~\bibnamefont
  {Ciafaloni}}\ and\ \bibinfo {author} {\bibfnamefont {G.}~\bibnamefont
  {Camici}},\ }\href {\doibase 10.1016/S0370-2693(98)00551-6} {\bibfield
  {journal} {\bibinfo  {journal} {Phys. Lett.}\ }\textbf {\bibinfo {volume}
  {B430}},\ \bibinfo {pages} {349} (\bibinfo {year} {1998})},\ \Eprint
  {http://arxiv.org/abs/hep-ph/9803389} {arXiv:hep-ph/9803389} \BibitemShut
  {NoStop}%
\bibitem [{\citenamefont {Balitsky}\ and\ \citenamefont
  {Chirilli}(2008)}]{Balitsky:2008zz}%
  \BibitemOpen
  \bibfield  {author} {\bibinfo {author} {\bibfnamefont {I.}~\bibnamefont
  {Balitsky}}\ and\ \bibinfo {author} {\bibfnamefont {G.~A.}\ \bibnamefont
  {Chirilli}},\ }\href@noop {} {\bibfield  {journal} {\bibinfo  {journal}
  {Phys. Rev.}\ }\textbf {\bibinfo {volume} {D77}},\ \bibinfo {pages} {014019}
  (\bibinfo {year} {2008})},\ \Eprint {http://arxiv.org/abs/arXiv:0710.4330
  [hep-ph]} {arXiv:0710.4330 [hep-ph]} \BibitemShut {NoStop}%
\bibitem [{\citenamefont {Balitsky}\ and\ \citenamefont
  {Chirilli}(2009)}]{Balitsky:2009xg}%
  \BibitemOpen
  \bibfield  {author} {\bibinfo {author} {\bibfnamefont {I.}~\bibnamefont
  {Balitsky}}\ and\ \bibinfo {author} {\bibfnamefont {G.~A.}\ \bibnamefont
  {Chirilli}},\ }\href {\doibase 10.1016/j.nuclphysb.2009.07.003} {\bibfield
  {journal} {\bibinfo  {journal} {Nucl.Phys.}\ }\textbf {\bibinfo {volume}
  {B822}},\ \bibinfo {pages} {45} (\bibinfo {year} {2009})},\ \Eprint
  {http://arxiv.org/abs/arXiv:0903.5326 [hep-ph]} {arXiv:0903.5326 [hep-ph]}
  \BibitemShut {NoStop}%
\bibitem [{\citenamefont {Balitsky}\ and\ \citenamefont
  {Chirilli}(2013{\natexlab{b}})}]{Balitsky:2013fea}%
  \BibitemOpen
  \bibfield  {author} {\bibinfo {author} {\bibfnamefont {I.}~\bibnamefont
  {Balitsky}}\ and\ \bibinfo {author} {\bibfnamefont {G.~A.}\ \bibnamefont
  {Chirilli}},\ }\href {\doibase 10.1103/PhysRevD.88.111501} {\bibfield
  {journal} {\bibinfo  {journal} {Phys.Rev.}\ }\textbf {\bibinfo {volume}
  {D88}},\ \bibinfo {pages} {111501} (\bibinfo {year} {2013}{\natexlab{b}})},\
  \Eprint {http://arxiv.org/abs/1309.7644} {arXiv:1309.7644 [hep-ph]}
  \BibitemShut {NoStop}%
\bibitem [{\citenamefont {Lublinsky}\ and\ \citenamefont
  {Mulian}(2017)}]{Lublinsky:2016meo}%
  \BibitemOpen
  \bibfield  {author} {\bibinfo {author} {\bibfnamefont {M.}~\bibnamefont
  {Lublinsky}}\ and\ \bibinfo {author} {\bibfnamefont {Y.}~\bibnamefont
  {Mulian}},\ }\href {\doibase 10.1007/JHEP05(2017)097} {\bibfield  {journal}
  {\bibinfo  {journal} {JHEP}\ }\textbf {\bibinfo {volume} {05}},\ \bibinfo
  {pages} {097} (\bibinfo {year} {2017})},\ \Eprint
  {http://arxiv.org/abs/1610.03453} {arXiv:1610.03453 [hep-ph]} \BibitemShut
  {NoStop}%
\bibitem [{\citenamefont {Salam}(1998)}]{Salam:1998tj}%
  \BibitemOpen
  \bibfield  {author} {\bibinfo {author} {\bibfnamefont {G.~P.}\ \bibnamefont
  {Salam}},\ }\href@noop {} {\bibfield  {journal} {\bibinfo  {journal} {JHEP}\
  }\textbf {\bibinfo {volume} {07}},\ \bibinfo {pages} {019} (\bibinfo {year}
  {1998})},\ \Eprint {http://arxiv.org/abs/hep-ph/9806482}
  {arXiv:hep-ph/9806482} \BibitemShut {NoStop}%
\bibitem [{\citenamefont {Beuf}(2014)}]{Beuf:2014uia}%
  \BibitemOpen
  \bibfield  {author} {\bibinfo {author} {\bibfnamefont {G.}~\bibnamefont
  {Beuf}},\ }\href {\doibase 10.1103/PhysRevD.89.074039} {\bibfield  {journal}
  {\bibinfo  {journal} {Phys. Rev.}\ }\textbf {\bibinfo {volume} {D89}},\
  \bibinfo {pages} {074039} (\bibinfo {year} {2014})},\ \Eprint
  {http://arxiv.org/abs/1401.0313} {arXiv:1401.0313 [hep-ph]} \BibitemShut
  {NoStop}%
\bibitem [{\citenamefont {Iancu}\ \emph
  {et~al.}(2015{\natexlab{a}})\citenamefont {Iancu}, \citenamefont {Madrigal},
  \citenamefont {Mueller}, \citenamefont {Soyez},\ and\ \citenamefont
  {Triantafyllopoulos}}]{Iancu:2015vea}%
  \BibitemOpen
  \bibfield  {author} {\bibinfo {author} {\bibfnamefont {E.}~\bibnamefont
  {Iancu}}, \bibinfo {author} {\bibfnamefont {J.~D.}\ \bibnamefont {Madrigal}},
  \bibinfo {author} {\bibfnamefont {A.~H.}\ \bibnamefont {Mueller}}, \bibinfo
  {author} {\bibfnamefont {G.}~\bibnamefont {Soyez}}, \ and\ \bibinfo {author}
  {\bibfnamefont {D.~N.}\ \bibnamefont {Triantafyllopoulos}},\ }\href {\doibase
  10.1016/j.physletb.2015.03.068} {\bibfield  {journal} {\bibinfo  {journal}
  {Phys. Lett.}\ }\textbf {\bibinfo {volume} {B744}},\ \bibinfo {pages} {293}
  (\bibinfo {year} {2015}{\natexlab{a}})},\ \Eprint
  {http://arxiv.org/abs/1502.05642} {arXiv:1502.05642 [hep-ph]} \BibitemShut
  {NoStop}%
\bibitem [{\citenamefont {Motyka}\ and\ \citenamefont
  {Stasto}(2009)}]{Motyka:2009gi}%
  \BibitemOpen
  \bibfield  {author} {\bibinfo {author} {\bibfnamefont {L.}~\bibnamefont
  {Motyka}}\ and\ \bibinfo {author} {\bibfnamefont {A.~M.}\ \bibnamefont
  {Stasto}},\ }\href {\doibase 10.1103/PhysRevD.79.085016} {\bibfield
  {journal} {\bibinfo  {journal} {Phys.Rev.}\ }\textbf {\bibinfo {volume}
  {D79}},\ \bibinfo {pages} {085016} (\bibinfo {year} {2009})},\ \Eprint
  {http://arxiv.org/abs/0901.4949} {arXiv:0901.4949 [hep-ph]} \BibitemShut
  {NoStop}%
\bibitem [{\citenamefont {Iancu}\ \emph
  {et~al.}(2015{\natexlab{b}})\citenamefont {Iancu}, \citenamefont {Madrigal},
  \citenamefont {Mueller}, \citenamefont {Soyez},\ and\ \citenamefont
  {Triantafyllopoulos}}]{Iancu:2015joa}%
  \BibitemOpen
  \bibfield  {author} {\bibinfo {author} {\bibfnamefont {E.}~\bibnamefont
  {Iancu}}, \bibinfo {author} {\bibfnamefont {J.~D.}\ \bibnamefont {Madrigal}},
  \bibinfo {author} {\bibfnamefont {A.~H.}\ \bibnamefont {Mueller}}, \bibinfo
  {author} {\bibfnamefont {G.}~\bibnamefont {Soyez}}, \ and\ \bibinfo {author}
  {\bibfnamefont {D.~N.}\ \bibnamefont {Triantafyllopoulos}},\ }\href {\doibase
  10.1016/j.physletb.2015.09.071} {\bibfield  {journal} {\bibinfo  {journal}
  {Phys. Lett.}\ }\textbf {\bibinfo {volume} {B750}},\ \bibinfo {pages} {643}
  (\bibinfo {year} {2015}{\natexlab{b}})},\ \Eprint
  {http://arxiv.org/abs/1507.03651} {arXiv:1507.03651 [hep-ph]} \BibitemShut
  {NoStop}%
\bibitem [{\citenamefont {Albacete}(2017)}]{Albacete:2015xza}%
  \BibitemOpen
  \bibfield  {author} {\bibinfo {author} {\bibfnamefont {J.~L.}\ \bibnamefont
  {Albacete}},\ }\href {\doibase 10.1016/j.nuclphysa.2016.07.008} {\bibfield
  {journal} {\bibinfo  {journal} {Nucl. Phys.}\ }\textbf {\bibinfo {volume}
  {A957}},\ \bibinfo {pages} {71} (\bibinfo {year} {2017})},\ \Eprint
  {http://arxiv.org/abs/1507.07120} {arXiv:1507.07120 [hep-ph]} \BibitemShut
  {NoStop}%
\bibitem [{\citenamefont {Lappi}\ and\ \citenamefont
  {Mäntysaari}(2016)}]{Lappi:2016fmu}%
  \BibitemOpen
  \bibfield  {author} {\bibinfo {author} {\bibfnamefont {T.}~\bibnamefont
  {Lappi}}\ and\ \bibinfo {author} {\bibfnamefont {H.}~\bibnamefont
  {Mäntysaari}},\ }\href {\doibase 10.1103/PhysRevD.93.094004} {\bibfield
  {journal} {\bibinfo  {journal} {Phys. Rev.}\ }\textbf {\bibinfo {volume}
  {D93}},\ \bibinfo {pages} {094004} (\bibinfo {year} {2016})},\ \Eprint
  {http://arxiv.org/abs/1601.06598} {arXiv:1601.06598 [hep-ph]} \BibitemShut
  {NoStop}%
\bibitem [{\citenamefont {Lappi}\ and\ \citenamefont
  {Mäntysaari}(2015)}]{Lappi:2015fma}%
  \BibitemOpen
  \bibfield  {author} {\bibinfo {author} {\bibfnamefont {T.}~\bibnamefont
  {Lappi}}\ and\ \bibinfo {author} {\bibfnamefont {H.}~\bibnamefont
  {Mäntysaari}},\ }\href {\doibase 10.1103/PhysRevD.91.074016} {\bibfield
  {journal} {\bibinfo  {journal} {Phys. Rev.}\ }\textbf {\bibinfo {volume}
  {D91}},\ \bibinfo {pages} {074016} (\bibinfo {year} {2015})},\ \Eprint
  {http://arxiv.org/abs/1502.02400} {arXiv:1502.02400 [hep-ph]} \BibitemShut
  {NoStop}%
\bibitem [{\citenamefont {Brodsky}\ \emph {et~al.}(1998)\citenamefont
  {Brodsky}, \citenamefont {Pauli},\ and\ \citenamefont
  {Pinsky}}]{Brodsky:1997de}%
  \BibitemOpen
  \bibfield  {author} {\bibinfo {author} {\bibfnamefont {S.~J.}\ \bibnamefont
  {Brodsky}}, \bibinfo {author} {\bibfnamefont {H.-C.}\ \bibnamefont {Pauli}},
  \ and\ \bibinfo {author} {\bibfnamefont {S.~S.}\ \bibnamefont {Pinsky}},\
  }\href {\doibase 10.1016/S0370-1573(97)00089-6} {\bibfield  {journal}
  {\bibinfo  {journal} {Phys. Rept.}\ }\textbf {\bibinfo {volume} {301}},\
  \bibinfo {pages} {299} (\bibinfo {year} {1998})},\ \Eprint
  {http://arxiv.org/abs/hep-ph/9705477} {arXiv:hep-ph/9705477} \BibitemShut
  {NoStop}%
\bibitem [{\citenamefont {Sterman}(1993)}]{Sterman:1994ce}%
  \BibitemOpen
  \bibfield  {author} {\bibinfo {author} {\bibfnamefont {G.~F.}\ \bibnamefont
  {Sterman}},\ }\href@noop {} {\emph {\bibinfo {title} {{An Introduction to
  quantum field theory}}}}\ (\bibinfo  {publisher} {Cambridge University
  Press},\ \bibinfo {year} {1993})\BibitemShut {NoStop}%
\bibitem [{\citenamefont {Collins}(2011)}]{Collins_TMD_book}%
  \BibitemOpen
  \bibfield  {author} {\bibinfo {author} {\bibfnamefont {J.~C.}\ \bibnamefont
  {Collins}},\ }\href@noop {} {\emph {\bibinfo {title} {{Foundations of
  Perturbative QCD}}}}\ (\bibinfo  {publisher} {Cambridge University Press,
  Cambridge},\ \bibinfo {year} {2011})\BibitemShut {NoStop}%
\bibitem [{\citenamefont {Peskin}\ and\ \citenamefont
  {Schroeder}(1995)}]{Peskin:1995ev}%
  \BibitemOpen
  \bibfield  {author} {\bibinfo {author} {\bibfnamefont {M.~E.}\ \bibnamefont
  {Peskin}}\ and\ \bibinfo {author} {\bibfnamefont {D.~V.}\ \bibnamefont
  {Schroeder}},\ }\href {http://www.slac.stanford.edu/~mpeskin/QFT.html} {\emph
  {\bibinfo {title} {{An Introduction to quantum field theory}}}}\ (\bibinfo
  {publisher} {Addison-Wesley},\ \bibinfo {address} {Reading, USA},\ \bibinfo
  {year} {1995})\BibitemShut {NoStop}%
\bibitem [{\citenamefont {Kogut}\ and\ \citenamefont
  {Soper}(1970)}]{Kogut:1969xa}%
  \BibitemOpen
  \bibfield  {author} {\bibinfo {author} {\bibfnamefont {J.~B.}\ \bibnamefont
  {Kogut}}\ and\ \bibinfo {author} {\bibfnamefont {D.~E.}\ \bibnamefont
  {Soper}},\ }\href {\doibase 10.1103/PhysRevD.1.2901} {\bibfield  {journal}
  {\bibinfo  {journal} {Phys.Rev.}\ }\textbf {\bibinfo {volume} {D1}},\
  \bibinfo {pages} {2901} (\bibinfo {year} {1970})}\BibitemShut {NoStop}%
\bibitem [{\citenamefont {Jeon}\ and\ \citenamefont
  {Venugopalan}(2004)}]{Jeon:2004rk}%
  \BibitemOpen
  \bibfield  {author} {\bibinfo {author} {\bibfnamefont {S.}~\bibnamefont
  {Jeon}}\ and\ \bibinfo {author} {\bibfnamefont {R.}~\bibnamefont
  {Venugopalan}},\ }\href {\doibase 10.1103/PhysRevD.70.105012} {\bibfield
  {journal} {\bibinfo  {journal} {Phys. Rev.}\ }\textbf {\bibinfo {volume}
  {D70}},\ \bibinfo {pages} {105012} (\bibinfo {year} {2004})},\ \Eprint
  {http://arxiv.org/abs/hep-ph/0406169} {arXiv:hep-ph/0406169 [hep-ph]}
  \BibitemShut {NoStop}%
\bibitem [{\citenamefont {Kovchegov}\ \emph {et~al.}(2004)\citenamefont
  {Kovchegov}, \citenamefont {Szymanowski},\ and\ \citenamefont
  {Wallon}}]{Kovchegov:2003dm}%
  \BibitemOpen
  \bibfield  {author} {\bibinfo {author} {\bibfnamefont {Y.~V.}\ \bibnamefont
  {Kovchegov}}, \bibinfo {author} {\bibfnamefont {L.}~\bibnamefont
  {Szymanowski}}, \ and\ \bibinfo {author} {\bibfnamefont {S.}~\bibnamefont
  {Wallon}},\ }\href {\doibase 10.1016/j.physletb.2004.02.036} {\bibfield
  {journal} {\bibinfo  {journal} {Phys. Lett.}\ }\textbf {\bibinfo {volume}
  {B586}},\ \bibinfo {pages} {267} (\bibinfo {year} {2004})},\ \Eprint
  {http://arxiv.org/abs/hep-ph/0309281} {arXiv:hep-ph/0309281 [hep-ph]}
  \BibitemShut {NoStop}%
\bibitem [{\citenamefont {Hatta}\ \emph {et~al.}(2005)\citenamefont {Hatta},
  \citenamefont {Iancu}, \citenamefont {Itakura},\ and\ \citenamefont
  {McLerran}}]{Hatta:2005as}%
  \BibitemOpen
  \bibfield  {author} {\bibinfo {author} {\bibfnamefont {Y.}~\bibnamefont
  {Hatta}}, \bibinfo {author} {\bibfnamefont {E.}~\bibnamefont {Iancu}},
  \bibinfo {author} {\bibfnamefont {K.}~\bibnamefont {Itakura}}, \ and\
  \bibinfo {author} {\bibfnamefont {L.}~\bibnamefont {McLerran}},\ }\href
  {\doibase 10.1016/j.nuclphysa.2005.05.163} {\bibfield  {journal} {\bibinfo
  {journal} {Nucl. Phys.}\ }\textbf {\bibinfo {volume} {A760}},\ \bibinfo
  {pages} {172} (\bibinfo {year} {2005})},\ \Eprint
  {http://arxiv.org/abs/hep-ph/0501171} {arXiv:hep-ph/0501171 [hep-ph]}
  \BibitemShut {NoStop}%
\bibitem [{\citenamefont {Jeon}\ and\ \citenamefont
  {Venugopalan}(2005)}]{Jeon:2005cf}%
  \BibitemOpen
  \bibfield  {author} {\bibinfo {author} {\bibfnamefont {S.}~\bibnamefont
  {Jeon}}\ and\ \bibinfo {author} {\bibfnamefont {R.}~\bibnamefont
  {Venugopalan}},\ }\href {\doibase 10.1103/PhysRevD.71.125003} {\bibfield
  {journal} {\bibinfo  {journal} {Phys. Rev.}\ }\textbf {\bibinfo {volume}
  {D71}},\ \bibinfo {pages} {125003} (\bibinfo {year} {2005})},\ \Eprint
  {http://arxiv.org/abs/hep-ph/0503219} {arXiv:hep-ph/0503219 [hep-ph]}
  \BibitemShut {NoStop}%
\bibitem [{\citenamefont {Lappi}\ \emph {et~al.}(2016)\citenamefont {Lappi},
  \citenamefont {Ramnath}, \citenamefont {Rummukainen},\ and\ \citenamefont
  {Weigert}}]{Lappi:2016gqe}%
  \BibitemOpen
  \bibfield  {author} {\bibinfo {author} {\bibfnamefont {T.}~\bibnamefont
  {Lappi}}, \bibinfo {author} {\bibfnamefont {A.}~\bibnamefont {Ramnath}},
  \bibinfo {author} {\bibfnamefont {K.}~\bibnamefont {Rummukainen}}, \ and\
  \bibinfo {author} {\bibfnamefont {H.}~\bibnamefont {Weigert}},\ }\href
  {\doibase 10.1103/PhysRevD.94.054014} {\bibfield  {journal} {\bibinfo
  {journal} {Phys. Rev.}\ }\textbf {\bibinfo {volume} {D94}},\ \bibinfo {pages}
  {054014} (\bibinfo {year} {2016})},\ \Eprint
  {http://arxiv.org/abs/1606.00551} {arXiv:1606.00551 [hep-ph]} \BibitemShut
  {NoStop}%
\bibitem [{\citenamefont {Lappi}\ and\ \citenamefont
  {Paatelainen}(2017)}]{Lappi:2016oup}%
  \BibitemOpen
  \bibfield  {author} {\bibinfo {author} {\bibfnamefont {T.}~\bibnamefont
  {Lappi}}\ and\ \bibinfo {author} {\bibfnamefont {R.}~\bibnamefont
  {Paatelainen}},\ }\href {\doibase 10.1016/j.aop.2017.02.002} {\bibfield
  {journal} {\bibinfo  {journal} {Annals Phys.}\ }\textbf {\bibinfo {volume}
  {379}},\ \bibinfo {pages} {34} (\bibinfo {year} {2017})},\ \Eprint
  {http://arxiv.org/abs/1611.00497} {arXiv:1611.00497 [hep-ph]} \BibitemShut
  {NoStop}%
\bibitem [{\citenamefont {Iancu}\ \emph {et~al.}(2016)\citenamefont {Iancu},
  \citenamefont {Mueller},\ and\ \citenamefont
  {Triantafyllopoulos}}]{Iancu:2016vyg}%
  \BibitemOpen
  \bibfield  {author} {\bibinfo {author} {\bibfnamefont {E.}~\bibnamefont
  {Iancu}}, \bibinfo {author} {\bibfnamefont {A.~H.}\ \bibnamefont {Mueller}},
  \ and\ \bibinfo {author} {\bibfnamefont {D.~N.}\ \bibnamefont
  {Triantafyllopoulos}},\ }\href {\doibase 10.1007/JHEP12(2016)041} {\bibfield
  {journal} {\bibinfo  {journal} {JHEP}\ }\textbf {\bibinfo {volume} {12}},\
  \bibinfo {pages} {041} (\bibinfo {year} {2016})},\ \Eprint
  {http://arxiv.org/abs/1608.05293} {arXiv:1608.05293 [hep-ph]} \BibitemShut
  {NoStop}%
\bibitem [{\citenamefont {Ciafaloni}(1988)}]{Ciafaloni:1987ur}%
  \BibitemOpen
  \bibfield  {author} {\bibinfo {author} {\bibfnamefont {M.}~\bibnamefont
  {Ciafaloni}},\ }\href {\doibase 10.1016/0550-3213(88)90380-X} {\bibfield
  {journal} {\bibinfo  {journal} {Nucl.Phys.}\ }\textbf {\bibinfo {volume}
  {B296}},\ \bibinfo {pages} {49} (\bibinfo {year} {1988})}\BibitemShut
  {NoStop}%
\bibitem [{\citenamefont {Andersson}\ \emph {et~al.}(1996)\citenamefont
  {Andersson}, \citenamefont {Gustafson},\ and\ \citenamefont
  {Samuelsson}}]{Andersson:1995ju}%
  \BibitemOpen
  \bibfield  {author} {\bibinfo {author} {\bibfnamefont {B.}~\bibnamefont
  {Andersson}}, \bibinfo {author} {\bibfnamefont {G.}~\bibnamefont
  {Gustafson}}, \ and\ \bibinfo {author} {\bibfnamefont {J.}~\bibnamefont
  {Samuelsson}},\ }\href {\doibase 10.1016/0550-3213(96)00114-9} {\bibfield
  {journal} {\bibinfo  {journal} {Nucl.Phys.}\ }\textbf {\bibinfo {volume}
  {B467}},\ \bibinfo {pages} {443} (\bibinfo {year} {1996})}\BibitemShut
  {NoStop}%
\bibitem [{\citenamefont {Kwiecinski}\ \emph {et~al.}(1996)\citenamefont
  {Kwiecinski}, \citenamefont {Martin},\ and\ \citenamefont
  {Sutton}}]{Kwiecinski:1996td}%
  \BibitemOpen
  \bibfield  {author} {\bibinfo {author} {\bibfnamefont {J.}~\bibnamefont
  {Kwiecinski}}, \bibinfo {author} {\bibfnamefont {A.~D.}\ \bibnamefont
  {Martin}}, \ and\ \bibinfo {author} {\bibfnamefont {P.}~\bibnamefont
  {Sutton}},\ }\href {\doibase 10.1007/s002880050206} {\bibfield  {journal}
  {\bibinfo  {journal} {Z.Phys.}\ }\textbf {\bibinfo {volume} {C71}},\ \bibinfo
  {pages} {585} (\bibinfo {year} {1996})},\ \Eprint
  {http://arxiv.org/abs/hep-ph/9602320} {arXiv:hep-ph/9602320 [hep-ph]}
  \BibitemShut {NoStop}%
\bibitem [{\citenamefont {Golec-Biernat}\ \emph {et~al.}(2002)\citenamefont
  {Golec-Biernat}, \citenamefont {Motyka},\ and\ \citenamefont
  {Stasto}}]{GolecBiernat:2001if}%
  \BibitemOpen
  \bibfield  {author} {\bibinfo {author} {\bibfnamefont {K.~J.}\ \bibnamefont
  {Golec-Biernat}}, \bibinfo {author} {\bibfnamefont {L.}~\bibnamefont
  {Motyka}}, \ and\ \bibinfo {author} {\bibfnamefont {A.~M.}\ \bibnamefont
  {Stasto}},\ }\href {\doibase 10.1103/PhysRevD.65.074037} {\bibfield
  {journal} {\bibinfo  {journal} {Phys. Rev.}\ }\textbf {\bibinfo {volume}
  {D65}},\ \bibinfo {pages} {074037} (\bibinfo {year} {2002})},\ \Eprint
  {http://arxiv.org/abs/hep-ph/0110325} {arXiv:hep-ph/0110325} \BibitemShut
  {NoStop}%
\bibitem [{\citenamefont {Rummukainen}\ and\ \citenamefont
  {Weigert}(2004)}]{Rummukainen:2003ns}%
  \BibitemOpen
  \bibfield  {author} {\bibinfo {author} {\bibfnamefont {K.}~\bibnamefont
  {Rummukainen}}\ and\ \bibinfo {author} {\bibfnamefont {H.}~\bibnamefont
  {Weigert}},\ }\href@noop {} {\bibfield  {journal} {\bibinfo  {journal} {Nucl.
  Phys.}\ }\textbf {\bibinfo {volume} {A739}},\ \bibinfo {pages} {183}
  (\bibinfo {year} {2004})},\ \Eprint {http://arxiv.org/abs/hep-ph/0309306}
  {hep-ph/0309306} \BibitemShut {NoStop}%
\bibitem [{\citenamefont {Albacete}\ \emph {et~al.}(2005)\citenamefont
  {Albacete}, \citenamefont {Armesto}, \citenamefont {Milhano}, \citenamefont
  {Salgado},\ and\ \citenamefont {Wiedemann}}]{Albacete:2004gw}%
  \BibitemOpen
  \bibfield  {author} {\bibinfo {author} {\bibfnamefont {J.~L.}\ \bibnamefont
  {Albacete}}, \bibinfo {author} {\bibfnamefont {N.}~\bibnamefont {Armesto}},
  \bibinfo {author} {\bibfnamefont {J.~G.}\ \bibnamefont {Milhano}}, \bibinfo
  {author} {\bibfnamefont {C.~A.}\ \bibnamefont {Salgado}}, \ and\ \bibinfo
  {author} {\bibfnamefont {U.~A.}\ \bibnamefont {Wiedemann}},\ }\href@noop {}
  {\bibfield  {journal} {\bibinfo  {journal} {Phys. Rev.}\ }\textbf {\bibinfo
  {volume} {D71}},\ \bibinfo {pages} {014003} (\bibinfo {year} {2005})},\
  \Eprint {http://arxiv.org/abs/hep-ph/0408216} {hep-ph/0408216} \BibitemShut
  {NoStop}%
\bibitem [{\citenamefont {Mueller}\ and\ \citenamefont
  {Triantafyllopoulos}(2002)}]{Mueller:2002zm}%
  \BibitemOpen
  \bibfield  {author} {\bibinfo {author} {\bibfnamefont {A.~H.}\ \bibnamefont
  {Mueller}}\ and\ \bibinfo {author} {\bibfnamefont {D.~N.}\ \bibnamefont
  {Triantafyllopoulos}},\ }\href@noop {} {\bibfield  {journal} {\bibinfo
  {journal} {Nucl. Phys.}\ }\textbf {\bibinfo {volume} {B640}},\ \bibinfo
  {pages} {331} (\bibinfo {year} {2002})},\ \Eprint
  {http://arxiv.org/abs/hep-ph/0205167} {hep-ph/0205167} \BibitemShut {NoStop}%
\bibitem [{\citenamefont {Munier}\ and\ \citenamefont
  {Peschanski}(2004)}]{Munier:2003sj}%
  \BibitemOpen
  \bibfield  {author} {\bibinfo {author} {\bibfnamefont {S.}~\bibnamefont
  {Munier}}\ and\ \bibinfo {author} {\bibfnamefont {R.}~\bibnamefont
  {Peschanski}},\ }\href@noop {} {\bibfield  {journal} {\bibinfo  {journal}
  {Phys. Rev.}\ }\textbf {\bibinfo {volume} {D69}},\ \bibinfo {pages} {034008}
  (\bibinfo {year} {2004})},\ \Eprint {http://arxiv.org/abs/hep-ph/0310357}
  {hep-ph/0310357} \BibitemShut {NoStop}%
\bibitem [{\citenamefont {Brodsky}\ \emph {et~al.}(1983)\citenamefont
  {Brodsky}, \citenamefont {Lepage},\ and\ \citenamefont
  {Mackenzie}}]{Brodsky:1982gc}%
  \BibitemOpen
  \bibfield  {author} {\bibinfo {author} {\bibfnamefont {S.~J.}\ \bibnamefont
  {Brodsky}}, \bibinfo {author} {\bibfnamefont {G.~P.}\ \bibnamefont {Lepage}},
  \ and\ \bibinfo {author} {\bibfnamefont {P.~B.}\ \bibnamefont {Mackenzie}},\
  }\href {\doibase 10.1103/PhysRevD.28.228} {\bibfield  {journal} {\bibinfo
  {journal} {Phys. Rev.}\ }\textbf {\bibinfo {volume} {D28}},\ \bibinfo {pages}
  {228} (\bibinfo {year} {1983})}\BibitemShut {NoStop}%
\bibitem [{\citenamefont {Ducloué}\ \emph {et~al.}(2017)\citenamefont
  {Ducloué}, \citenamefont {Lappi},\ and\ \citenamefont
  {Zhu}}]{Ducloue:2017mpb}%
  \BibitemOpen
  \bibfield  {author} {\bibinfo {author} {\bibfnamefont {B.}~\bibnamefont
  {Ducloué}}, \bibinfo {author} {\bibfnamefont {T.}~\bibnamefont {Lappi}}, \
  and\ \bibinfo {author} {\bibfnamefont {Y.}~\bibnamefont {Zhu}},\ }\href
  {\doibase 10.1103/PhysRevD.95.114007} {\bibfield  {journal} {\bibinfo
  {journal} {Phys. Rev.}\ }\textbf {\bibinfo {volume} {D95}},\ \bibinfo {pages}
  {114007} (\bibinfo {year} {2017})},\ \Eprint
  {http://arxiv.org/abs/1703.04962} {arXiv:1703.04962 [hep-ph]} \BibitemShut
  {NoStop}%
\end{thebibliography}%


\end{document}